\documentclass[12pt,leqno]{article}
\usepackage[left=1.25in,right=1.25in,top=1.25in,bottom=1.25in]{geometry}
\usepackage{amssymb,amsmath,amsthm}
\usepackage{amsfonts,bbm}
\usepackage{blindtext}
\usepackage{booktabs}
\usepackage{bm}
\usepackage{caption}
\usepackage{comment}
\usepackage{enumitem}
\usepackage{float}
\usepackage{graphicx}
\usepackage{bbm}
\graphicspath{ {../figures/} }
\usepackage{indentfirst}
\usepackage{mathtools}
\usepackage{mathrsfs}
\usepackage{multirow}
\usepackage{upgreek}
\usepackage{soul}
\usepackage{xcolor}
\usepackage[bibstyle=numeric, citestyle=authoryear, natbib=true,doi=false, url=true, backend=biber, maxbibnames=6, maxcitenames=4, uniquelist=false, uniquename=false, sorting=nyt]{biblatex}
\renewbibmacro{in:}{}
\DeclareNameAlias{default}{last-first/first-last}
\DeclareMathOperator*{\argmin}{arg\,min}

\usepackage{array}
\newcolumntype{L}[1]{>{\raggedright\arraybackslash}m{#1}}
\usepackage{rotating}
\usepackage{setspace}
\usepackage{subcaption}
\usepackage{tikz}
\usepackage{ctable}
\usepackage{hyperref}
\usepackage{cleveref}
\crefname{enumi}{}{} 
\usepackage{dcolumn}
\usepackage{float}
\usepackage{pdflscape}

\usepackage[utf8]{inputenc}
\usepackage{longtable}
\usepackage{threeparttable}

\makeatletter

\newcommand{\Rmnum}[1]{\expandafter\@slowromancap\romannumeral #1@}

\makeatother

\newtheorem{theorem}{Theorem}

\newtheorem{lemma}{Lemma}

\newtheorem{assumption}{Assumption}

\newtheorem{remark}{Remark}
\newtheorem{proposition}{Proposition}
\newtheorem{condition}{Condition}
\newtheorem{algorithm}{Algorithm}

\crefname{figure}{Figure}{Figures}
\crefname{assumption}{assumption}{assumptions}
\crefname{lemma}{Lemma}{Lemmata}
\crefname{table}{Table}{Tables}
\crefname{condition}{Condition}{Conditions}

\numberwithin{theorem}{section}
\numberwithin{corollary}{section}
\numberwithin{stheorem}{section}
\numberwithin{lemma}{section}
\numberwithin{result}{section}
\numberwithin{definition}{section}
\numberwithin{assumption}{section}
\numberwithin{sassumption}{section}
\numberwithin{remark}{section}
\numberwithin{proposition}{section}
\numberwithin{condition}{section}
\numberwithin{equation}{section}
\numberwithin{sassumption}{section}
\numberwithin{stheorem}{section}

\newcommand\numberthis{\addtocounter{equation}{1}\tag{\theequation}}
\newcommand{\X}{{\bf X}}
\newcommand{\x}{{\bm{x}}}
\newcommand{\z}{{\bm{z}}}

\newcommand{\m}{{\bm{m}}}
\newcommand{\bbeta}{\bm{\beta}}
\newcommand{\bdelta}{\bm{\delta}}
\newcommand{\E}{\mathbb{E}}
\newcommand{\R}{\mathbb{R}}

\newfloat{sfigure}{tbp}{sfig}
\DeclarePairedDelimiter{\ceil}{\lceil}{\rceil}

\begin{document}
	\title{Clustered Covariate Regression\thanks{This paper benefited from the valuable support of Brantly Callaway, Enoch Fedah, Weige Huang, Chengye Jia, Feiyu Jiang, Sonia Karami, Manikou-anoun Somda, Mark Robinson, Oleg Rytchkov, Vira Semenova, Rosnel Sessinou, Rahul Singh, and Dai Zusai. This paper is based on the second chapter of author 2's PhD. dissertation at Temple University.}}
	\date{\today }
	\author{Abdul-Nasah Soale\footnote{email: abdul-nasah.soale@case.edu, Department of Mathematics, Applied Mathematics, and Statistics, Case Western Reserve University} \and Emmanuel Selorm Tsyawo\footnote{Corresponding author, email: estsyawo@gmail.com, Department of Economics, Finance and Legal Studies, Culverhouse College of Business, University of Alabama}}
	\maketitle
\begin{abstract}\small
High covariate dimensionality is increasingly occurrent in model estimation, and existing techniques to address this issue typically require sparsity or discrete heterogeneity of the \emph{unobservable} parameter vector. However, neither restriction may be supported by economic theory in some empirical contexts, leading to severe bias and misleading inference. The clustering-based grouped parameter estimator (GPE) introduced in this paper drops both restrictions and maintains the natural one that the parameter support be bounded. GPE exhibits robust large sample properties under standard conditions and accommodates both sparse and non-sparse parameters whose support can be bounded away from zero. Extensive Monte Carlo simulations demonstrate the excellent performance of GPE in terms of bias reduction and size control compared to competing estimators. An empirical application of GPE to estimating price and income elasticities of demand for gasoline highlights its practical utility.

\vspace{0.5cm}

\noindent \textit{Keywords:} high-dimension, non-sparsity, parameter heterogeneity, clustering, approximation error

\noindent \textit{JEL classification: C01, C55} 

\end{abstract}
	
\newpage
\begin{refsection}	
\section{Introduction}
Let $y$ be a scalar outcome and $\x\in\R^p$ be a vector of covariates. Consider the standard linear model
\begin{align}
    y = \x\bbeta_o + \varepsilon,
    \label{eqn:basemodel}
\end{align}
where the random noise $\varepsilon$ satisfies $\E[\varepsilon|\x] = 0$ almost surely ($a.s.$). To estimate $\bbeta_o$ with a finite sample of size $n$, one typically uses Ordinary Least Squares (OLS) when $p < n$. However, when $p$ is large relative to $n$, OLS gives highly imprecise estimates when $p<n$ or is infeasible when $p > n$. Unfortunately, the problem of high dimensionality in model estimation is increasingly becoming common in empirical practice. 
 
A popular approach to tackling the problem of high dimensionality is to assume sparsity where only a few covariates are relevant and determine them using a regularisation technique, e.g., LASSO \citep{tibshirani1996regression}.\footnote{Sparsity-at-zero on the absolute values of parameters is the notion of sparsity considered in this paper for simplicity. While it is conceivable to view clustered structures as some form of sparsity on pairwise differences between parameters, e.g., \`a la \citet{ke2015homogeneity}, this risks belabouring the terminology.} Several sparsity-dependent estimators have been proposed in the literature, e.g., the post-LASSO estimators of \citet{belloni2014inference,belloni2017program}. The Dantzig selector \citep{candes2007dantzig,bickel2009simultaneous,chernozhukov2022biased} constitutes another important category of sparsity-dependent estimators. The rationale behind the aforementioned methods is that the penalty promotes sparsity in the estimate. Thus, a single parameter value, typically zero, can be assigned to a large subset of the covariates while parameters of the remaining covariates are ``freely" estimated. As \citet{li-muller-2021linear} points out, (approximate) sparsity in social science applications is not \textit{a priori} obvious. Moreover, sparsity is not invariant to linear reparametrisations of covariates \citep{li-muller-2021linear,giannone2021economic}.\footnote{ For example, sparsity in the rotated space of principal components, e.g, the first principal component only, corresponds to density in the un-transformed space of covariates \citep[2412]{giannone2021economic}.}

Clustering is another useful technique used to deal with high dimensionality. Discrete heterogeneity, i.e., finite number of support points in $\bbeta_o$, and well-separated groups which are commonly assumed in clustering-based estimators, improve upon the sparsity assumption, see, e.g., \citet{bonhomme2015grouped,ke2015homogeneity,sarafidis2012cross,su2016identifying,cheng2021clustering}. While the aforementioned methods improve upon sparsity-dependent methods by achieving group-wise heterogeneity, they typically assume discrete heterogeneity of $\bbeta_o$ (characterised by a finite number of latent-types) in panel data settings with a large time dimension.\footnote{To have a sense of the improvement, note that a clustering-based estimator is consistent under the discrete-non-sparse configuration with $\bbeta_o=(1,\ldots,1)'$ using at least one group of covariates while sparsity-dependent estimators fail.} Notable exceptions include \citet{bonhomme2022discretizing} which allows continuous heterogeneity and \citet{ke2015homogeneity} which considers a cross-sectional setting but imposes discrete heterogeneity. Besides some game-theoretic settings, discrete heterogeneity does not appear to enjoy much support from, e.g., economic theory \citep{hahn2010panel}. 

The grouped parameter estimator (GPE) is fundamentally different from the aforementioned clustering-based estimators. (1) In this paper, each element of $\bbeta_o$ corresponds to a different covariate. This contrasts with the literature on the above clustering-based estimators, which typically imposes clustering on a single parameter---such as an intercept or coefficients of particular covariates---that would otherwise be treated as homogeneous across units. (2) Our objective is to flexibly estimate a potentially high-dimensional regression function, in line with non-parametric approaches. In contrast to the literature on clustering-based estimators, which focuses on identifying discrete groupings and conducting inference on cluster-specific parameters, our goal is not to uncover latent classifications of units. (3) Clustering is a pure approximation tool in this paper as discrete heterogeneity is \emph{not assumed}, groups need not be separated, and the resulting approximation error is controlled via a consistent choice of the number of groups. Thus, this paper is correctly viewed as complementing clustering-based estimators by not assuming discrete heterogeneity, not requiring well-separated groups, characterising the resulting approximation error, and choosing the number of groups such that the estimator is asymptotically normal in the presence of approximation error.



This paper bears some similarity with the recent paper of \citet{chernozhukov2023inference} as both do not impose sparsity and treat the estimand as a product of two matrix-valued parameters. However, the focus of \citet{chernozhukov2023inference} differs substantially from ours. (1) Their parameter is high dimensional $(n\times p)$ matrix-valued while $\bbeta_o$ is $p\times 1$. Thus, we do not need to impose a rank condition on $\bbeta_o$. The product representation in this paper comes from assigning $\bbeta_o$ (up to an approximation bias) to $k\leq p$ groups. (2) It is not obvious how a standard textbook linear model such as \eqref{eqn:basemodel} with cross-sectional data can be cast in their proposed framework since their outcome, covariate matrix, parameter, and random noise is each $n\times p$ matrix-valued -- see \citet[p. 1309]{chernozhukov2023inference}. (3) They necessarily require $p \rightarrow \infty$ while this paper does not. (4) Unlike this paper, Assumption 4.3 of \citet{chernozhukov2023inference} requires a bounded support of $\x$ with substantial restrictions on its variation -- see \citet[Assumption 4.3(ii)]{chernozhukov2023inference}.

Since $\bbeta_o$ is never observed in practice, wrongly assumed restrictions can result in misleading inference. For this reason, there is an emerging need in the literature for methods that are agnostic of and robust to different degrees of (non)-sparsity and heterogeneity of $\bbeta_o$. The Grouped Parameter Estimator (GPE) introduced in this paper is developed using clustering as an approximation tool under the natural condition that the support of $\bbeta_o$ be bounded. Relative to existing estimators, GPE is robust to pairwise configurations of (1) sparsity, approximate sparsity, or non-sparsity on the one hand and (2) continuous, mixed, or discrete heterogeneity of $\bbeta_o$ on the other hand. Therefore, GPE eliminates the burden of assuming specific configurations of the \emph{unobservable} $\bbeta_o$, e.g., sparsity, and the attendant consequences for inference. A simple data-driven selection rule is proposed in this paper to determine the number of groups in order to control approximation error. Under mild conditions, the large sample properties of GPE are established under both fixed and increasing $p$ asymptotics. 

The remainder of the paper is laid out as follows. \Cref{Sect:The_GPE} presents the GPE estimand and estimator. The large sample properties of the grouped parameter estimator are established in  \Cref{Sect:Asymp_Theory}.  \Cref{Sect_MC_Experiments} examines the finite sample performance of GPE relative to competing estimators using simulation experiments. In \Cref{Sect:Empirical_Application}, we apply GPE in the analyses of price and income elasticities of demand for gasoline and conclude in \Cref{Sect:Conclusion}. The proofs of all theoretical results are collected in the appendix. An online appendix contains extra theoretical and simulation results.

\section{The Grouped Parameter Estimator}\label{Sect:The_GPE}

The Grouped Parameter Estimator (GPE) exploits proximity between elements of $\bbeta_o$ in order to decrease the size of the approximating model from $p$ to $k$ where $k\in \{\underline{k}_p,\ldots,\bar{k}_p\} $ and $1\leq \underline{k}_p\leq \bar{k}_p \leq p$. For instance, if $\beta_{oj}\approx \beta_{oj'}\approx \beta_{oj''}$ ``sufficiently", one can assign a common parameter to the corresponding $(j,j',j'')$'th covariates in $\x$.\footnote{Technically, ``sufficiently" as used here requires that the resulting approximation error not dominate the estimation error.} This implies that for $p\geq 2$, one can consider partitioning $\bbeta_o$ into $k$ groups. Owing to the linear index $\x\bbeta_o$ in \eqref{eqn:basemodel}, grouping parameters translates into grouping covariates by design. Covariates in each group are aggregated and assigned their respective group parameters $\{\delta_{ol}, 1\leq l\leq k\} $. In essence, the idea behind GPE is to fit \eqref{eqn:basemodel} using $k$ instead of $p$ parameters. GPE simplifies to the OLS estimator when $k=p$.

\subsection{A population-level treatment}

Let $\m_o \in \mathcal{M}_k$ be a binary matrix that represents the optimal assignment (in the $k$-means clustering sense) of elements in $\bbeta_o$ to $k$ groups, where $ \mathcal{M}_k \subset \R^{p\times k} $ denotes a space of $ p\times k $ binary matrices. The $(j,l)$'th entry of $\m_o$ equals $1$ if $\beta_{oj}$ belongs to group $l$ and zero otherwise. Since each row of $\m_o$ contains a single one and $k-1$ zeros, $\m_o'\m_o \in \R^{k\times k}$ is a diagonal matrix with group sizes on the diagonal and $\mathrm{tr}(\m_o'\m_o) = p$, where $\mathrm{tr}(A)$ denotes the trace of matrix $A$. Also, because each element of $\bbeta_o$ belongs to only one group and empty groups are not allowed, the inner product of any two columns of $\m_o$ is zero. 

For a fixed $k \in \{\underline{k}_p,\bar{k}_p\} $, $\bbeta_o$ is decomposed as
\begin{align}
    \bbeta_o = \m_o\bdelta_o + \bm{b}_k,
    \label{beta_decomp}
\end{align}
where $\bm{b}_k: = \bbeta_o - \m_o\bdelta_o $ is a non-stochastic \emph{approximation bias}. The GPE estimand, viz. $ \m_o\bdelta_o $ solves the following clustering problem:
\begin{equation}
	\min_{\m \in \mathcal{M}_k,\ \bdelta \in \Delta }\lVert \m\bdelta - \bbeta_o \rVert^2
 \label{eqn:clustering}
\end{equation}

\noindent where $\lVert \cdot \rVert$ denotes the Euclidean norm when applied to a vector and $\Delta \subset \mathbb{R}^k $. We introduce the following standard assumption in order to characterise the bound on the approximation bias $\bm{b}_k$.

\begin{assumption}
\label{Ass:Bounded_Support_B} $\bbeta \in [-C_b,C_b]^p $ for some positive constant $C_b<\infty$ that does not depend on $p$.
\end{assumption}

\noindent \Cref{Ass:Bounded_Support_B} is a standard assumption in the literature, which is used to bound the approximation bias $\bm{b}_k$ -- cf. \citet[Assumption A1(iii)]{su2016identifying}, \citet[Assumption 1(a)]{bonhomme2015grouped}, \citet[Assumption R(ii)]{cheng2021clustering}, and \citet[Lemma 2]{bonhomme2022discretizing}.

The GPE estimand $\m_o\bdelta_o$ is unique since it is invariant to the relabelling/permutation of groups. Moreover, neither $\m_o$ nor $\bdelta_o$ is of independent interest in this paper. We, nonetheless, follow the literature, e.g., \citet{bonhomme2015grouped,cheng2021clustering}, in defining $\m\in\mathcal{M}_k$ up to permutations that leave $\m\bdelta_o$ unchanged. To avoid overly dominant or empty groups that can induce ill-conditioning in the matrix $A_m:=\m'\mathbb{E}[\x'\x]\m$ for all $ \m \in \mathcal{M}_k $, we maintain the following condition throughout the paper.
\begin{condition}\label{Cond:m}
The group sizes induced by all \( \m \in \mathcal{M}_k \) are uniformly bounded above by some constant \( M \) and below by 1, uniformly in \( p \).
\end{condition}

\noindent Group sizes induced by the group assignment matrix $\m\in \mathcal{M}_k$ constitute the diagonal elements of $\m'\m$. Since the columns of $\m$ are orthogonal, group sizes are also the eigenvalues of $\m'\m$. Thus, \Cref{Cond:m} implies that $||\m|| $ is bounded for all $\m\in \mathcal{M}_k$ where $||\cdot||$ denotes the spectral norm when applied to matrices. In practice, \Cref{Cond:m} can \emph{always be guaranteed} by splitting up overly dominant groups and ensuring each group has at least one element. \Cref{Cond:m} hence requires that the number of groups increases if $p$ is increasing. This also means that the lower and upper bounds on $k$ ($\underline{k}_p$ and $\bar{k}_p$), which are $p$-indexed, ought to increase under increasing-$p$ asymptotics. To alleviate concerns of scaling as covariates are aggregated into groups, we implicitly assume throughout the paper that $\E[\x]=0$.

For any two sequences of non-negative numbers $\{a_n:n\geq 1\}$ and $\{b_n:n\geq 1\}$, $ a_n \lesssim b_n $ means $ a_n \leq cb_n $ for some finite $ c>0 $ and $a_n \asymp b_n$ means $a_n\lesssim b_n$ and $b_n\lesssim a_n$. Also, let $|\mathrm{supp}(\bbeta_o)|$ denote the number of support points of $\bbeta_o$. The following result derives the bound on $ ||\bm{b}_k|| $ in terms of $p$ and $k$.

\begin{proposition}\label{Prop:Rate_ApprxBias}
Let \Cref{Ass:Bounded_Support_B} hold, then $ ||\bm{b}_k||^2 \leq 4C_b^2M^2p/k^2 $ if $k<|\mathrm{supp}(\bbeta_o)|$ and $ ||\bm{b}_k||^2 = 0 $ if $k\geq |\mathrm{supp}(\bbeta_o)|$.
\end{proposition}

\noindent \Cref{Prop:Rate_ApprxBias} shows $ ||\bm{b}_k||$ is decreasing in $k$ but increasing in $p$. Thus, the more granular the partitions of $\bbeta_o$, the smaller the approximation error. \Cref{Prop:Rate_ApprxBias} is a univariate \emph{non-asymptotic} and \emph{non-stochastic} analogue of the stochastic bound on the $k$-means clustering approximation bias in \citet[Theorem 5.3]{graf2002rates} -- cf. \citet[Lemma 1]{bonhomme2022discretizing}.

\begin{remark}\label{Rem:ApproxBias_suppbeta}
    \Cref{Prop:Rate_ApprxBias} is a worst-case bound under continuous heterogeneity and increasing $p$ asymptotics.\footnote{$ ||\bm{b}_k||^2 \asymp k^{-2} $ if $p$ is fixed and heterogeneity is continuous.} Under discrete or mixed heterogeneity of $\bbeta_o$ with $|\mathrm{supp}(\bbeta_o) |<p$, approximation bias $\bm{b}_k$ is zero when $k\geq |\mathrm{supp}(\bbeta_o) |$.
\end{remark}

Next, we motivate GPE from a regression point of view. Define the following least squares criterion:
\begin{equation}
	Q_o(\m\bdelta): = \E[( y - \x\m\bdelta)^2]/p.
	\label{eqn:Exp_ObjFunF}
\end{equation}

\noindent Given $\m\in\mathcal{M}_k$, $\bdelta_o(\m):=(\m'\E[\x'\x]\m)^{-1}\m'\E[\x'y]$ is a deterministic function of $\m$ that minimises $ Q_o(\m\bdelta) $ with respect to $\bdelta\in\Delta$ holding $\m\in\mathcal{M}_k$ fixed. Define $ \mathcal{S}_p := \{ \uptau \in \R^p: ||\uptau||=1 \} $ as the space of vectors with unit length. The following standard assumptions are imposed in order to characterise the identification of the GPE estimand $\m_o\bdelta_o$.
\begin{assumption}
	\label{Ass:ZeroCorr_x.eps} 
 $ \mathbb{E}[\varepsilon|\x] = 0 \ a.s. $
\end{assumption}

\begin{assumption}\label{Ass:Bnds_x_eps} 
There exist constants $ B > 0 $ and $ d > 2 $ such that uniformly in $ n $, $ \mathbb{E}[(|\varepsilon|^d)|\x] \leq B \ a.s. $, $\E[|\x\uptau_p|^d] \leq B $, and $B^{-1} \leq \E[|\x\uptau_p|^2] $  for all $\uptau_p\in \mathcal{S}_p$.
\end{assumption} 

\noindent \Cref{Ass:ZeroCorr_x.eps} is a standard exogeneity condition that precludes neither non-Gaussianity nor arbitrary heteroskedasticity of the model error $\varepsilon$. \Cref{Ass:Bnds_x_eps} requires that the conditional variance be bounded, i.e., $ \mathbb{E}[\varepsilon^2|\x] \leq B \ a.s. $ The conditional moment restriction, $ \mathbb{E}[|\varepsilon|^d|\x] \leq B \ a.s. $, is fairly weak, cf. \citet[Assumption A.3]{sarafidis2012cross}. It is weaker than the sub-Gaussian tail condition imposed on $\varepsilon$ in \citet{ke2015homogeneity} but slightly stronger than Condition SM(i) of \citet{belloni2012sparse}. Let $\rho_{\mathrm{\max}}(A)$ denote the largest eigenvalue of the square matrix $A$.

\begin{remark}\label{Rem:Bnds_Eig_XX}
    \Cref{Ass:Bnds_x_eps} implies that the eigenvalues of $ \E[\x'\x] $ are bounded above and away from zero uniformly in $n$ since for any $ \uptau_p \in \mathcal{S}_p $, $ \E[|\x\uptau_p|^2] = \uptau_p'\E[\x'\x]\uptau_p $ and 
    \[ B^{-1} \leq \underline{\uptau}_p'\E[\x'\x]\underline{\uptau}_p=:\rho_{\mathrm{\min}}(\mathbb{E}[\x'\x]) \leq \uptau_p'\E[\x'\x]\uptau_p \leq \rho_{\mathrm{\max}}(\mathbb{E}[\x'\x]):= \bar{\uptau}_p'\E[\x'\x]\bar{\uptau}_p \leq B \] where $\underline{\uptau}_p \in \mathcal{S}_p  $ and $\bar{\uptau}_p \in \mathcal{S}_p $ denote the eigenvectors associated with the smallest and largest eigenvalues of $\mathbb{E}[\x'\x]$, respectively.
\end{remark}
\noindent In view of \Cref{Rem:Bnds_Eig_XX}, \Cref{Ass:Bnds_x_eps} rules out perfect or very high collinearity in $ \x $ -- cf. \citet[Condition A.2]{belloni2015some} and \citet[Condition SM(i)]{belloni2012sparse}. It also ensures that $A_m:=\m'\mathbb{E}[\x'\x]\m$ is positive-definite for all $\m\in\mathcal{M}_k$ and $k\in \{\underline{k}_p,\bar{k}_p\} $.

By the decomposition \eqref{beta_decomp}, $Q_o(\m\bdelta)$ in \eqref{eqn:Exp_ObjFunF} consists of terms involving $\bm{b}_k$ and terms not dependent on $\bm{b}_k$. Define the auxiliary criterion 
\begin{align}
    \widetilde{Q}_o(\m\bdelta):= (\m\bdelta -\m_o\bdelta_o)'\E[\x'\x](\m\bdelta -\m_o\bdelta_o)/p + \sigma^2/p
    \label{aux_lscriterion}
\end{align}
where $ \sigma^2:= \E[\varepsilon^2] $. $\widetilde{Q}_o(\m\bdelta)$ is uniquely minimised at $ \m\bdelta  = \m_o\bdelta_o $. Thus, we can characterise the identification of the GPE estimand $\m_o\bdelta_o $ up to the bias term $||\bm{b}_k||$.

\begin{theorem}\label{Theorem:Identification}
Under \Cref{Ass:ZeroCorr_x.eps,Ass:Bnds_x_eps,Ass:Bounded_Support_B}, $ Q_o(\m\bdelta) = \widetilde{Q}_o(\m\bdelta) + O(||\bm{b}_k||/\sqrt{p}) $ and $\widetilde{Q}_o(\m\bdelta)$ is uniquely minimised at $ \m_o\bdelta_o $.
\end{theorem}

\noindent By \Cref{Theorem:Identification}, the identification of $ \m_o\bdelta_o $ holds exactly if $||\bm{b}_k||=0$ or approximately if $||\bm{b}_k||/\sqrt{p}=o(1)$.

\begin{proposition}\label{Prop:Clustering} Suppose \Cref{Ass:ZeroCorr_x.eps,Ass:Bnds_x_eps} hold, then \\
(a) the minimiser of $Q_o(\m\bdelta)$ solves the parameter clustering problem: \\
$\displaystyle  \min_{m \in \mathcal{M}_k,\ \bdelta \in \Delta }||\m\bdelta - \bbeta_o||_{\E[\x'\x]}^2 $ and \\
(b) $\displaystyle ||\bm{b}_k||^2 \asymp \min_{\m \in \mathcal{M}_k,\ \bdelta \in \Delta }||\m\bdelta - \bbeta_o||_{\E[\x'\x]}^2$, \\
where $ ||\Upsilon||_W^2:= \Upsilon'W\Upsilon  $ is the weighted Euclidean norm of a vector $ \Upsilon $ for some conformable positive definite matrix $W$.
\end{proposition}

\noindent\Cref{Prop:Clustering} has two important takeaways: (1) the GPE estimand solves a (weighted) univariate $k$-means parameter clustering problem, and (2) the approximation biases from the weighted and unweighted clustering problems are of the same order. The latter is exploited in this paper to lower the computational cost.

\subsection{Computation}\label{SubSect:Computation}
Let $\{(\x_i,y_i), 1\leq i \leq n \}$ be a random sample of $(\x,y)$ and $\E_n[\cdot]$ denote the sample mean. Define $ Q_n(\m\bdelta) :=\E_n[( y_i - \x_i\m\bdelta)^2]/p$ as the sample version of \eqref{eqn:Exp_ObjFunF}.  GPE is given by $ \widehat{\bbeta}_n := \hat{\m}_n\hat{\bdelta}_n $, where
\begin{equation}\label{eqn:CCR_Lin_Estimator}
	\hat{\m}_n\hat{\bdelta}_n = \argmin_{\{\m\bdelta :\ \m \in \mathcal{M}_k, \bdelta \in \Delta\}} Q_n(\m\bdelta). 
\end{equation}

Unlike in the population, a two-step estimation of $\hat{\m}_n$ and $\hat{\bdelta}_n$ cannot be followed in finite samples because (1) $ \bbeta_o $ is unknown; (2) $\hat{\m}_n$ and $\hat{\bdelta}_n$ are interdependent and cannot be computed separately; and (3) the computational cost of an exhaustive search is prohibitive, especially for large $p$. In light of the foregoing, we propose the following iterative algorithm for GPE. 

\begin{algorithm}\label{Alg:Lloyd_GPE}
	\begin{enumerate}[label=(\alph*)]\quad\vspace{-.2cm}
		\item Fix $k \leq \min\{p,n-2\}$, initialise $ \widehat{\bbeta}_n^{(0)}$, compute $ \{\hat{\m}^{(0)}, \hat{\bdelta}_n^{(0)}\}$, and set counter $ s = 1 $.
		\item\label{Alg:Step:Update_M} For each $ j \in \{1,\ldots,p\} $, compute $ \hat{\beta}_{n,j}^{(s)} $ and update the $ j $'th row of $ \hat{\m}^{(s)} $.
		\item\label{Alg:Step:Update_Delta} Compute $\displaystyle \hat{\bdelta}_n^{(s)} = \argmin_{\bdelta \in \Delta} Q_n(\hat{\m}^{(s)}\bdelta)   $.
		\item Check for numerical convergence, else set $ s \leftarrow s+1 $ and return to Step \ref{Alg:Step:Update_M}.
	\end{enumerate}
\end{algorithm}

$k\leq (n-2)$ ensures there is at least one degree of freedom for estimation. $\hat{\beta}_{n,j}^{(s)} $ in Step \ref{Alg:Step:Update_M} is the OLS slope estimate of $ y-\x_{-j}\widehat{\bbeta}_{n,-j}^{(s-1)} $ regressed on the $j$'th covariate $ x_j $ where $\widehat{\bbeta}_{n,-j}^{(s-1)}:=(\hat{\beta}_{n,1}^{(s)},\ldots,\hat{\beta}_{n,j-1}^{(s)},\hat{\beta}_{n,j+1}^{(s-1)},\ldots,\hat{\beta}_{n,p}^{(s-1)})'$ and $\x_{-j}'$ is a $(p-1)\times 1$ vector formed from $\x$ by removing the $j$'th covariate. Updating the $ j $'th row of $ \hat{\m}^{(s)} $ involves assigning $ \hat{\beta}_{n,j}^{(s)} $ to the nearest element (in absolute value) in $ \hat{\bdelta}_n^{(s-1)} $, and  updating the group mean. This is the step that differs from, e.g., \citet[Algorithm 1]{bonhomme2015grouped}; it is a faster yet (asymptotically) equivalent step thanks to \Cref{Prop:Clustering}(b). Assigning $ \hat{\beta}_{n,j}^{(s)} $ to the nearest group in Step \Cref{Alg:Step:Update_M} achieves the same goal up to the same order of approximation bias as obtaining the group assignment of $x_j$ which minimises the objective function. The latter involves computing the objective function $p-1$ times while the former only requires computing $k-1$ scalar absolute differences.

\begin{remark}\label{Rem:Alg_Weighted_Clus}
    \Cref{Theorem:Identification} shows that the objective function is a weighted within-group sum of squared deviations up to an approximation bias term. One observes from \Cref{Prop:Clustering} that unweighted group assignment saves much computational cost as it avoids recomputing the objective function at the group assignment step of \Cref{Alg:Lloyd_GPE}. Unlike typical clustering-based regression algorithms, e.g., \citet[Algorithm 1]{bonhomme2015grouped}, $\x$ can be non-binary, and panel data are not \emph{sine qua non}.
\end{remark}

It is well known in the literature that the solution to iterative algorithms like \Cref{Alg:Lloyd_GPE} is sensitive to starting values. Hence, it is essential to have reliable starting schemes. 
Our starting values are obtained from subgroup analyses based on the regression model as proposed by \citet{ma2017concave}, except that in our case, we apply the concave penalty function to the pairwise differences of the model parameters instead of the intercepts. We proceed to solve the constrained optimisation problem using a modified alternating direction method of multipliers (ADMM) algorithm of \citet{boyd2011distributed} -- see \Cref{SubSect:Starting_Schemes} for details. Following the literature, e.g., \citet{bonhomme2022discretizing}, our asymptotic theory focuses on the global minimum (indexed by $k$) while abstracting away from optimisation error.

\begin{remark}
A researcher may sometimes be interested in the partial effects of a finite set of covariates and may not want to group them. Such a restriction is easily incorporated in $ \m $ by adding columns corresponding to singleton covariate groups whose parameters are updated in Step \Cref{Alg:Step:Update_Delta} while group assignments are held fixed in Step \Cref{Alg:Step:Update_M} across iterations. The intercept is handled in this way.
\end{remark}

\section{Large Sample Properties}\label{Sect:Asymp_Theory}
\subsection{Convergence of $\hat{\m}_n$}
We consider a sequence of models indexed by $ n $ with observed data $ \{(\x_i,y_i), 1\leq i\leq n \} $, $ y_i = y_{i,n} $, $ \x_i = \x_{i,n} $, and $p=p_n$. The large sample properties of GPE are developed under the following additional assumptions.
\begin{assumption}
	\label{Ass:Sampling} Observed data $ \{(\x_i,y_i), 1\leq i\leq n\}$  are independent and identically distributed $(iid)$ for each $n$.
\end{assumption}

\begin{assumption}
    \label{Ass:Rate_p_n} $ p\log(p) = o(n) $.
\end{assumption}

\noindent The dependence of the sampling process on $ n $ in \Cref{Ass:Sampling} allows $ p $ to grow with $ n $ while allowing for fixed-$p$ asymptotics as well. \Cref{Ass:Rate_p_n} characterises the rate at which $p$ is allowed to grow with $n$. As, \Cref{Ass:Rate_p_n} is an asymptotic rate condition, it does not rule out $p>n$ in finite samples.

\begin{remark}\label{Rem:p_n_growthrate}
    The rate condition in \Cref{Ass:Rate_p_n} required for GPE is more restrictive relative to LASSO's $\log(p)=o(n^{1/3})$,  which allows up to an exponential growth of $p$ in $n$ -- see e.g., \citet{belloni2012sparse,belloni2014inference,farrell2015robust}. Feature-screening methods, e.g., \citet{li2012feature,shao2014martingale} used in a first stage to weed out irrelevant covariates -- as do sparsity-dependent methods -- can complement GPE in handling ultra-high dimensional models.\footnote{For concerns of space and scope, such a pursuit is left for future work.}
\end{remark}

\noindent By \Cref{Rem:p_n_growthrate}, the slower growth rate of $p$ in $n$ allowed by GPE, unlike (post)-LASSO estimators, is the price to pay for not explicitly exploiting sparsity (if it holds) which eliminates irrelevant covariates in a first step. 

The following provides a characterisation of the convergence rate of $ ||\hat{\m}_n - \m_o|| $. 

\begin{theorem}\label{Theorem:Consis_Rate_M}
Suppose \Cref{Ass:Bounded_Support_B,Ass:ZeroCorr_x.eps,Ass:Sampling,Ass:Bnds_x_eps} hold, then $ ||\hat{\m}_n-\m_o|| = O_p(||\bm{b}_k||^2/p) + O_p(n^{-1})$.
\end{theorem}

\noindent From \Cref{Theorem:Consis_Rate_M} above, one observes that the convergence of $\hat{\m}_n$ to $\m_o$ depends on $||\bm{b}_k||$, $n$, and $p$.

\subsection{Consistency and Asymptotic Normality}
For a given $ \m \in \mathcal{M}_k $, the corresponding estimator is 
\begin{equation*}
\hat{\bdelta}_n(\m) = (\m'\E_n[\x_i'\x_i]\m)^{-1}\m'\E_n[\x_i'y_i].
\end{equation*}
The closed-form expression of $\hat{\bdelta}_n(\m)$ is useful in concentrating out $\bdelta$ from the objective function $Q_n(\m\bdelta)$.
\noindent Define $ \hat{A}_m:= \m'\E_n[\x_i'\x_i]\m $ as the sample analogue of $ A_m$. A consistency condition is imposed on $ \hat{A}_{m_o} $ in the following assumption.
\begin{assumption}\label{Ass:Convergence_Am}
	$ ||\hat{A}_{m_o} - A_{m_o}|| = o_p(1) $.
\end{assumption} 

\noindent \Cref{Ass:Convergence_Am} is a high-level assumption  -- cf. \citet{tropp2012user}, \citet[Corrolary 4.1]{chen2015optimal}, and \citet[Lemma 6.2]{belloni2015some}. \Cref{Lem:Consistency_Am} in the Online Appendix verifies \Cref{Ass:Convergence_Am} under sufficient conditions that allow an unbounded support of $\x$ and possibly increasing $p$. Also, observe that for a given $k$, \Cref{Ass:Convergence_Am} is imposed on $\m_o$ and not on the entirety of $\mathcal{M}_k$.

Using the decomposition of $\widehat{\bbeta}_n - \bbeta_o$ into the estimation error $ \widehat{\mathcal{A}}_k $ and the approximation error $\widehat{\mathcal{B}}_k$, namely $\widehat{\bbeta}_n - \bbeta_o = \widehat{\mathcal{A}}_k + \widehat{\mathcal{B}}_k $ where
\begin{align*}
    \widehat{\mathcal{A}}_k: = \m_o\hat{A}_{m_o}^{-1}\m_o'\E_n[\x_i'\varepsilon_i]\ \text{ and } \ 
    \widehat{\mathcal{B}}_k: = \widehat{\bbeta}_n - \m_o\hat{\bdelta}_n(\m_o) + (\mathrm{I}_p + \m_o\hat{A}_{m_o}^{-1}\m_o'\E_n[\x_i'\x_i])\bm{b}_k,
\end{align*}

\noindent we characterise the convergence rate of GPE $\widehat{\bbeta}_n$ and its asymptotic normality for any $ \uptau_p \in \mathcal{S}_p $. Define $\sigma_\theta: = (\uptau_p'\m_oA_{m_o}^{-1}\m_o'\E[\x'\x\varepsilon^2]\m_oA_{m_o}^{-1}\m_o'\uptau_p)^{1/2}$ where the dependence of $\sigma_\theta$ on $\uptau_p\in\mathcal{S}_p$ is suppressed for notational ease.

\begin{theorem}[Consistency and Asymptotic Normality]\label{Theorem:Consis_AsympNorm}
Suppose \Cref{Ass:Bnds_x_eps,Ass:Bnds_x_eps,Ass:Bounded_Support_B,Ass:Convergence_Am,Ass:Sampling,Ass:ZeroCorr_x.eps,Ass:Rate_p_n} hold, then for all $\uptau_p \in \mathcal{S}_p $, (a) $\uptau_p'\widehat{\mathcal{B}}_k = O_p(||\bm{b}_k||) + o_p(n^{-1/2}) $; (b) $\uptau_p'(\widehat{\bbeta}_n - \bbeta_o - \widehat{\mathcal{B}}_k) = O_p(n^{-1/2}) $; and (c) $\sigma_\theta^{-1} \sqrt{n}\uptau_p'(\widehat{\bbeta}_n - \bbeta_o - \widehat{\mathcal{B}}_k) \xrightarrow{d} \mathcal{N}(0,1)$.
\end{theorem}

\subsection{Selection rule}
The underlying principle of GPE is that \eqref{eqn:basemodel} be estimable with fewer than $p$ parameters and thus conserve degrees of freedom. Although \Cref{Prop:Rate_ApprxBias} could be used to specify a selection rule of $k$ that ensures that the approximation error is controlled, such a choice rule would lack flexibility and be poorly adapted to the configuration of $\bbeta_o$. For instance, under discrete or mixed heterogeneity of $\bbeta_o$, a small $k$ should suffice to control approximation error. Further, simulations under a continuous-non-sparse configuration of $\bbeta_o$ in \Cref{Sect_MC_Experiments} suggest that a small number of groups $k$ suffices to control approximation error. Therefore, instead of using a large value of $k$ to drive approximation bias almost to zero at the cost of imprecise estimates and poor inference, we choose the smallest $k$ in a data-driven way to control approximation error -- see \citet[p. 630]{bonhomme2022discretizing} for a selection rule based on a similar concept.

We select the number of groups per the rule 
$$
\hat{k}_n:= \min_{k\geq 1} \{ k: n\widehat{\varphi}_n(k)\leq C  \} 
$$ where  
$$
\widehat{\varphi}_n(k):= \frac{\mathbb{E}_n[(y_i - \x_i\widehat{\bbeta}_n^k)^2] - \mathbb{E}_n[(y_i - \x_i\widehat{\bbeta}_n^{k+1})^2]}{\mathbb{E}_n[(y_i - \x_i\widehat{\bbeta}_n^{k+1})^2]},
$$ $\widehat{\bbeta}_n^k$ is GPE with $k$ groups (for notational emphasis), and $ C > 0 $ is a user-defined constant.\footnote{We recommend using $C=2.7$ -- see \Cref{SubSect:Calibration_gamma} for details.} As $\mathbb{E}_n[(y_i - \x_i\widehat{\bbeta}_n^k)^2]$ is non-increasing in $k$, $\widehat{\varphi}_n(k)$ is non-negative.\footnote{To see why, it suffices to create a $(k+1)$'th singleton group using the covariate whose parameter estimate generates the largest within-group absolute deviation.} Also, $\hat{k}_n$ is scale-invariant since $\widehat{\varphi}_n(k)$ is a ratio. The above selection rule bears an interesting similarity to model selection criteria such as the Bayesian Information Criterion (BIC) -- see, e.g., \citet[p. 279]{sarafidis2015partially}. Unlike the BIC which has a penalty term to avoid over-fitting, setting $C$ to a constant and choosing $\hat{k}_n$ as the minimum admissible $k$ guards against over-fitting in our case. The following result shows that this simple selection rule is consistent, i.e., it selects $k$ such that the approximation error $\widehat{\mathcal{B}}_k$ in \Cref{Theorem:Consis_AsympNorm} does not dominate the estimation error $\widehat{\mathcal{A}}_k$. This is similar in spirit to the choice of bandwidth in non-parametric methods. In view of \Cref{Prop:Rate_ApprxBias} and \Cref{Theorem:Consis_AsympNorm}(a), let $\{a_k: k\geq 1\}$ be a decreasing sequence of positive numbers such that $||\widehat{\mathcal{B}}_k|| = O_p(n^{-1/a_k}) + o_p(n^{-1/2})$.
    
\begin{theorem}\label{Theorem:Select_k}
Suppose \Cref{Ass:Bnds_x_eps,Ass:Bnds_x_eps,Ass:ZeroCorr_x.eps,Ass:Bounded_Support_B,Ass:Sampling,Ass:Rate_p_n} hold, then (a) $ n\widehat{\varphi}_n(k) = O_p(1) $ if $a_k\leq 2$; and (b) $\displaystyle n\widehat{\varphi}_n(k) \rightarrow \infty $ as $n\rightarrow \infty$ otherwise. 
\end{theorem}

\noindent An important implication of \Cref{Theorem:Select_k} is that $ n\widehat{\varphi}_n(k) = O_p(1) $ implies $||\widehat{\mathcal{B}}_k|| = O_p(n^{-1/2})$. $k=\hat{k}_n$ is chosen to ensure inference is not contaminated by the approximation error. Thus, $C > 0$ is set to a suitable constant under $a_k<2$, so that the approximation error is dominated by the estimation error, i.e., $||\widehat{\mathcal{B}}_k|| = o_p(n^{-1/2})$.\footnote{$C=2.7$ is the $90$’th percentile of the $\chi_1^2$ limiting distribution of $n\widehat{\varphi}_n(k)$ under two technical conditions and $||\widehat{\mathcal{B}}_k|| = o_p(n^{-1/2})$ -- see \Cref{SubSect:Calibration_gamma}.}

It is instructive to draw parallels between $k$ and its approximate analogue in the post-LASSO estimator of \citet{belloni2014inference} (pLASSO hereafter), namely the sparsity index $s$, which is the bound on the number of covariates in a sparse model whose corresponding elements in $\bbeta_o$ are non-zero.

\begin{remark}\label{Rem:k_growthrate}
 In addition to requiring that zero be an atom of $\bbeta_o$, pLASSO imposes the rate condition $ s^2(\log (p \vee n))^2/n = o(1) $. As GPE remains consistent even when $\mathrm{supp}(\bbeta_o)$ is bounded away from zero and $\hat{k}_n < |\mathrm{supp}(\bbeta_o)|$ is allowed as long as approximation error is controlled, pLASSO is more restrictive in this sense. Unlike, e.g., pLASSO -- see \citet[Condition ASTE (ii)]{belloni2014inference} -- which attains valid inference by assuming the approximation error is bounded, an important element to GPE, in light of \Cref{Prop:Rate_ApprxBias} and \Cref{Theorem:Select_k}, is that $k$ is chosen to control approximation error.
\end{remark}

\noindent The choice of $k$ is user-determined in a data-driven way whereas the pLASSO rate condition on $s$ is an assumption imposed on the unobservable $\bbeta_o$. \Cref{Rem:p_n_growthrate,Rem:k_growthrate} highlight the complementarity and trade-off between sparsity-dependent estimators and GPE in handling high covariate dimensionality. The consistent choice $k_n$ controls the approximation error and avoids \emph{unverifiable} assumptions on the approximation error. In contrast, panel-data clustering-based methods, e.g., \citet{bonhomme2015grouped,ke2015homogeneity,sarafidis2015partially} assume zero approximation error with well-separated groups subject to correctly choosing $k$, and regularisation-based methods such as \citet{belloni2014inference,chernozhukov2023inference} impose order conditions on the approximation error.

\section{Simulation Experiment}\label{Sect_MC_Experiments}
This section focuses on the finite sample performance of GPE using simulated data. We compare the bias and rejection rates of competing estimators under Continuous-non-Sparse (CnS), Continuous-approximately-Sparse (CaS), and Discrete-Sparse (D-S) configurations of $ \bbeta_o $.\footnote{See \Cref{App:Sect:Taxonomy_Config} of the Online Appendix for details on the possible support configurations of $ \bbeta_o $.} Estimators considered include (1) the proposed GPE; (2) the post-LASSO estimator of \citet{belloni2014inference} (pLASSO); (3) post-Generalised Dantzig Selector (GDS) estimator of \citet{chernozhukov2022biased} (pGDS); (4) OLS; (5) an ``oracle" estimator which is OLS on sample size $3n$ (Orac.OLS); and an infeasible GPE that uses $\bbeta_o$ as the starting value in \Cref{Alg:Lloyd_GPE} (Orac.GPE).\footnote{pLASSO and pGDS estimators are based on tuning parameters recommended in \citet{belloni2014inference,chernozhukov2022biased}, respectively.}$^,$\footnote{Simulations in an earlier draft suggest that pLASSO and pGDS with the tuning parameter selected via 10-fold cross-validation do not provide meaningful inference in any setting considered.} The infeasible Orac.OLS provides a useful benchmark for examining the gains in bias reduction and inference using a feasible estimator relative to OLS if the sample size were thrice as large.\footnote{In the context of experimental data, a practical consideration is whether a researcher would use, e.g., GPE or pLASSO to estimate a high-dimensional model or incur the added cost of collecting twice as much data in order to apply OLS.} The inclusion of Orac.GPE is helpful in gauging the performance of \Cref{Alg:Lloyd_GPE} and the proposed selection rule of $\hat{k}_n$. In comparing it to GPE, one is also able to examine the reliability of the starting scheme.

$ \theta_o:= p^{-1/2}\sum_{j=1}^{p}\beta_{oj} $ is the scalar-valued parameter of interest in this simulation exercise. The comparison of competing estimators is based on the following metrics: (1) mean bias (MnB) namely $ ||\widehat{\bbeta}_n - \bbeta_o||/\sqrt{p} $ averaged over all simulated samples; (2) median absolute deviation 
(MAD) of $ \hat{\theta}_n $; (3) root mean squared error (RMSE) of $ \hat{\theta}_n $; (4) the rejection rate (Rej.) of a 5\%-level $t$-test $ \mathbb{H}_o: \theta - \theta_o = 0 $; and (5) the median size of the approximating model across simulated samples ($ \mathrm{med}(\hat{k}_n) $). In computing the standard errors of $\hat{\theta}_n$ and rejection rates of pLASSO and pGDS, parameters shrunk to zero in the first step are treated as constants.

Define $ \Phi^{-1}(\tau_j) $ as the $ \tau_j $'th quantile of the standard normal distribution where $ \tau_j = 0.9(j-1)/(p-1) + 0.05 $. \eqref{eqn:basemodel} is the data-generating process with $ \varepsilon=U\sqrt{(1+x_1^2)/2}$ and the following specifications of $\bbeta_o$:
\begin{enumerate}
	\item[] DGP CnS: $ \beta_{oj} = 2 + 4(j-1)/(p-1) \in [2,4] $;
        \item[] DGP $\mathrm{CaS_1}$: $\beta_{oj} = \Phi^{-1}(\tau_j) \in [-1.645, 1.645] $; 
	\item[] DGP $\mathrm{CaS_2}$: $ \beta_{oj} = 0.7^{j-1} \in (0,1] $; and
	\item[] DGP D-S$_1$: $\beta_{oj} = \mathrm{I}(j\leq 5) \in \{0,1\} $.
\end{enumerate}

\noindent The vector of covariates is generated as $\x \sim \mathcal{N}(\bm 0, \bm \Sigma)$, with $\bm \Sigma_{jj'} = 0.5^{|j-j'|}$. $ U \sim (\chi_1^2-1)/\sqrt{2} $ in DGPs CnS and $\mathrm{CaS_1}$ while $ U \sim \mathcal{N}(0,1) $ in DGPs $\mathrm{CaS_2}$ and D-S$_1$. The configuration in DGP CnS is dense; heterogeneity is continuous and the support is bounded away from zero. DGP $\mathrm{CaS_1}$ introduces approximate sparsity by allowing zero to fall within the bounds of $\mathrm{supp}(\bbeta_o)$. DGPs $\mathrm{CaS_2}$ and D-S$_1$ are adapted from \citet{belloni2012sparse} in order to compare GPE to pLASSO under approximately sparse and sparse configurations, respectively. Save DGP D-S$_1$ where heterogeneity is discrete, all other DGPs have continuous heterogeneity.\footnote{See \Cref{Sect:Additional_Details} for simulation results on other discrete and mixed heterogeneity configurations.} Heteroskedasticity is imposed in all DGPs at pairs $ (n,p) \in\{100,400\} \times \{75,150\} $. As OLS is not feasible when $ p>n $, OLS results are not reported for $(n,p)=(100,150)$. All results are based on 1000 simulated samples.  

\begin{table}[!htbp]
	\setlength{\tabcolsep}{4pt}
	\caption{DGP CnS}
	\footnotesize
		\centering
	\begin{minipage}{.5\linewidth}
	\label{Tab:DGP_CnS}
	\begin{tabular}{lccccc}	
		\toprule
		& \multicolumn{5}{c}{$ p = 75 $}  \\ \cmidrule(rl){2-6}
		$ n=100 $ & MnB & MAD & RMSE & Rej. & $ \mathrm{med}(\hat{k}_n) $ \\
		\midrule
GPE   &0.647 &0.240 &0.355 &0.056 & 3    \\ 
pLASSO &1.662  &2.302  &2.496  &0.947  &64     \\ 
pGDS  &1.839 &2.615 &2.719 &0.958 &62    \\ 
OLS   &0.259 &0.070 &0.118 &0.000 &75    \\ 
Orac.OLS &0.085    &0.028    &0.039    &0.019    &75       \\ 
Orac.GPE &0.136    &0.081    &0.123    &0.051    & 7       \\ 

		\midrule
		$ n=400 $ &  &  &  &  &  \\
		\midrule
GPE   &0.106 &0.029 &0.046 &0.065 &12    \\ 
pLASSO &0.071  &0.023  &0.033  &0.030  &75     \\ 
pGDS  &0.071 &0.023 &0.033 &0.030 &75    \\ 
OLS   &0.071 &0.023 &0.033 &0.030 &75    \\ 
Orac.OLS &0.038    &0.012    &0.018    &0.039    &75       \\ 
Orac.GPE &0.127    &0.036    &0.057    &0.050    & 7       \\ 

		\bottomrule
	\end{tabular}
\end{minipage}%
\begin{minipage}{.5\linewidth}
	\begin{tabular}{rccccc}
		\toprule
		& \multicolumn{5}{c}{$ p = 150 $}  \\ \cmidrule(rl){2-6}
		& MnB & MAD & RMSE & Rej. & $ \mathrm{med}(\hat{k}_n) $ \\
		\midrule
 &0.652 &0.402 &0.630 &0.054 &  4   \\ 
 &3.072  &9.635  &9.761  &0.998  & 81    \\ 
 &3.505 &9.760 &9.868 &0.933 & 85   \\ 
 & --  & --  & --  & --  & --  \\ 
 &0.104    &0.031    &0.048    &0.004    &150      \\ 
 &0.094    &0.076    &0.123    &0.046    &  9      \\

		\midrule
		&  &  &  &  &  \\
		\midrule
 &0.229 &0.053 &0.083 &0.057 &  9   \\ 
 &0.081  &0.025  &0.039  &0.007  &150    \\ 
 &0.133 &0.034 &0.112 &0.118 &150   \\ 
 &0.081 &0.025 &0.037 &0.006 &150   \\ 
 &0.040    &0.012    &0.018    &0.038    &150      \\ 
 &0.058    &0.027    &0.043    &0.048    & 18      \\ 
		\bottomrule
		\end{tabular}
	\end{minipage}
\end{table}

\begin{table}[!htbp]
	\setlength{\tabcolsep}{4pt}
	\caption{DGP $\mathrm{CaS_1}$}
	\footnotesize
	\centering
	\begin{minipage}{.5\linewidth}
		\label{Tab:DGP_CaS1}
		\begin{tabular}{lccccc}
			\toprule
			& \multicolumn{5}{c}{$ p = 75 $}  \\ \cmidrule(rl){2-6}
			$ n=100 $ & MnB & MAD & RMSE & Rej. & $ \mathrm{med}(\hat{k}_n) $ \\
			\midrule
     GPE   &0.825 &0.259 &0.388 &0.054 & 3    \\ 
pLASSO &0.525  &0.176  &0.255  &0.159  &36     \\ 
pGDS  &0.492 &0.146 &0.222 &0.076 &43    \\ 
OLS   &0.259 &0.070 &0.118 &0.000 &75    \\ 
Orac.OLS &0.085    &0.028    &0.039    &0.019    &75       \\ 
Orac.GPE &0.776    &0.396    &0.591    &0.042    & 5       \\ 
                
			\midrule
			$ n=400 $ &  &  &  &  &  \\
			\midrule
			
    GPE   &0.107 &0.026 &0.039 &0.042 & 19    \\ 
pLASSO &0.118  &0.034  &0.051  &0.147  &60     \\ 
pGDS  &0.099 &0.025 &0.038 &0.054 &69    \\ 
OLS   &0.071 &0.023 &0.033 &0.030 &75    \\ 
Orac.OLS &0.038    &0.012    &0.018    &0.039    &75       \\ 
Orac.GPE &0.673    &0.182    &0.271    &0.050    & 7       \\ 

			\bottomrule
		\end{tabular}
	\end{minipage}%
	\begin{minipage}{.5\linewidth}
		\begin{tabular}{rccccc}
			\toprule
			& \multicolumn{5}{c}{$ p = 150 $}  \\ \cmidrule(rl){2-6}
			& MnB & MAD & RMSE & Rej. & $ \mathrm{med}(\hat{k}_n) $ \\
			\midrule
 &0.874 &0.474 &0.683 &0.054 &  3   \\ 
 &0.738  &0.321  &0.462  &0.151  & 52    \\ 
 &0.748 &0.293 &0.450 &0.033 & 66   \\ 
 & --  & --  & -- & -- & -- \\ 
 &0.104    &0.031    &0.048    &0.004    &150      \\ 
 &0.768    &0.687    &0.941    &0.063    &  4      \\

			\midrule
			&  &  &  &  &  \\
			\midrule
			
 &0.365 &0.116 &0.174 &0.053 &  7   \\ 
 &0.194  &0.055  &0.083  &0.130  &110    \\ 
 &0.222 &0.050 &0.077 &0.097 &115   \\ 
 &0.081 &0.025 &0.037 &0.006 &150   \\ 
 &0.040    &0.012    &0.018    &0.038    &150      \\ 
 &0.724    &0.288    &0.431    &0.053    &  4      \\ 
			\bottomrule
		\end{tabular}
	\end{minipage}
\end{table}

\begin{table}[!htbp]
	\setlength{\tabcolsep}{4pt}
	\caption{DGP $\mathrm{CaS}_2$}
	\footnotesize
	\centering
	\begin{minipage}{.5\linewidth}
		\label{Tab:DGP_CaS2}
		\begin{tabular}{lccccc}
			\toprule
			& \multicolumn{5}{c}{$ p = 75 $}  \\ \cmidrule(rl){2-6}
			$ n=100 $ & MnB & MAD & RMSE & Rej. & $ \mathrm{med}(\hat{k}_n) $ \\
			\midrule
	GPE   &0.089 &0.054 &0.079 &0.047 & 2    \\ 
pLASSO &0.069  &0.087  &0.095  &0.912  & 3     \\ 
pGDS  &0.052 &0.037 &0.048 &0.397 & 5    \\ 
OLS   &0.263 &0.080 &0.121 &0.001 &75    \\ 
Orac.OLS &0.086    &0.027    &0.040    &0.029    &75       \\ 
Orac.GPE &0.045    &0.043    &0.061    &0.034    & 7       \\ 

			\midrule
			$ n=400 $ &  &  &  &  &  \\
			\midrule
GPE   &0.046 &0.020 &0.031 &0.054 & 4    \\ 
pLASSO &0.035  &0.043  &0.046  &0.909  & 5     \\ 
pGDS  &0.027 &0.016 &0.022 &0.325 & 8    \\ 
OLS   &0.071 &0.021 &0.033 &0.024 &75    \\ 
Orac.OLS &0.038    &0.012    &0.018    &0.050    &75       \\ 
Orac.GPE &0.024    &0.019    &0.030    &0.052    &9       \\ 

			\bottomrule
		\end{tabular}
	\end{minipage}%
	\begin{minipage}{.5\linewidth}
		\begin{tabular}{rccccc}
			\toprule
			& \multicolumn{5}{c}{$ p = 150 $}  \\ \cmidrule(rl){2-6}
			& MnB & MAD & RMSE & Rej. & $ \mathrm{med}(\hat{k}_n) $ \\
			\midrule

 &0.091 &0.066 &0.098 &0.056 &  2   \\ 
 &0.050  &0.064  &0.069  &0.903  &  3    \\ 
 &0.037 &0.027 &0.036 &0.434 &  5   \\ 
 & --  & --  & --  & --  & --  \\ 
 &0.105    &0.034    &0.049    &0.007    &150      \\ 
 &0.032    &0.044    &0.065    &0.050    &  7      \\ 

			\midrule
			&  &  &  &  &  \\
			\midrule

 &0.024 &0.021 &0.031 &0.055 &  5   \\ 
 &0.025  &0.031  &0.033  &0.918  &  5    \\ 
 &0.019 &0.012 &0.016 &0.331 &  7   \\ 
 &0.081 &0.027 &0.039 &0.019 &150   \\ 
 &0.040    &0.011    &0.017    &0.027    &150      \\ 
 &0.017    &0.021    &0.031    &0.050    & 9      \\ 

			\bottomrule
		\end{tabular}
	\end{minipage}
\end{table}

\begin{table}[!htbp]
	\setlength{\tabcolsep}{4pt}
	\caption{DGP D-S$_1$}
	\footnotesize
	\centering
	\begin{minipage}{.5\linewidth}
		\label{Tab:DGP_D_S1}
		\begin{tabular}{lccccc}
			\toprule
			& \multicolumn{5}{c}{$ p = 75 $}  \\ \cmidrule(rl){2-6}
			$ n=100 $ & MnB & MAD & RMSE & Rej. & $ \mathrm{med}(\hat{k}_n) $ \\
			\midrule

GPE   &0.014 &0.040 &0.059 &0.047 & 2    \\ 
pLASSO &0.035  &0.016  &0.025  &0.077  & 5     \\ 
pGDS  &0.044 &0.021 &0.035 &0.086 & 8    \\ 
OLS   &0.263 &0.080 &0.121 &0.001 &75    \\ 
Orac.OLS &0.086    &0.027    &0.040    &0.029    &75       \\ 
Orac.GPE &0.014    &0.040    &0.059    &0.043    & 2       \\ 
                
			\midrule
			$ n=400 $ &  &  &  &  &  \\
			\midrule
			
GPE   &0.007 &0.018 &0.029 &0.053 & 2    \\ 
pLASSO &0.016  &0.007  &0.010  &0.046  & 5     \\ 
pGDS  &0.023 &0.011 &0.018 &0.046 & 11    \\ 
OLS   &0.071 &0.021 &0.033 &0.024 &75    \\ 
Orac.OLS &0.038    &0.012    &0.018    &0.050    &75       \\ 
Orac.GPE &0.007    &0.019    &0.029    &0.053    & 2       \\ 

			\bottomrule
		\end{tabular}
	\end{minipage}%
	\begin{minipage}{.5\linewidth}
		\begin{tabular}{rccccc}
			\toprule
			& \multicolumn{5}{c}{$ p = 150 $}  \\ \cmidrule(rl){2-6}
			& MnB & MAD & RMSE & Rej. & $ \mathrm{med}(\hat{k}_n) $ \\
			\midrule
			
 &0.203 &0.095 &0.146 &0.052 &  2   \\ 
 &0.025  &0.011  &0.018  &0.080  &  5    \\ 
 &0.035 &0.017 &0.030 &0.068 &  10   \\ 
 & --  & --  & --  & --  & -- \\ 
 &0.105    &0.034    &0.049    &0.007    &150      \\ 
 &0.009    &0.042    &0.060    &0.048    &  2      \\ 

			\midrule
			&  &  &  &  &  \\
			\midrule
			
 &0.004 &0.021 &0.030 &0.057 &  2   \\ 
 &0.011  &0.005  &0.007  &0.040  &  5    \\ 
 &0.020 &0.010 &0.016 &0.060 & 17   \\ 
 &0.081 &0.027 &0.039 &0.019 &150   \\ 
 &0.040    &0.011    &0.017    &0.027    &150      \\ 
 &0.005    &0.021    &0.030    &0.057    &  2      \\ 

			\bottomrule
		\end{tabular}
	\end{minipage}
\end{table}

\begin{figure}[!htbp]
	\centering
	\caption{DGPs Cns \& $\mathrm{CaS_1}$ at $(n,p)=(100,75)$ }
	\begin{minipage}{.5\linewidth}
		\includegraphics[width=3.0in]{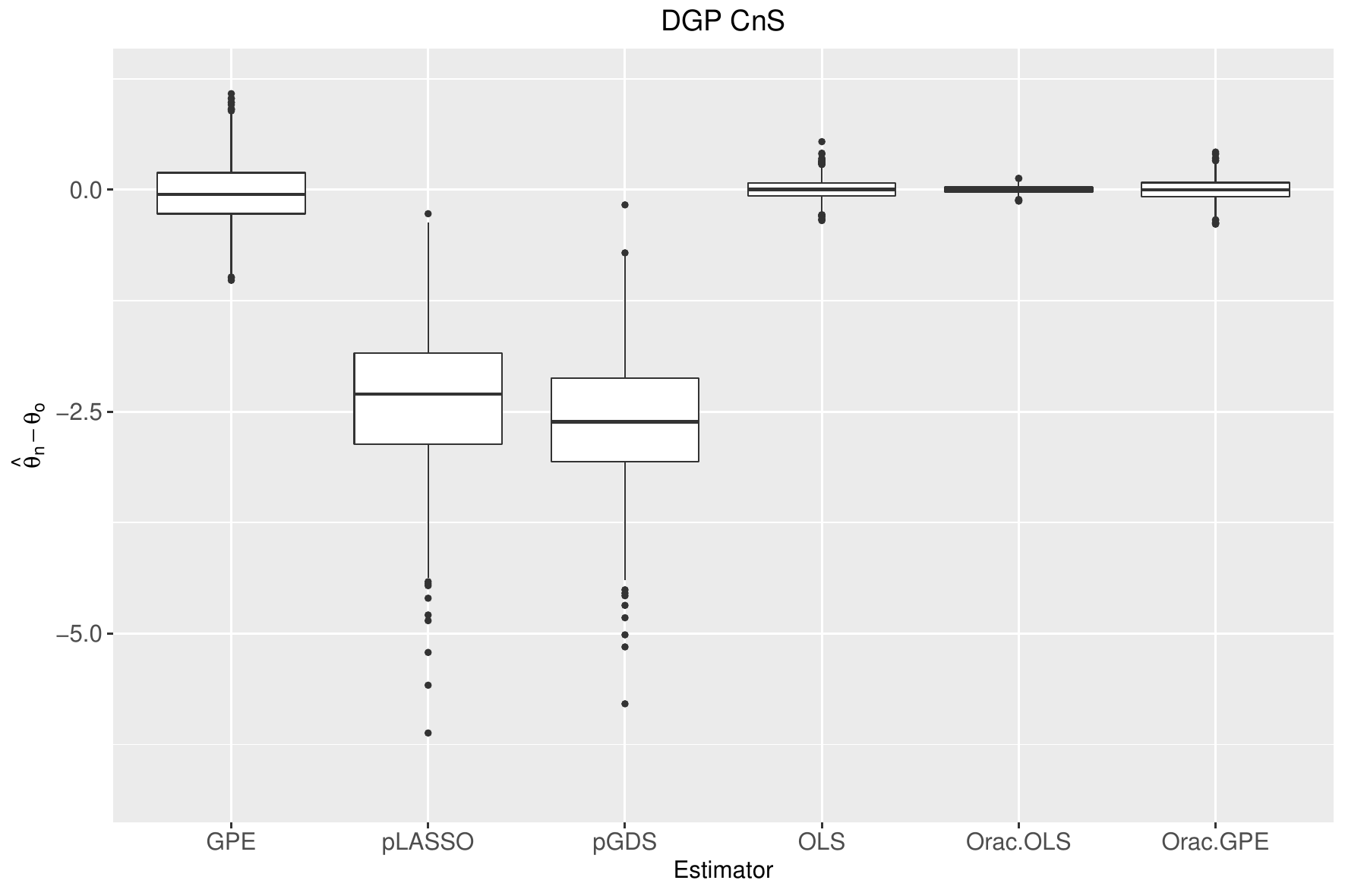}
        \label{fig:boxplot1}
	\end{minipage}%
        \begin{minipage}{.5\linewidth}
		\includegraphics[width=3.0in]{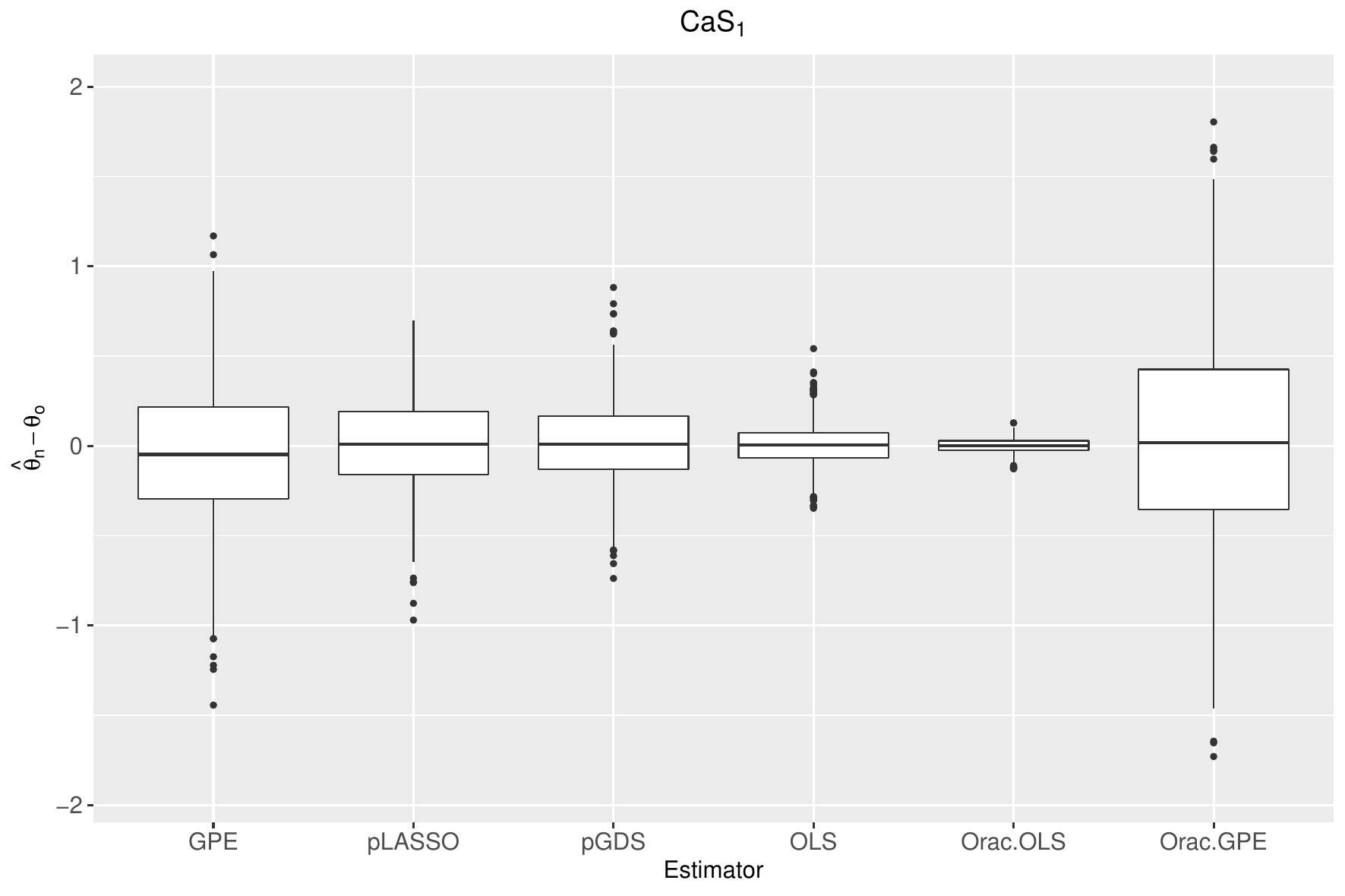}
	\end{minipage}
	\begin{minipage}{.5\linewidth}
		\includegraphics[width=3.0in]{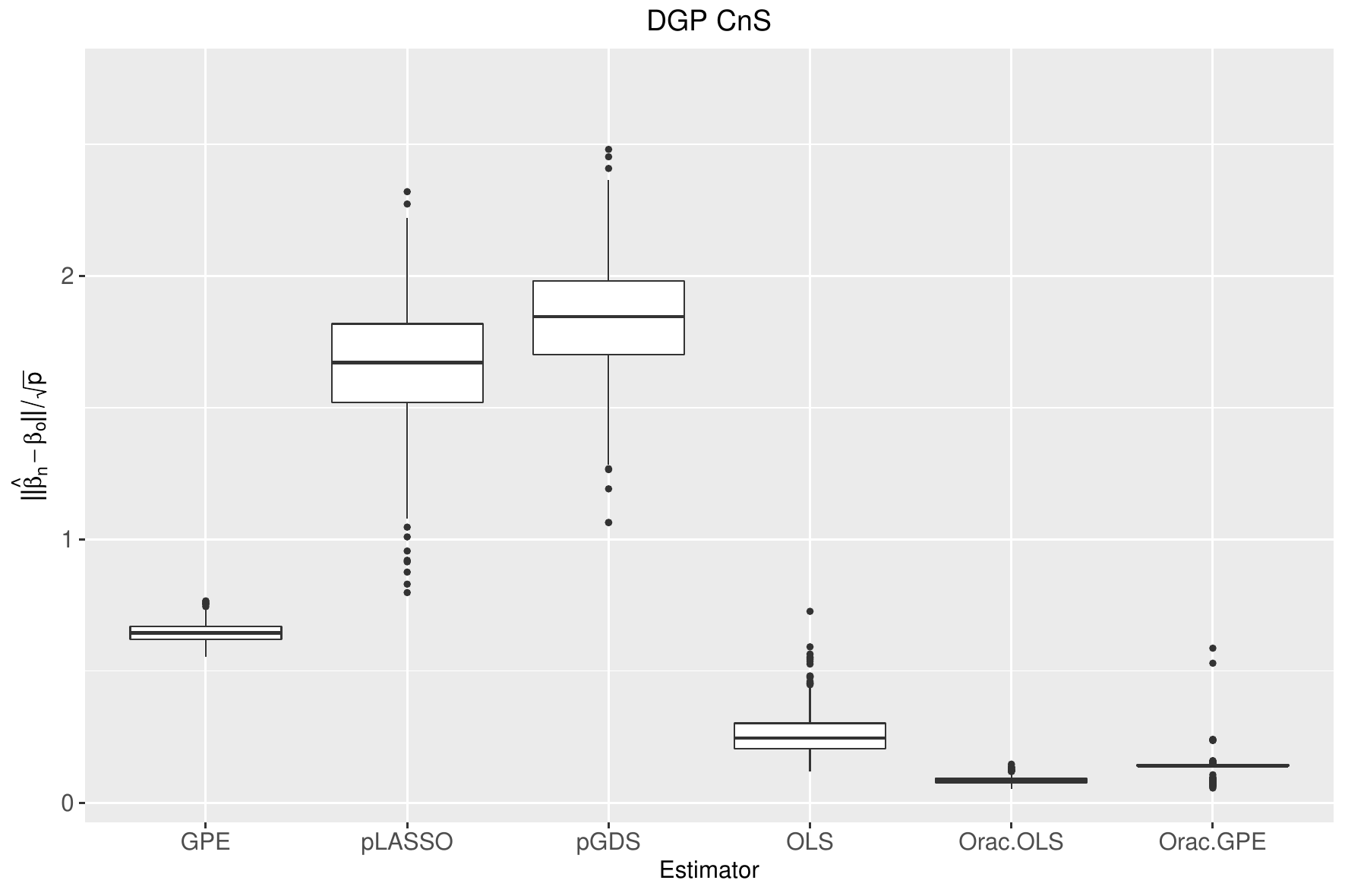}
	\end{minipage}%
	\begin{minipage}{.5\linewidth}
		\includegraphics[width=3.0in]{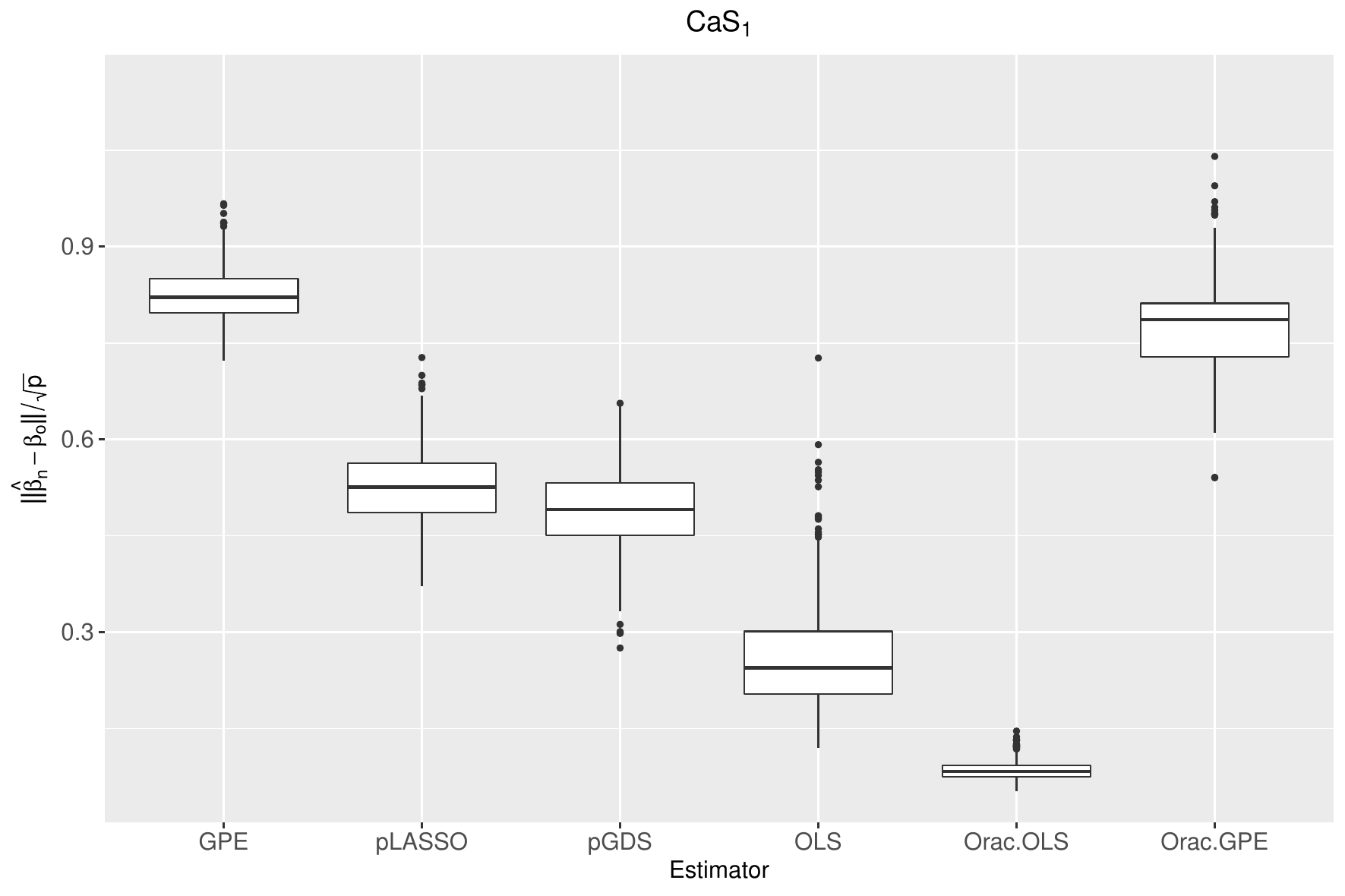}
	\end{minipage}
        \begin{minipage}{.5\linewidth}
		\includegraphics[width=3.0in]{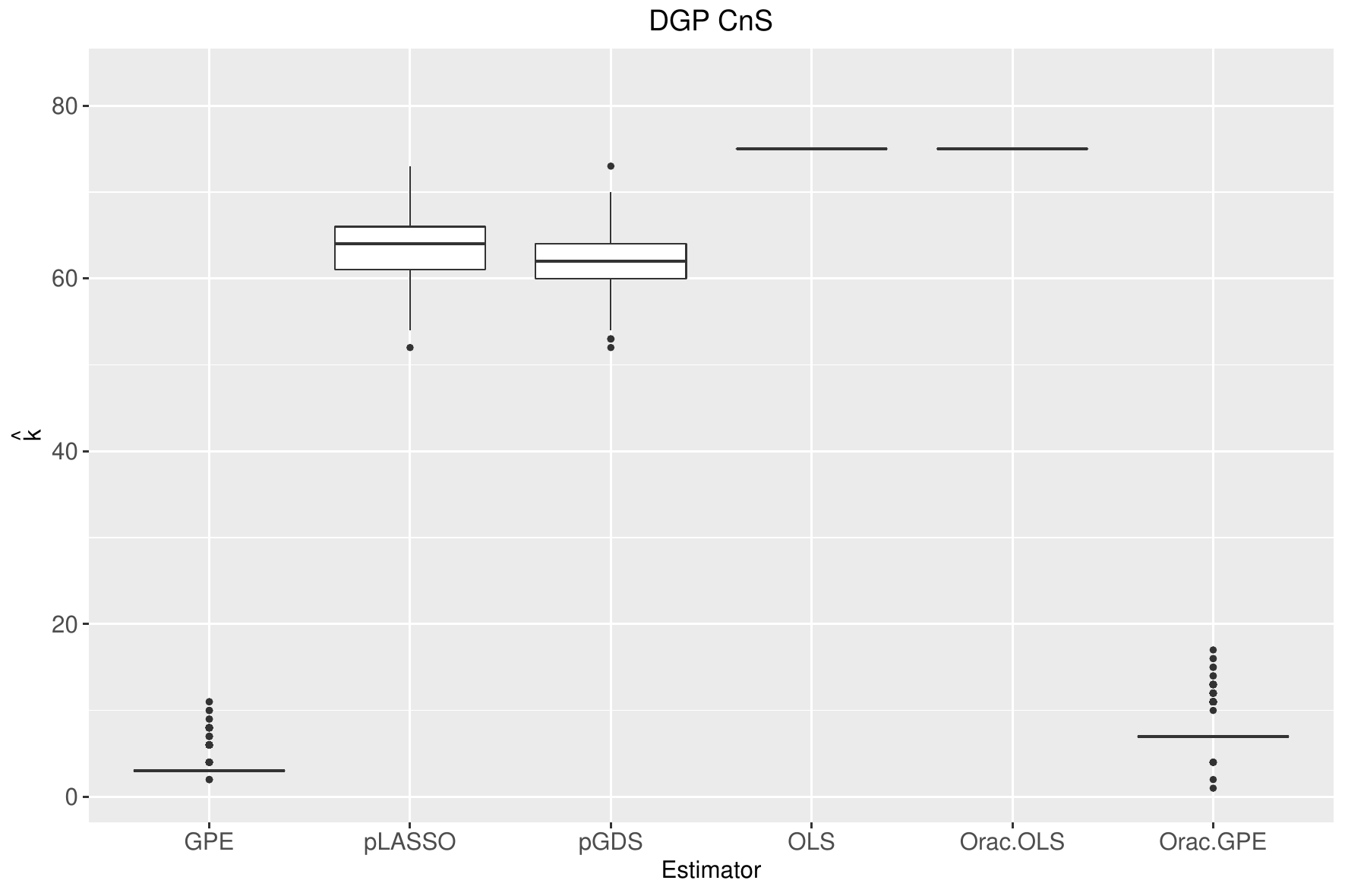}
        \label{fig:DGP4A}
	\end{minipage}%
        \begin{minipage}{.5\linewidth}
		\includegraphics[width=3.0in]{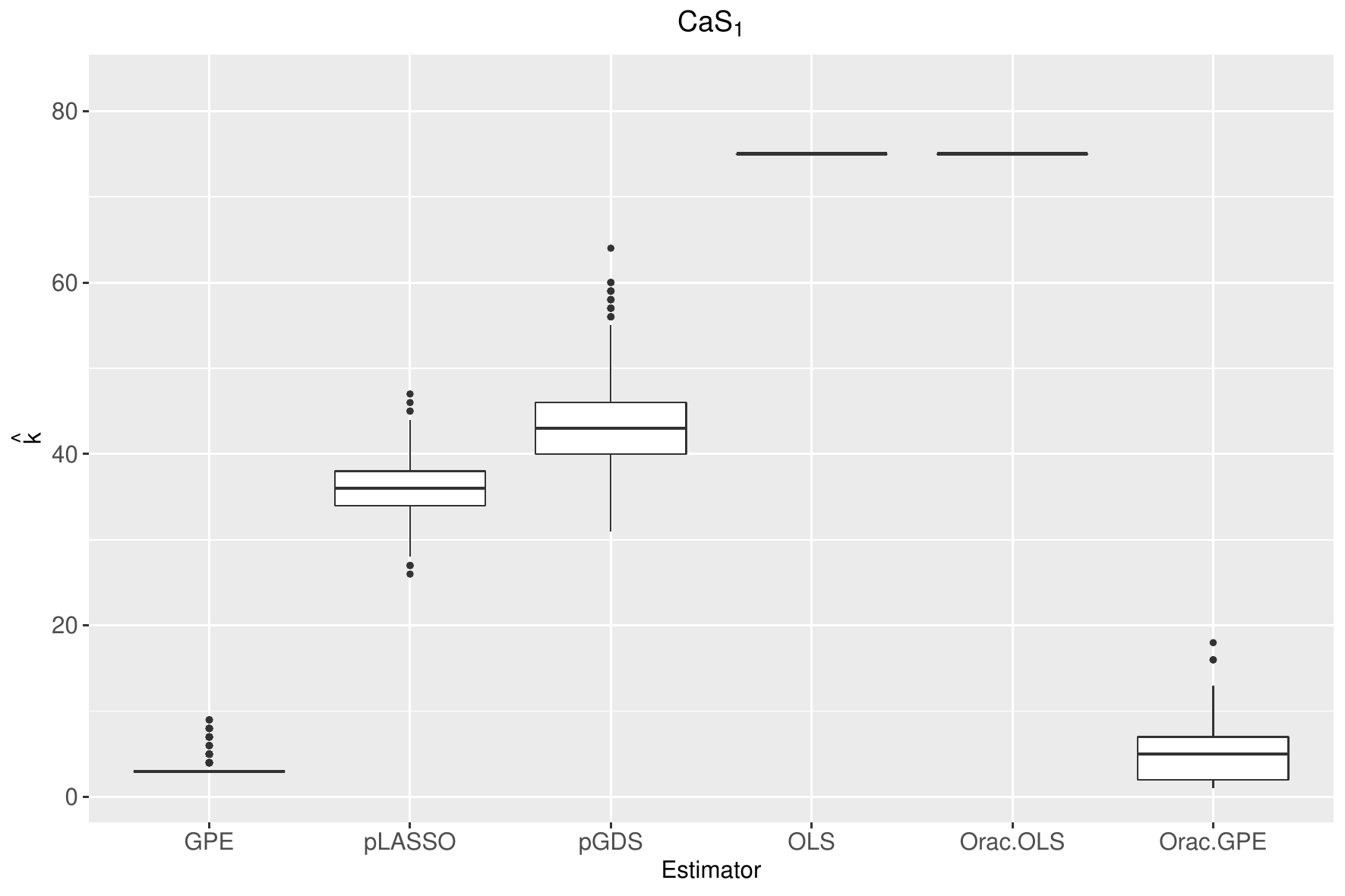}
	\end{minipage}
\end{figure}

\begin{figure}[!htbp]
	\centering
	\caption{DGPs $\mathrm{CaS_2}$ \& D-S$_1$ at $(n,p)=(100,75)$ }
	\begin{minipage}{.5\linewidth}
		\includegraphics[width=3.0in]{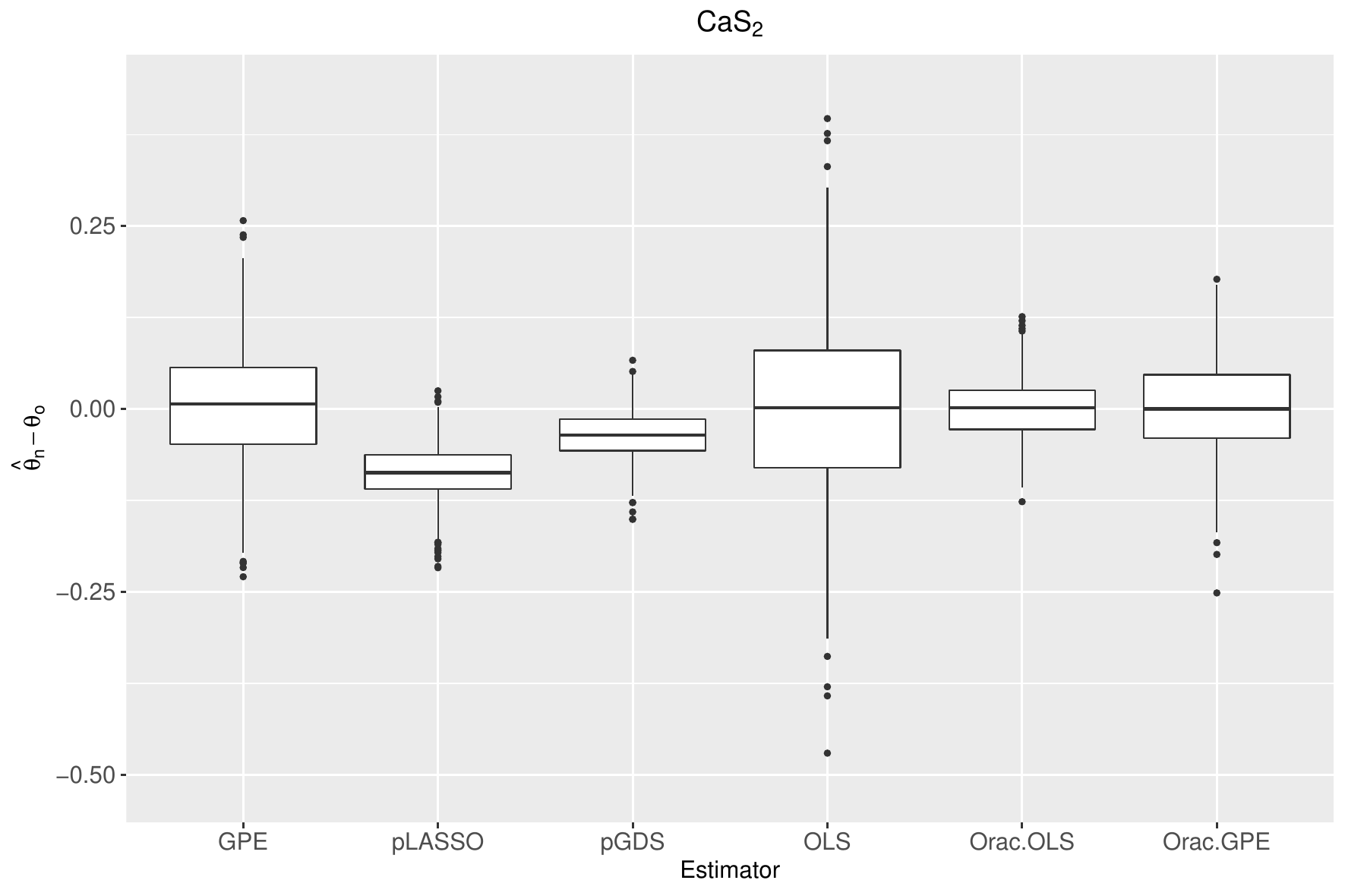}
        \label{fig:boxplot2}
	\end{minipage}%
        \begin{minipage}{.5\linewidth}
		\includegraphics[width=3.0in]{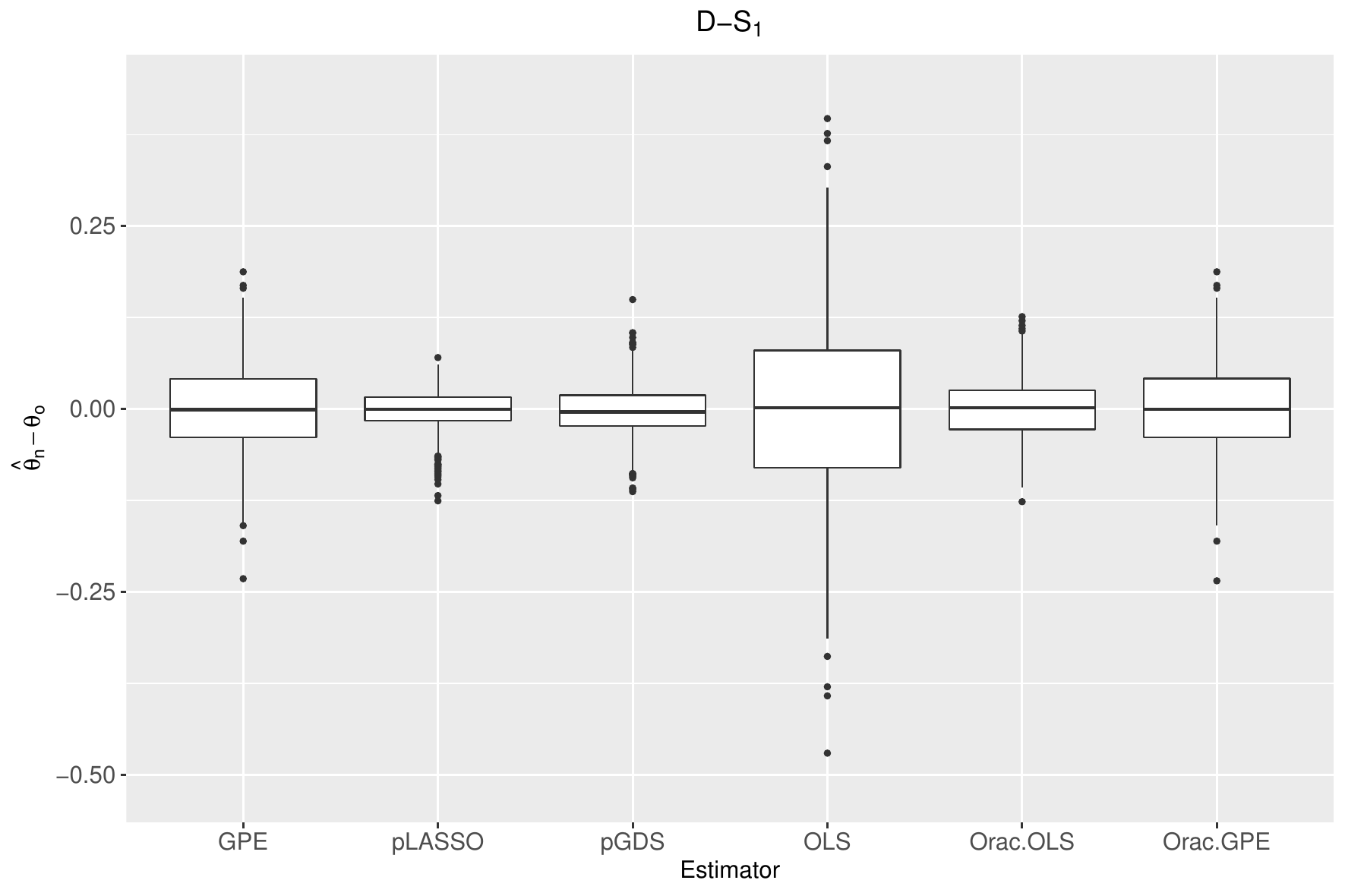}
	\end{minipage}
	\begin{minipage}{.5\linewidth}
		\includegraphics[width=3.0in]{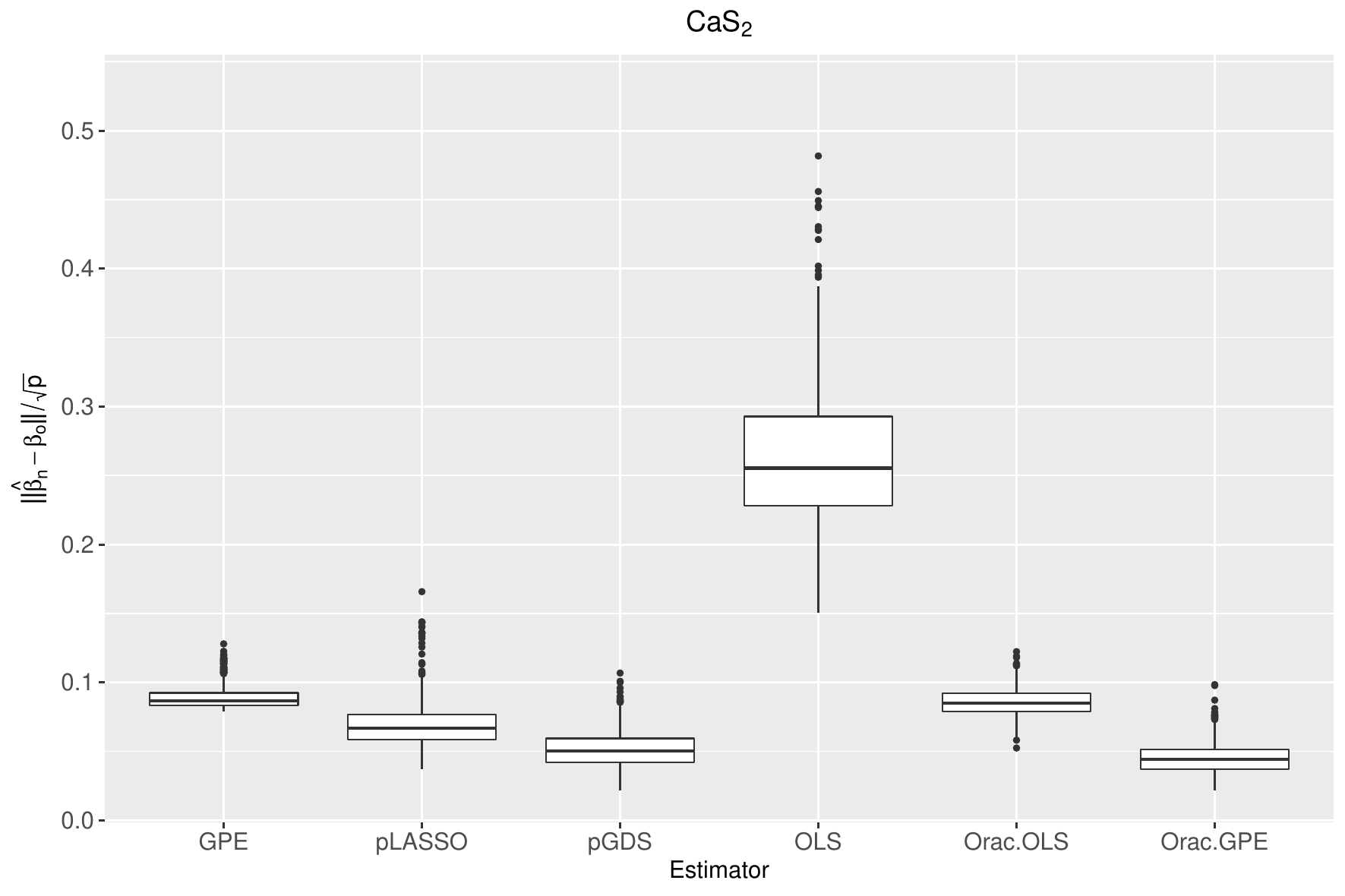}
	\end{minipage}%
	\begin{minipage}{.5\linewidth}
		\includegraphics[width=3.0in]{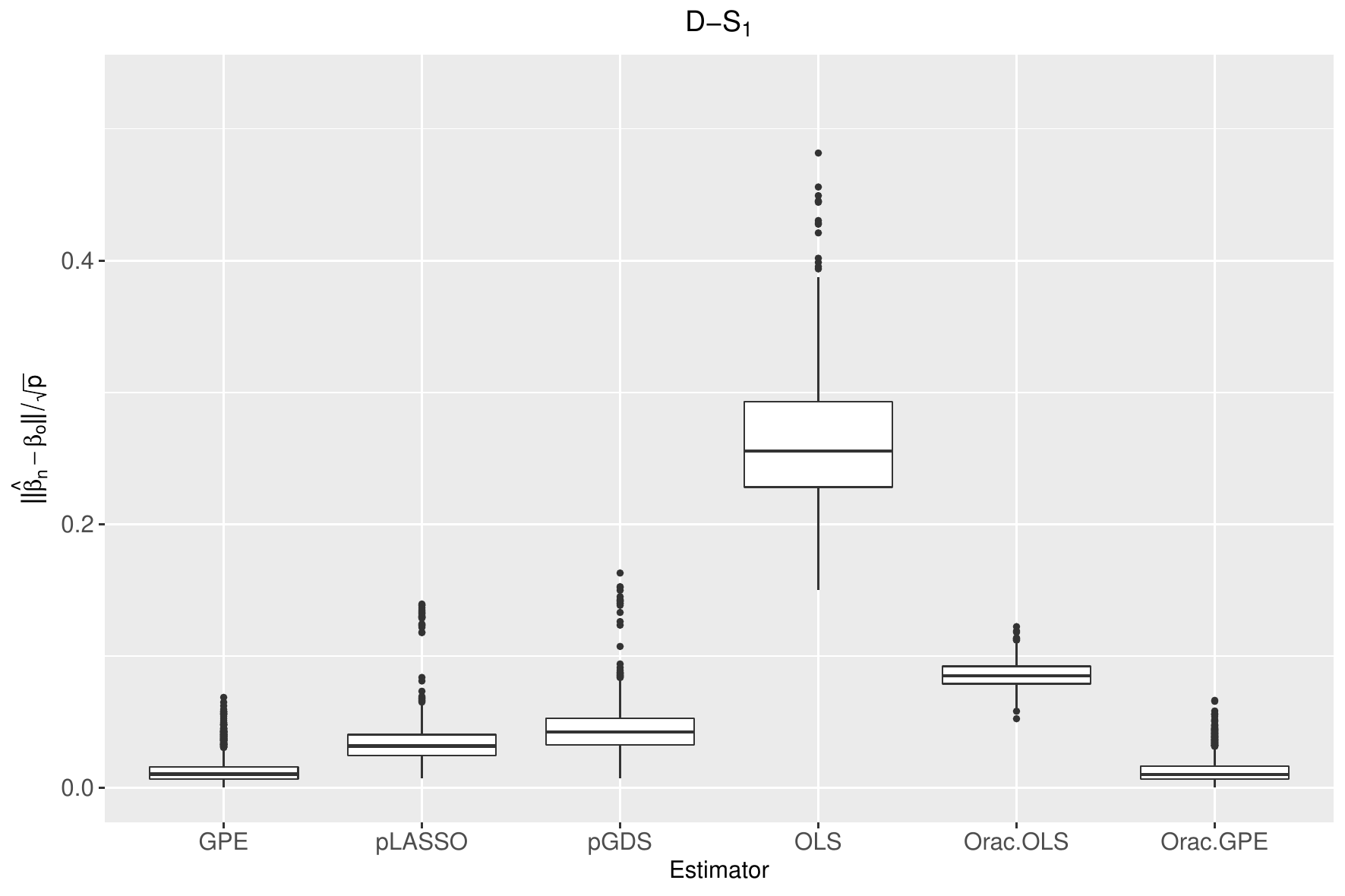}
	\end{minipage}
        \begin{minipage}{.5\linewidth}
		\includegraphics[width=3.0in]{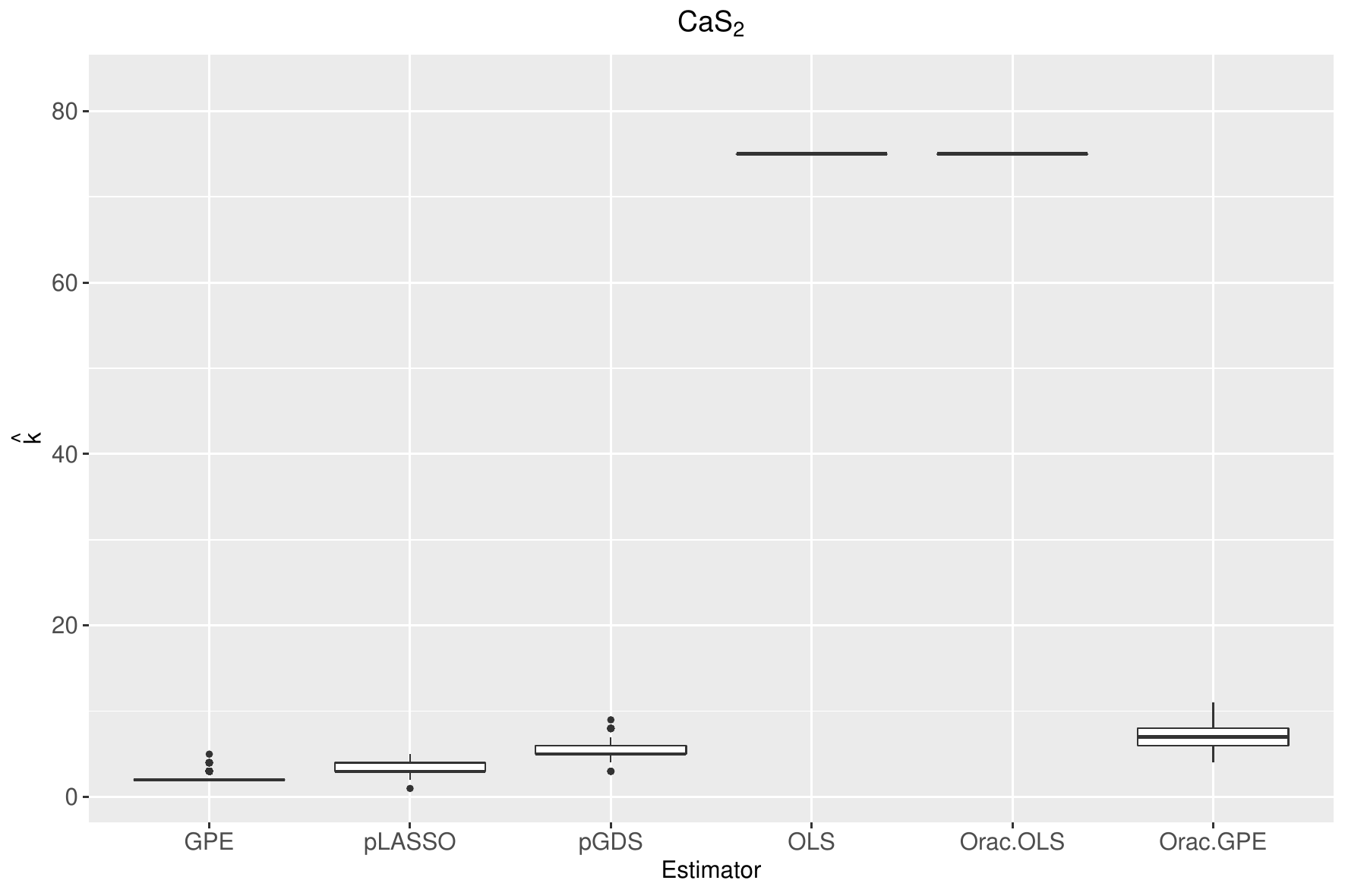}
        \label{fig:DGP4A}
	\end{minipage}%
        \begin{minipage}{.5\linewidth}
		\includegraphics[width=3.0in]{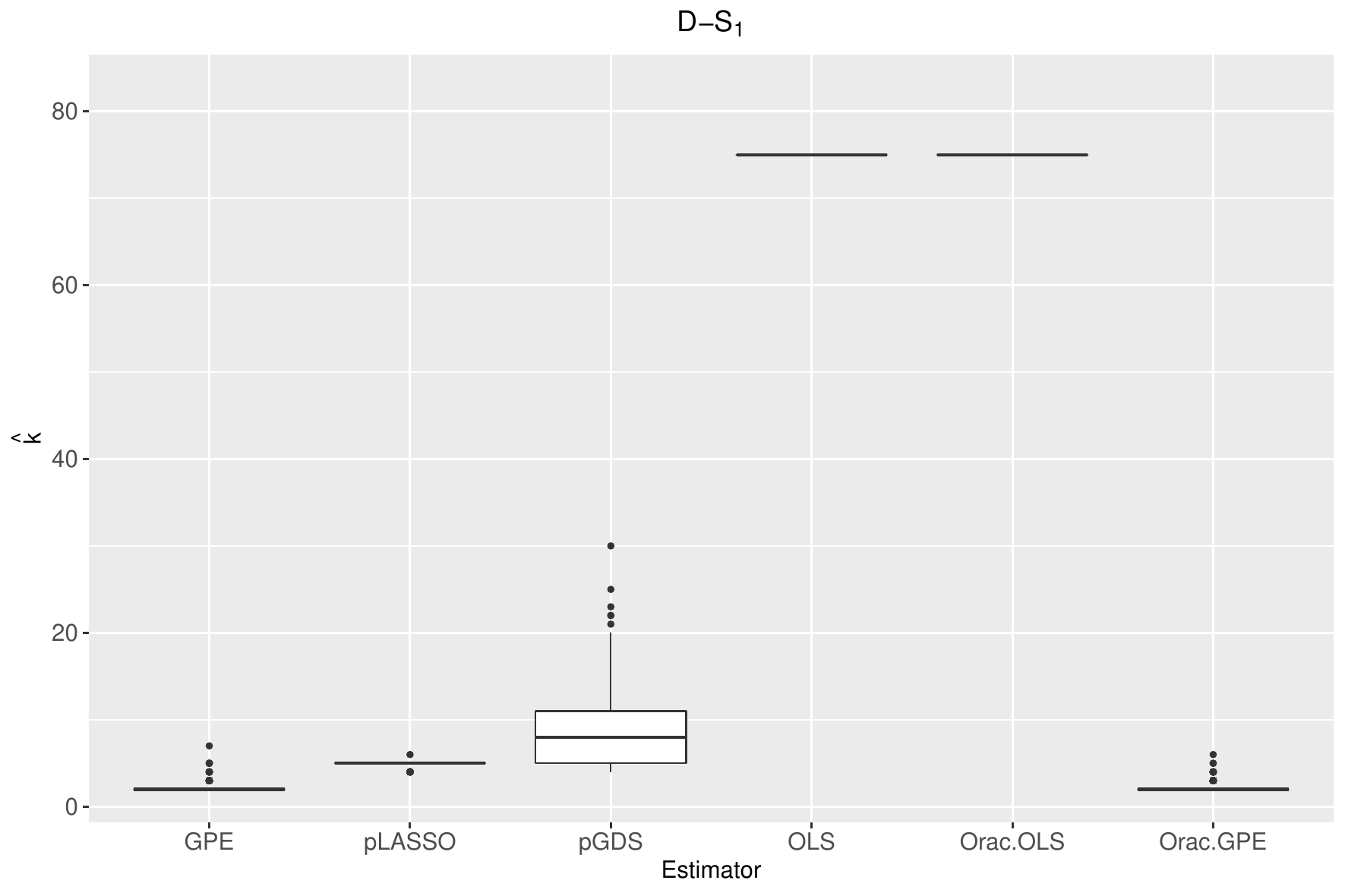}
	\end{minipage}
\end{figure}

\begin{figure}[!htbp]
	\centering
	\begin{minipage}{.5\linewidth}
        \caption{Power Curve: CnS}
		\includegraphics[width=3.0in]{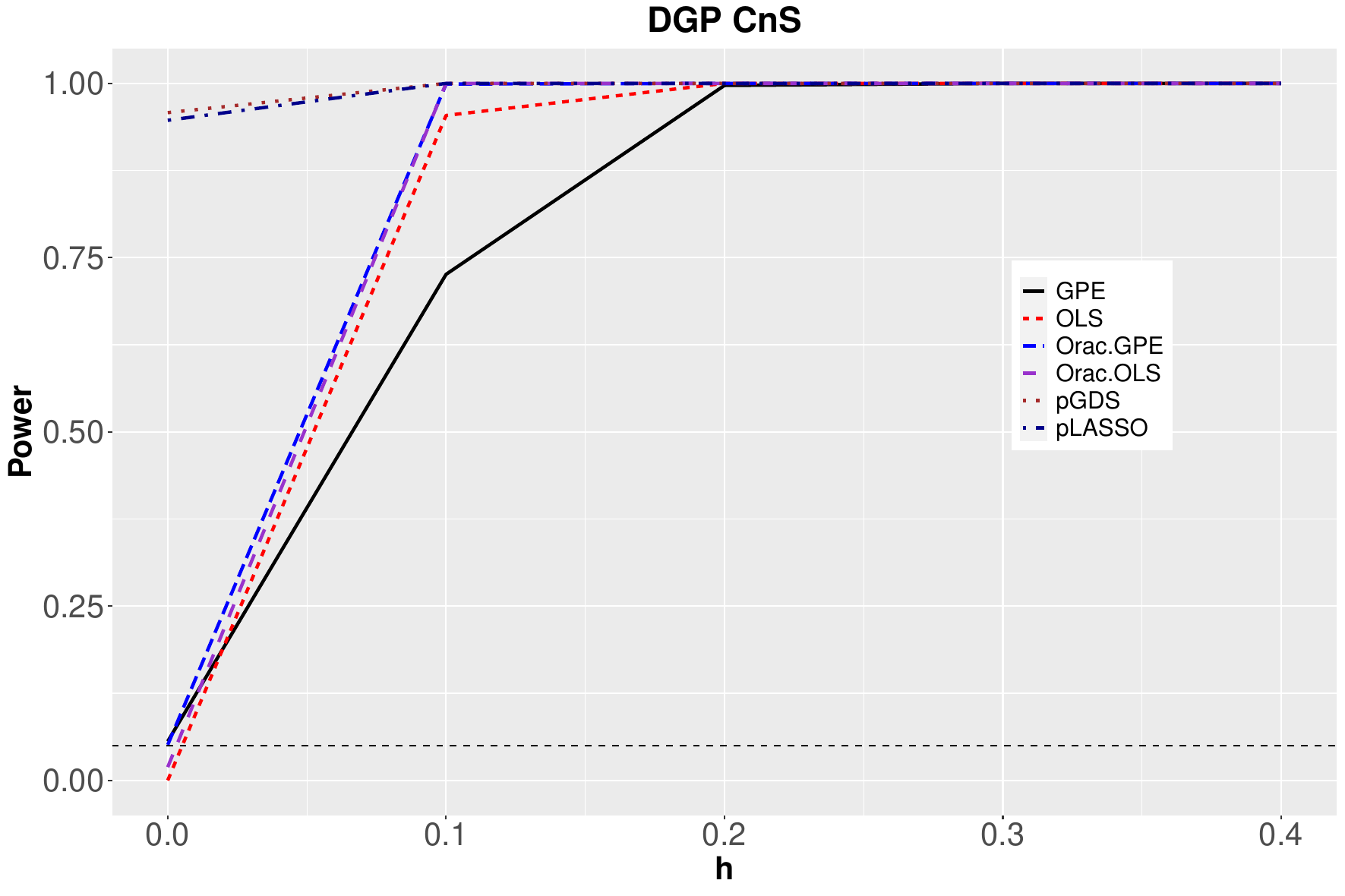}
        \label{fig.pow_CnS}
	\end{minipage}%
        \begin{minipage}{.5\linewidth}
        \caption{Power Curve: DGP CaS$_1$}
		\includegraphics[width=3.0in]{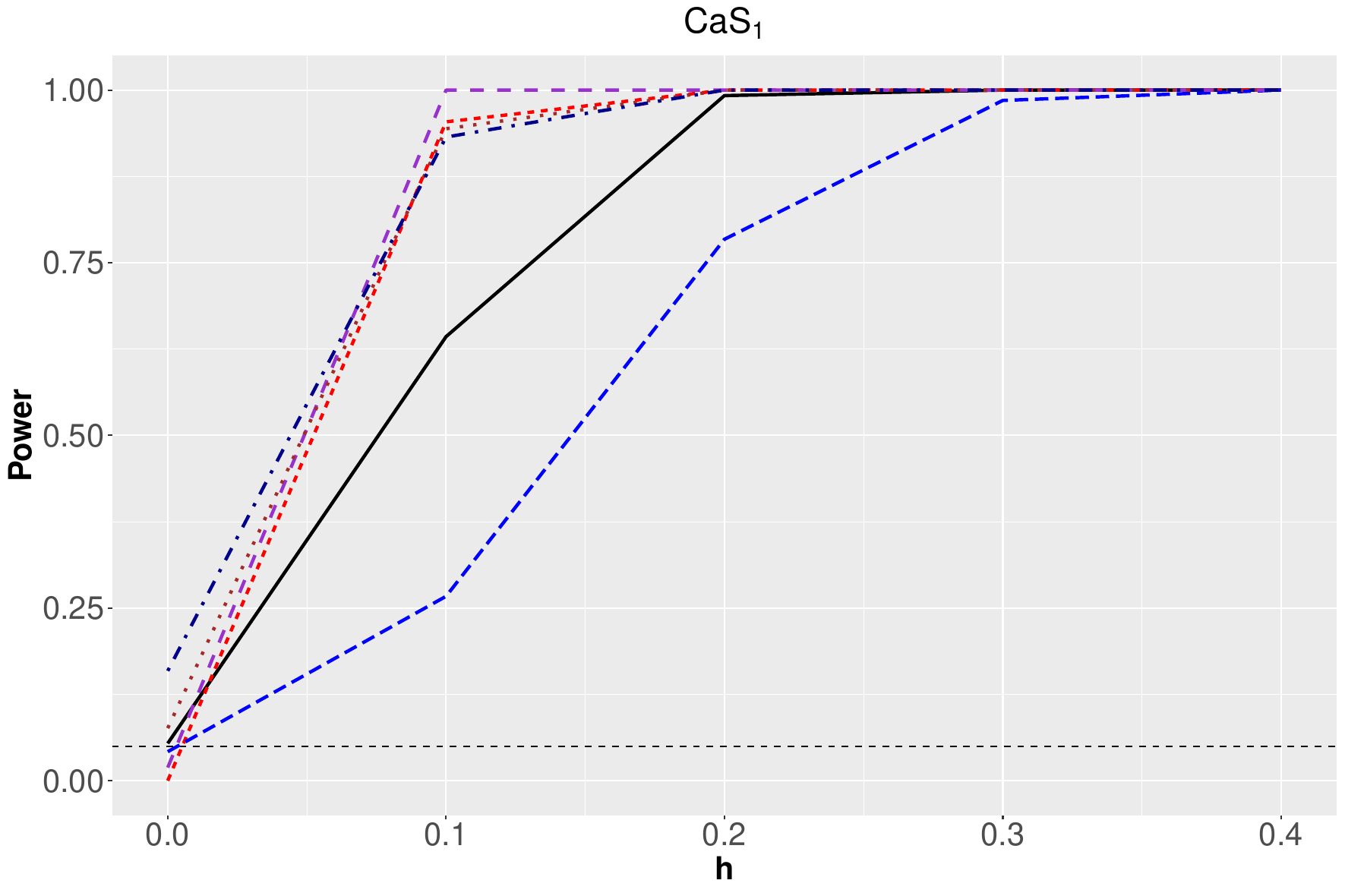}
        \label{fig.pow_CaS1}
	\end{minipage}
        \begin{minipage}{.5\linewidth}
        \caption{Power Curve: DGP CaS$_2$}
		\includegraphics[width=3.0in]{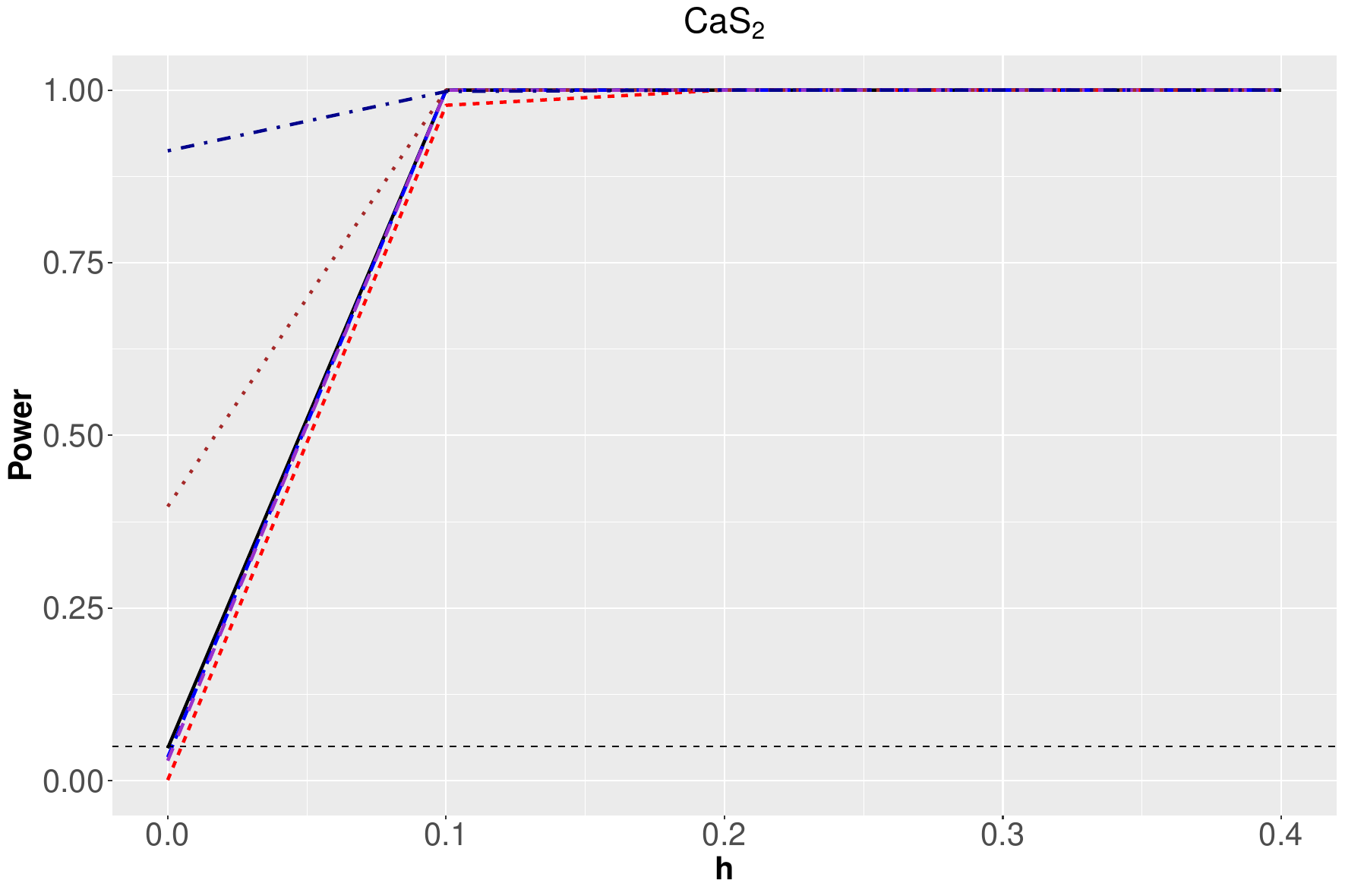}
        \label{fig.pow_CaS2}
	\end{minipage}%
        \begin{minipage}{.5\linewidth}
        \caption{Power Curve: DGP D-S$_1$}
		\includegraphics[width=3.0in]{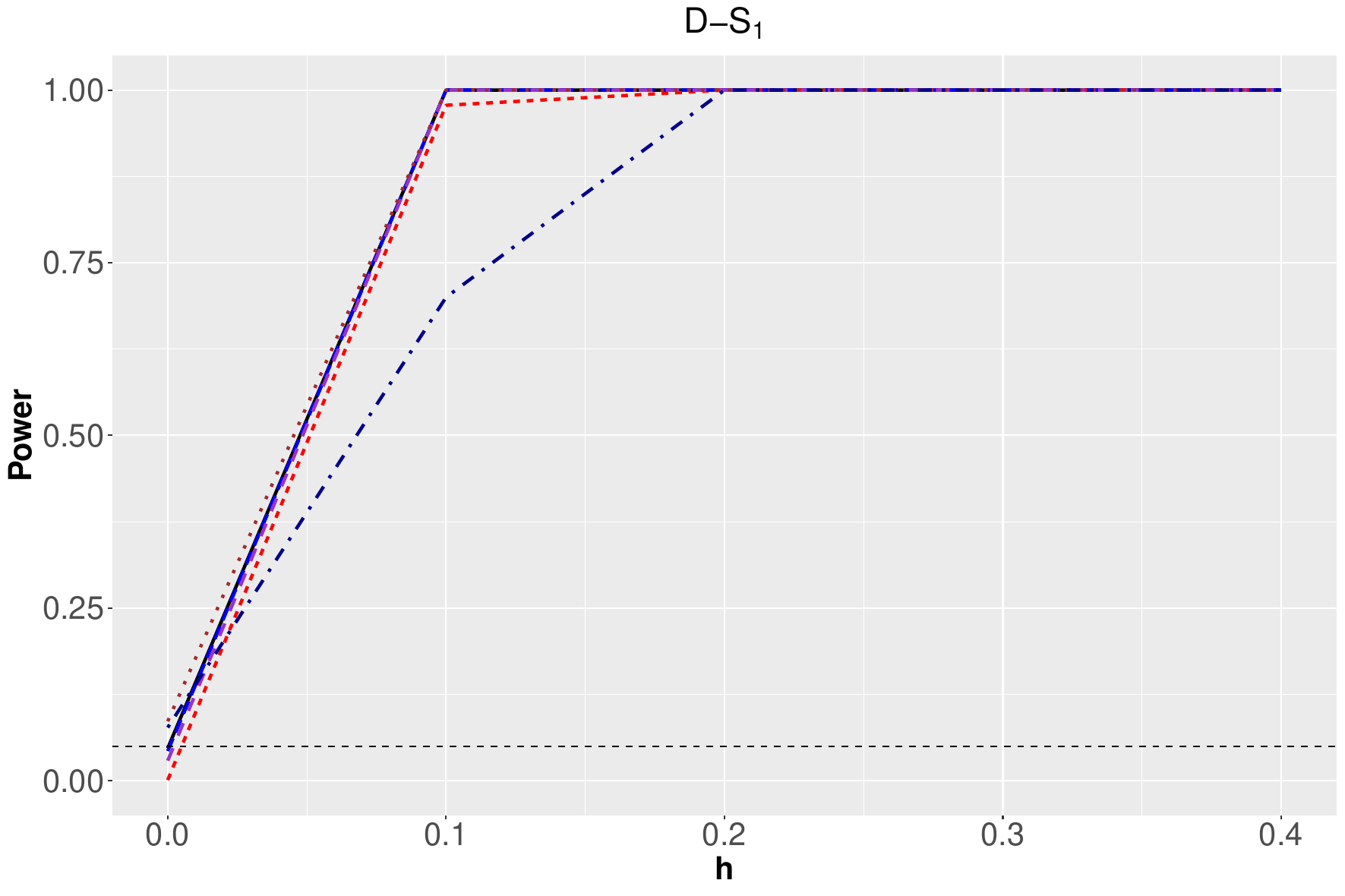}
        \label{fig.pow_D_S1}
	\end{minipage}

 {\footnotesize
    \textit{Notes}: The above figures plot power curves on $\theta$ at $(n,p) = (100,75)$. The deviation from the null is generated as $ \theta - \theta_o = -h\sqrt{p}, \ h\in [0,0.4] $.}
\end{figure}

\Cref{Tab:DGP_CnS,Tab:DGP_CaS1,Tab:DGP_CaS2,Tab:DGP_D_S1} and \Cref{fig:boxplot1,fig:boxplot2,fig.pow_CnS,fig.pow_CaS1,fig.pow_CaS2,fig.pow_D_S1} present the findings of the simulation exercise. Fixing $ p $, one observes a general decrease in MnB, MAD, and RMSE for all estimators as $ n $ increases in \Cref{Tab:DGP_CnS,Tab:DGP_CaS1,Tab:DGP_CaS2,Tab:DGP_D_S1}. Unsurprisingly, one observes a slight deterioration in performance as $ p $ is doubled and $ n $ is held fixed. Rejection rates are useful in showing which estimators provide reliable inference -- this is crucial in model estimation and hypothesis tests. Out of the four feasible estimators, only GPE delivers meaningful inference over all $ (n,p) $ configurations in DGPs Cns, CaS$_1$, and CaS$_2$. Orac.OLS, in spite of being run on samples of size $3n$, does not have good rejection rates across all $(n,p)$ configurations -- the Orac.OLS severely under-rejects in \Cref{Tab:DGP_CnS,Tab:DGP_CaS1,Tab:DGP_CaS2,Tab:DGP_D_S1} at $(n,p)=(100,150)$.

Even under approximate sparsity (\Cref{Tab:DGP_CaS2}), neither of the sparsity-dependent methods (pLASSO and pGDS), delivers reliable inference. GPE, in contrast, continues to perform reasonably. It is under exact sparsity (\Cref{Tab:DGP_D_S1}) that inference based on pLASSO and pGDS becomes reliable. This appears to suggest that the performance of sparsity-dependent estimators can be very sensitive to almost negligible deviations from sparsity.\footnote{A similar observation emerges in \Cref{Tab:DGP_DaS2} of the Online Appendix where sparsity is perturbed with a $O(n^{-1/2})$ term.} As sparsity is not easily verifiable in typical empirical applications, this constitutes a major drawback of pLASSO and pGDS for model estimation and inference. It ought to be borne in mind, however, that pLASSO and pGDS, among other sparsity-dependent methods, provide good approximations to the optimal instruments in the first stage of instrumental variable methods -- see \citet{belloni2012sparse,farrell2015robust,belloni2017program,hansen2014instrumental,carrasco2015regularized} or the effect of a few covariates of interest in approximately sparse high-dimensional settings, e.g., \citet{belloni2014inference,chernozhukov2022biased}. 

Interestingly, $ \mathrm{med}(\hat{k}_n) = 2$ for GPE in DGP D-S$_1$ where $\bbeta_o$ has exactly $2$ support points -- see \Cref{Tab:DGP_D_S1} and the bottom-right box-plot in \Cref{fig:boxplot2}. In the other DGPs where $\bbeta_o$ has continuous heterogeneity, one observes that $\mathrm{med}(\hat{k}_n)$ increases weakly in $n$ but not in $p$. Turning to the box plots in \Cref{fig:boxplot1,fig:boxplot2}, one observes that GPE and OLS' $\hat{\theta}_n-\theta_o$ appear well centred around zero across all four DGPs while that of pLASSO and pGDS are well-centred only in DGP D-S$_1$. This property of OLS does not translate into good size control; OLS severely under-rejects at $(n,p)=(100,75)$ and $(n,p)=(400,150)$ in all DGPs. Although tripling sample size and using OLS reduces bias relative to GPE, rejection rates of Orac.OLS can sometimes be very low, e.g., \Cref{Tab:DGP_CaS2} at $(n,p)=(100,150)$. The second rows of box plots in \Cref{fig:boxplot1,fig:boxplot2} suggest GPE, relative to OLS, better estimates $\bbeta_o$ (judging by the $||\widehat{\bbeta}_n - \bbeta_o||/\sqrt{p}$) as the configuration gets more sparse. With respect to the sparsity-dependent pLASSO and pGDS, a clear pattern does not emerge. The fairly small number of groups, see \Cref{fig:boxplot1,fig:boxplot2}, used by GPE while having a comparable or sometimes better performance than competing estimators suggests that its parsimony is not achieved at the expense of substantial bias or unreliable inference.

\Cref{fig.pow_CnS,fig.pow_CaS1,fig.pow_CaS2,fig.pow_D_S1} demonstrate that the good size control of GPE in \Cref{Tab:DGP_CnS,Tab:DGP_CaS1,Tab:DGP_CaS2,Tab:DGP_D_S1} is not at the expense of power. GPE is the only feasible estimator considered in this section that controls size meaningfully and has non-trivial power across all four DGPs. In \Cref{fig.pow_CnS,fig.pow_CaS1,fig.pow_CaS2} for example, OLS has rejection rates equal to zero, and it has power under the alternative. pLASSO and pGDS control size meaningfully in \Cref{fig.pow_D_S1} and possess non-trivial power under the alternative. It is interesting to note that GPE in \Cref{fig.pow_D_S1} remains competitive in terms of size control and power under the alternative although the setting is favourable to pLASSO and pGDS. 

In sum, this simulation exercise shows very robust performance in terms of bias reduction and good size control of GPE with its data-driven selection rule under (1) different empirically relevant but unknowable configurations of $\bbeta_o$; (2) heteroskedastic $\varepsilon$; (3) Gaussian and non-Gaussian $\varepsilon$; and (4) skewed and symmetrically distributed $\varepsilon$. Robustness to all the aforementioned features is an important property of an estimator especially in the high dimensional setting as the configuration of $\bbeta_o$ is typically unbeknownst to the practitioner or difficult to ascertain via economic theory. The exercise also shows a very high sensitivity of the  sparsity-dependent pLASSO and pGDS to almost negligible deviations from sparsity.

\section{Empirical Application}\label{Sect:Empirical_Application}
In this section, we estimate price and income elasticities of demand for gasoline following \citet{yatchew2001household,chernozhukov2022biased,semenova2021debiased}. We use data from the 1994-1996 Canadian National Private Vehicle Use Survey.\footnote{The data set is available on Professor Yatchew's website: \url{https://economics.utoronto.ca/yatchew/}.} The outcome variable is $\log$ consumption and the covariates of interest are $\log$ price and $\log$ income. The sparsity assumption for the GPE is not required, unlike \citet{chernozhukov2022biased,semenova2021debiased}, which rely on it in this empirical setting.

\begin{figure}
	\centering
	\begin{minipage}{.5\linewidth}
        \caption{Price Elasticity of demand}
		\includegraphics[width=3.0in]{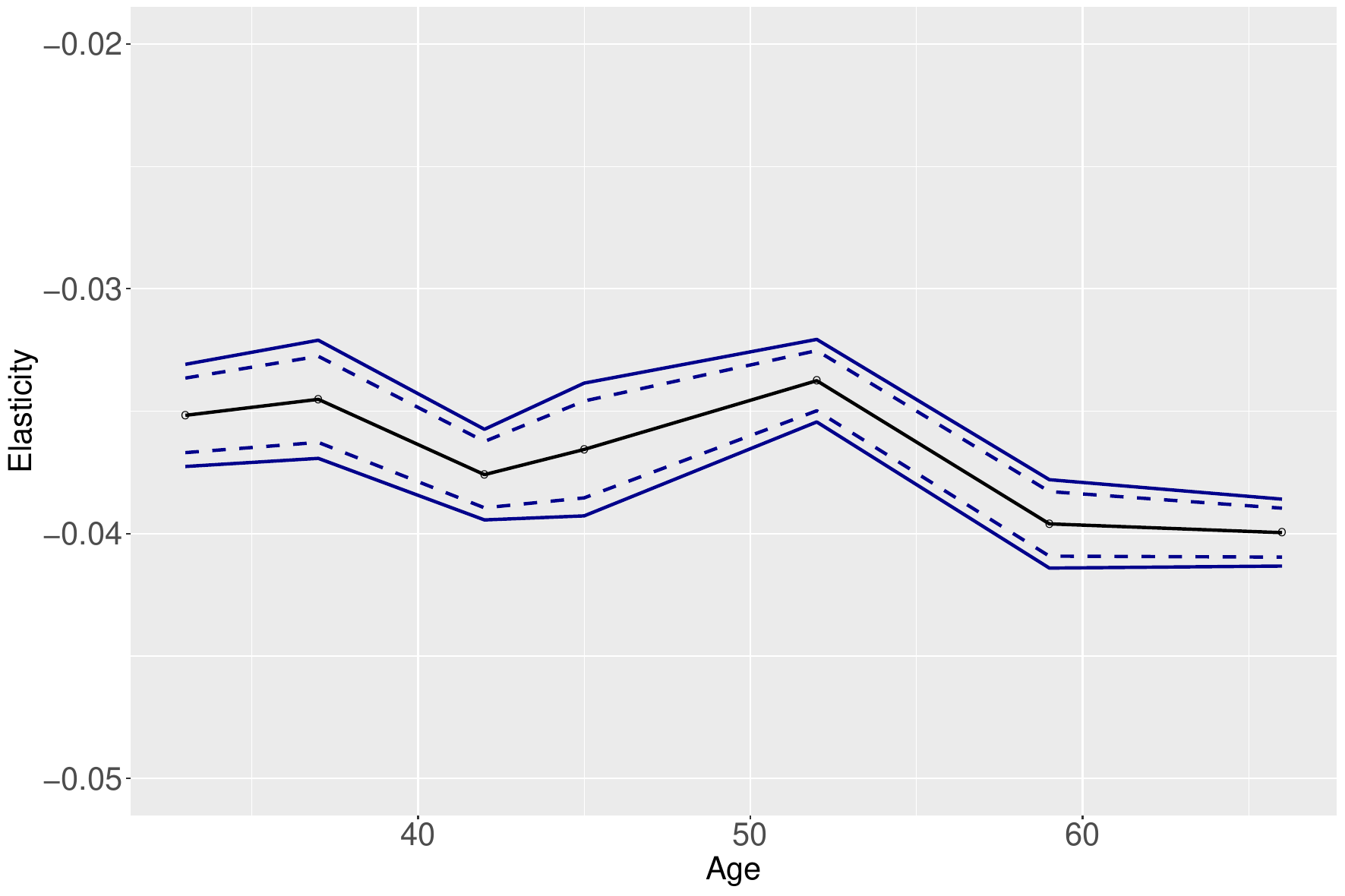}
        \label{fig:Price_Elasticity}
	\end{minipage}%
        \begin{minipage}{.5\linewidth}
        \caption{Income Elasticity of demand}
		\includegraphics[width=3.0in]{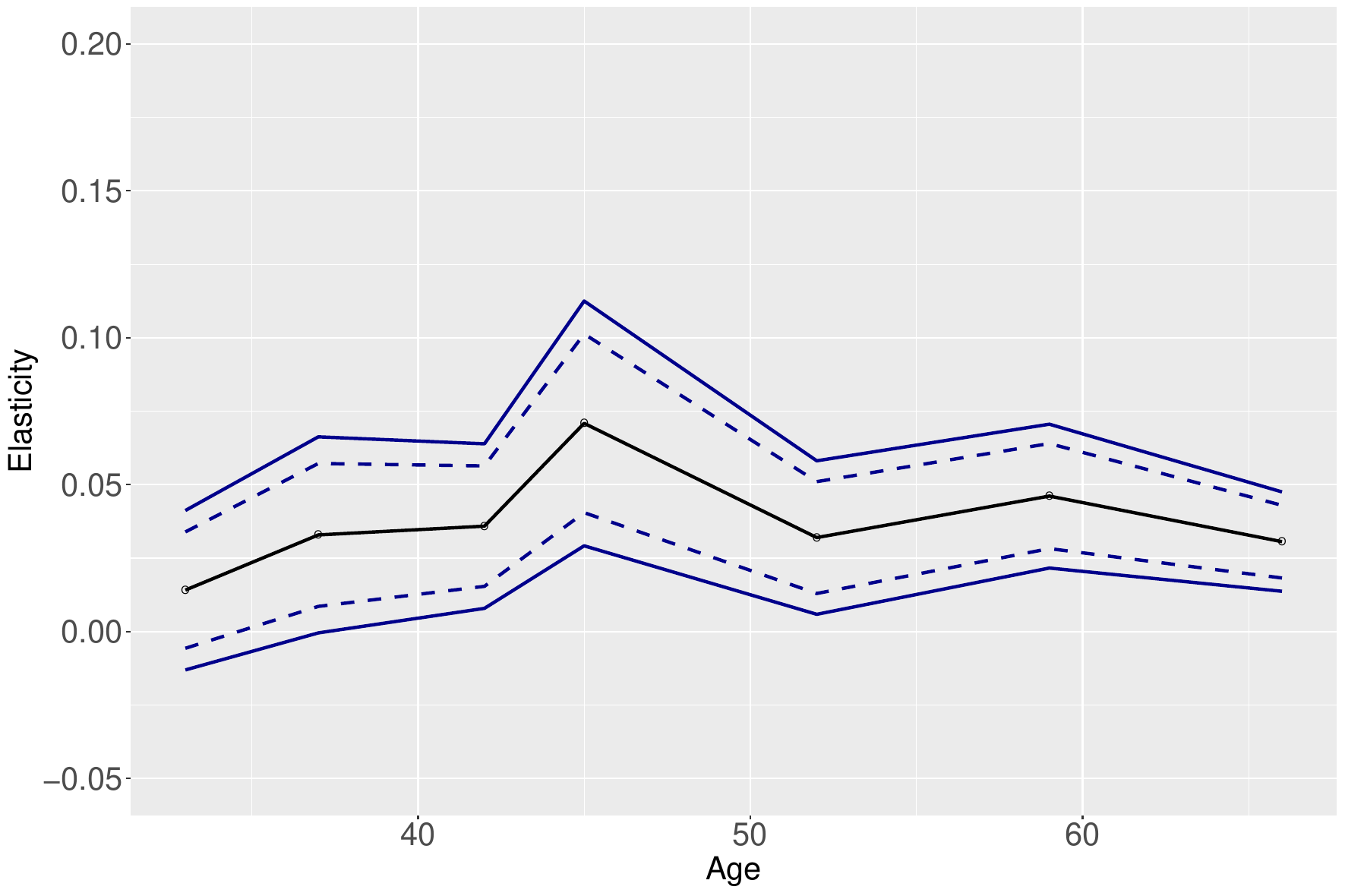}
        \label{fig:Income_Elasticity}
	\end{minipage}
 
{
 \footnotesize
    \textit{Notes}: The thick black line is the estimate, 95\% point-wise confidence bands are dashed and the thick outermost lines are the 95\% simultaneous confidence bands using the \citet{montiel2019simultaneous} approach.
}
\end{figure}

The approach taken in constructing the covariate vector $\x$ follows \citet{semenova2021debiased,chernozhukov2022biased}. The covariates include $\log $ price, $\log $ price squared, $\log $ income, $\log $ income squared, distance driven, distance driven squared, and 27 time, geographical, and household composition dummies. The aforementioned covariates are augmented with an interaction of the dummies with $\log$ price and its square in the first set of results (\Cref{fig:Price_Elasticity}), and $\log$ income and its square in the second set of results (\Cref{fig:Income_Elasticity}), respectively, resulting in $87$ covariates for each model. In this empirical exercise, we focus on price and income elasticities within sub-samples defined by age groups in \Cref{Tab:PED_IED}. This approach is useful in learning the variation in consumers' price and income elasticities of demand for gasoline by age group.

\begin{table}[!htbp]
\caption{Price and Income Elasticities of demand}
\footnotesize
\centering
\label{Tab:PED_IED}
\begin{tabular}{lccccc}
\toprule
        & &\multicolumn{2}{c}{Price Elasticity of demand} & \multicolumn{2}{c}{Income Elasticity of demand} \\ \midrule
        Age-group & \# Obs. & GPE & pLASSO & GPE  & pLASSO \\\midrule
        All    &5001   &-0.023 (0.002)  &-0.198 (0.053)  &0.045 (0.009) &0.070 (0.009) \\ 
        20 - 33 & 713    &-0.035 (0.001) &-0.044 (0.135) &0.014 (0.010) &0.029 (0.026) \\ 
        34 - 37 & 552    &-0.035 (0.001) &-0.279 (0.165) &0.033 (0.012) &0.037 (0.032) \\ 
        38 - 42 & 751    &-0.038 (0.001) &-0.385 (0.129) &0.036 (0.010) &0.075 (0.026) \\ 
        43 - 45 & 372    &-0.037 (0.001) &-0.320 (0.216) &0.071 (0.016) &0.129 (0.039) \\ 
        46 - 52 & 821    &-0.034 (0.001) &-0.311 (0.139) &0.032 (0.010) &0.020 (0.023) \\ 
        53 - 59 & 604    &-0.040 (0.001) &-0.275 (0.137) &0.046 (0.009) &0.072 (0.024) \\ 
        60 - 66 &1188    &-0.040 (0.001) &-0.198 (0.104) &0.031 (0.006) &0.056 (0.021) \\
        \bottomrule
   \end{tabular}
   
   {\footnotesize
    \textit{Notes}: Heteroskedasticity-robust standard errors are in parentheses. For price elasticity of demand, GPE selects 11 groups for the full sample and 2 groups for each sub-sample. pLASSO's first stage selects only 2 covariates (which exclude $\log$ price and $\log$ income) in the full sample and all sub-samples. For income elasticity of demand, GPE selects 5 groups for the full sample and 2 groups for each sub-sample. pLASSO's first stage selects only 2 covariates (which exclude $\log$ price and $\log$ income) in the full sample, a single covariate in the first two sub-samples, and 2 covariates in the remaining sub-samples.
    }
\end{table}

\Cref{fig:Price_Elasticity,fig:Income_Elasticity}, respectively, present the price and income elasticities of demand for gasoline by age group. One observes from \Cref{fig:Price_Elasticity} that the price elasticity of demand is strictly negative and statistically different from zero at the 5\% level for all age groups. Demand is, however, price inelastic for all age groups. The price elasticity of demand is non-monotone in age; consumers are most sensitive to price in the last two age groups where the price elasticity of demand does not vary (at the 5\% level), unlike younger age groups that exhibit significant variation by age group in the price elasticity of demand.

As expected, the income elasticity of demand is non-negative. Save for the first age group where the income elasticity of demand is not statistically significant (point-wise) at the 5\% level, the income elasticity of demand is point-wise significant for all other age groups. One cannot reject the hypothesis that income elasticity of demand equals a constant $c \in [0.029, 0.041] $ for all age groups. Moreover, the income elasticity of demand is not significantly different from zero for all age groups. One, however, rejects the hypothesis that consumers' demand for gasoline is not sensitive to income at all age groups. The $95\%$ simultaneous confidence bands suggest that gasoline is a necessity for all age groups.

Lastly, \Cref{Tab:PED_IED} offers a comparison of price and income elasticities of demand using GPE and pLASSO. Although both GPE and pLASSO estimates are qualitatively similar, GPE estimates are more precise. It is also worth emphasising that $\log$ price and $\log$ income are not selected by the first stages of pLASSO in the full sample or any sub-sample; they are added in the second stage as the amelioration set -- see \citet[p. 615]{belloni2014inference}.

\section{Conclusion}\label{Sect:Conclusion}
As the specific configuration of a high-dimensional $\bbeta_o$ cannot be known for sure in several empirical settings, there is a growing need for reliable, robust, and flexibly-adapted estimators. Although sparsity-dependent estimators such as pLASSO constitute the workhorse of econometric and applied work in high-dimensional settings and their theoretical properties are now well understood, their reliance on sparsity remains a major drawback. Existing clustering-based estimators are largely limited to fixed effects models with large $T$ panel data and do typically impose discrete heterogeneity. The aforementioned assumptions are restrictive and hard to verify or justify theoretically as they are imposed on the unobservable $\bbeta_o$. 

This paper makes a useful addition to the practitioner's toolkit of high-dimensional model estimation methods. The proposed GPE eliminates the burden on the practitioner of imposing typically unverifiable sparsity or discrete heterogeneity conditions on $\bbeta_o$ while maintaining the natural requirement that $\bbeta_o$ be defined on a bounded support. GPE is thus agnostic of and robust to different configurations of $\bbeta_o$ -- it accommodates various degrees of (non)-sparsity and heterogeneity. Thanks to a simple data-driven selection rule, GPE achieves parsimony and adaptability to different empirically relevant configurations.

This paper develops and applies GPE to the linear model with exogenous covariates in an $i.i.d.$ setting. To enhance GPE's usefulness to the practitioner, some important extensions remain. It will be interesting to consider GPE under endogeneity with possibly many (weak) instruments, clustered or weakly dependent data, and non-linear models with both smooth and non-smooth objective functions.

\appendix

\begin{center}
		\Large{\bf APPENDIX}
\end{center}

The proofs of the results in the main text are first presented in \Cref{Sect:Proof_Main_Results}, followed by supporting lemmata in \Cref{Sect:Useful_Lemmata}.
\section{Theoretical Results}\label{Sect:Proof_Main_Results}
\subsection{Proof of \Cref{Prop:Rate_ApprxBias}}

Let $ \tilde{\beta}(l), \ l\in \{1,\ldots,k\} $ denote the $ l $'th group average of $ \bbeta $, and denote group assignment indicators by $ g_j \in \{1,\ldots,k\} $ for $j=1,\ldots,p$. Then for $k<|\mathrm{supp}(\bbeta_o)|$,

\begin{align*}
    ||\m_o\bdelta_o - \bbeta_o||^2 &=: \min_{ \{\tilde{\bdelta}, g_1,\ldots,g_p\} }\sum_{j=1}^p(\tilde{\beta}(g_j) - \beta_{oj})^2 \\
    &=: \sum_{j=1}^p(\tilde{\beta}_o(g_j^*) - \beta_{oj})^2 \leq p\max_j(\tilde{\beta}_o(g_j^*) - \beta_{oj})^2 \\ 
    &\leq 4C_b^2M^2\frac{p}{k^2}
\end{align*}

\noindent by \Cref{Ass:Bounded_Support_B}, \Cref{Cond:m}, and \Cref{Apndix:Useful:Lem_Max_K_b} where $ \tilde{\bdelta} := (\tilde{\beta}(1),\ldots,\tilde{\beta}(k))' \in \Delta \subset \R^k $. 


$ ||\m_o\bdelta_o - \bbeta_o||^2 = 0 $ if $k\geq |\mathrm{supp}(\bbeta_o)| $, i.e., the approximation bias is exactly zero if $k$ is at least equal to the number of support points of $\bbeta_o$. This concludes the proof.
\qed

\subsection{Proof of \Cref{Theorem:Identification}}
First, let us obtain the following useful expression of $Q_o(\m\bdelta)$ in terms of the approximation bias $ \bm{b}_k:= \bbeta_o - \m_o\bdelta_o$ and $ \sigma^2 $.

\begin{align}\label{eqn:bk_dom_bk2rp}
	Q_o(\m\bdelta) &= \mathbb{E}[(\varepsilon - \x(\m\bdelta -\m_o\bdelta_o - \bm{b}_k) )^2]/p\nonumber\\ 
		&= \sigma^2/p + (\m\bdelta -\m_o\bdelta_o -  \bm{b}_k)'\E[\x'\x](\m\bdelta - \m_o\bdelta_o -  \bm{b}_k)/p\nonumber\\
		&= (\m\bdelta -\m_o\bdelta_o)'\E[\x'\x](\m\bdelta -\m_o\bdelta_o)/p + \sigma^2/p \nonumber\\ 
        &- 2(\m\bdelta -\m_o\bdelta_o)'\E[\x'\x]\bm{b}_k/p + \bm{b}_k'\E[\x'\x]\bm{b}_k/p \nonumber\\
		&=\widetilde{Q}_o(\m\bdelta) + O(||\bm{b}_k||/\sqrt{p}) + O(||\bm{b}_k||^2/p) \nonumber\\ 
        &= \widetilde{Q}_o(\m\bdelta) + O(||\bm{b}_k||/\sqrt{p})
\end{align}
 
\noindent The second equality follows from \Cref{Ass:ZeroCorr_x.eps}, and  the fourth equality follows from \Cref{Ass:Bnds_x_eps,Ass:Bounded_Support_B} noting that by the Schwartz inequality, 
\begin{align*}
    |(\m\bdelta -\m_o\bdelta_o)'\E[\x'\x]\bm{b}_k|/p \leq ||\E[\x'\x]||(||\m\bdelta -\m_o\bdelta_o||/\sqrt{p})(||\bm{b}_k||/\sqrt{p}) \lesssim ||\bm{b}_k||/\sqrt{p}.
\end{align*}
\noindent The last equality follows from \Cref{Prop:Rate_ApprxBias} as $||\bm{b}_k||/\sqrt{p} \asymp 1/k $ whence $||\bm{b}_k||^2/p \asymp 1/k^2 \leq 1/k \asymp ||\bm{b}_k||/\sqrt{p} $ if $k <|\mathrm{supp}(\bbeta_o)|$ and zero otherwise, i.e., the $O(||\bm{b}_k||/\sqrt{p})$ term (weakly) dominates the $O(||\bm{b}_k||^2/p)$ term.

Now, at a fixed $k$ notice that \begin{align*}
    \widetilde{Q}_o(\m\bdelta) &= (\m\bdelta -\m_o\bdelta_o)'\E[\x'\x](\m\bdelta -\m_o\bdelta_o)/p + \sigma^2/p \\
    &= \lVert \bm\E[\x'\x]^{1/2}(\m\bdelta -\m_o\bdelta_o) \rVert^2/p + \sigma^2/p.  
\end{align*}
The term $\lVert \bm\E[\x'\x]^{1/2} (\m\bdelta -\m_o\bdelta_o) \rVert^2 \geq 0$ with equality holding if and only if $\m\bdelta = \m_o\bdelta_o$ as $\E[\x'\x] $ is positive definite by \Cref{Ass:Bnds_x_eps} -- see also \Cref{Rem:Bnds_Eig_XX}. Thus, $ \widetilde{Q}_o(\m\bdelta)$ is minimised at $\m\bdelta = \m_o\bdelta_o$ which concludes the proof.
\qed

\subsection{Proof of \Cref{Prop:Clustering}}
 \textbf{Part (a)}:	Under \Cref{Ass:ZeroCorr_x.eps}, the expected value of the objective function \eqref{eqn:Exp_ObjFunF} is 
	\begin{align*}
		Q_o(\m\bdelta) &= \mathbb{E}[(y - \x\m\bdelta)^2]/p = \mathbb{E}[(\varepsilon - \x(\m\bdelta - \bbeta_o))^2]/p\\ 
		&= \E[\varepsilon^2]/p + (\m\bdelta - \bbeta_o)'\mathbb{E}[\x'\x](\m\bdelta - \bbeta_o)/p \\
		&= \sigma^2/p + ||\m\bdelta - \bbeta_o ||_{\E[\x'\x]}^2/p
	\end{align*}
 \noindent which is a (weighted) clustering criterion for assigning elements of $ \bbeta_o $ to $ k $ groups in order to minimise the squared weighted norm $ ||\cdot||_{\E[\x'\x]}^2 $. 

 \textbf{Part (b)}:	By \Cref{Ass:Bnds_x_eps}, $ B^{-1}\leq \rho_{\mathrm{\min}}(\E[\x'\x]) \leq \rho_{\mathrm{\max}}(\E[\x'\x]) \leq B $ whence
	\begin{align*}
		||\m\bdelta - \bbeta_o||_{\E[\x'\x]}^2 & = ||\m\bdelta - \bbeta_o||^2 \frac{(\m\bdelta - \bbeta_o)'\E[\x'\x](\m\bdelta - \bbeta_o)}{||\m\bdelta - \bbeta_o||^2} = \tilde{\uptau}_p'\E[\x'\x]\tilde{\uptau}_p\cdot||\m\bdelta - \bbeta_o||^2\\ 
		&\asymp ||\m\bdelta - \bbeta_o||^2
	\end{align*}
 
 \noindent for $(\m\bdelta - \bbeta_o)/||\m\bdelta - \bbeta_o||=:\tilde{\uptau}_p\in\mathcal{S}_p$ whenever $ ||\m\bdelta - \bbeta_o||\neq 0 $ and $ 0 = ||\m\bdelta - \bbeta_o||^2  = ||\m\bdelta - \bbeta_o||_{\E[\x'\x]}^2 $ otherwise.
\qed

\subsection{Proof of \Cref{Theorem:Consis_Rate_M}}
Let $Y$ be the $n\times 1$ vector of the outcome and $\X$ be the $n\times p$ matrix of covariates. Fixing $ \m \in \mathcal{M}_k $, one has (using matrix notation)
\[
    \hat{\bdelta}_n(\m): = \argmin_{\bdelta \in \Delta}Q_n(\m\bdelta) = (\m'\X'\X\m)^{-1}(\m'\X'Y).
\]
	
\noindent Denote the $n\times 1$ vector of residuals by $ \mathcal{E}_n(\m) := Y - \X\m\hat{\bdelta}_n(\m) =  (\mathrm{I}_n - \X\m(\m'\X'\X\m)^{-1}\m'\X')Y =: \mu_n(\m)Y$ where $ \mathrm{I}_n $ is the $ n \times n $ identity matrix. Also, define $ \mathcal{E}:= (\varepsilon_1,\ldots,\varepsilon_n)' $. Substitute in the DGP \eqref{eqn:basemodel}  to obtain $ \mathcal{E}_n(\m) = \mu_n(\m)\X\bbeta_o + \mu_n(\m)\mathcal{E} $. Using the decomposition $\bbeta_o = \m_o\bdelta_o + \bm{b}_k$, observe that at $\m = \m_o$, one has 
		$ \mathcal{E}_n(\m_o) = \mu_n(\m_o)(\mathcal{E} + \X \bm{b}_k) $ as
		$$ 
		\mu_n(\m_o)\X\m_o\bdelta_o = (\mathrm{I}_n - \X\m_o(\m_o'\X'\X\m_o)^{-1}\m_o'\X')\X\m_o\bdelta_o = 0. 
		$$

\noindent Define $\hat{\mathcal{E}}_{m}:= \mu_n(\m)\mathcal{E} $ as the $n\times 1$ vector of residuals from regressing $\mathcal{E}$ on $\X\m$. Given a finite sample, $\hat{\m}_n\neq\m_o$ implies that the objective function at $\m=\hat{\m}_n$, namely $Q_n(\hat{\m}_n\hat{\bdelta}_n(\hat{\m}_n))$, yields a smaller sum of squared residuals relative to the objective function at $\m=\m_o$. Thus for any $\eta_m >0 $, there exists an $\eta_\varepsilon>0 $ such that 
\begin{align}\label{eqn:mnhat_prob}
	&\mathbb{P}(||\hat{\m}_n-\m_o||\geq \eta_m) \leq \mathbb{P}(Q_n(\m_o\hat{\bdelta}_n(\m_o)) - Q_n(\hat{\m}_n\hat{\bdelta}_n(\hat{\m}_n)) \geq \eta_\varepsilon\big)\\  
		&= \mathbb{P}(\mathcal{E}_n(\m_o)'\mathcal{E}_n(\m_o) -\mathcal{E}_n(\hat{\m}_n)'\mathcal{E}_n(\hat{\m}_n) \geq \eta_\varepsilon np\big)\nonumber\\ 
		&= \mathbb{P}\big(\hat{\mathcal{E}}_{m_o}'\hat{\mathcal{E}}_{m_o} - \hat{\mathcal{E}}_{\hat{m}_n}'\hat{\mathcal{E}}_{\hat{m}_n} + 2\bm{b}_k'\X'\hat{\mathcal{E}}_{m_o} - 2\bbeta_o'\X'\hat{\mathcal{E}}_{\hat{m}_n} + \bm{b}_k'\X'\mu_n(\m_o)\X \bm{b}_k - ||\mu_n(\hat{\m}_n)\X\bbeta_o||^2 \geq \eta_\varepsilon np \big)\nonumber\\
		&\leq \max_{\m\in\mathcal{M}_k} \mathbb{P}\Big( \E_n[\hat{\varepsilon}_{m_o,i}^2 - \hat{\varepsilon}_{m,i}^2] + 2\bm{b}_k'\mathbb{E}_n[\x_i'\hat{\varepsilon}_{m_o,i}] - 2\bbeta_o'\mathbb{E}_n[\x_i'\hat{\varepsilon}_{m,i}] + \bm{b}_k'\mathbb{E}_n[\x_i'\x_i]\bm{b}_k \geq \eta_\varepsilon p \Big)\nonumber\\
        &=: \mathbb{P}\Big( \E_n[\hat{\varepsilon}_{m_o,i}^2 - \hat{\varepsilon}_{m^*,i}^2] + 2\bm{b}_k'\mathbb{E}_n[\x_i'\hat{\varepsilon}_{m_o,i}]- 2\bbeta_o'\mathbb{E}_n[\x_i'\hat{\varepsilon}_{m^*,i}] + \bm{b}_k'\mathbb{E}_n[\x_i'\x_i]\bm{b}_k \geq \eta_\varepsilon p \Big)\nonumber\\
        &\leq \mathbb{P}\Big( \max\Big\{ \E_n[\hat{\varepsilon}_{m_o,i}^2 - \hat{\varepsilon}_{m^*,i}^2], \ 2\bm{b}_k'\mathbb{E}_n[\x_i'\hat{\varepsilon}_{m_o,i}], \ -2\bbeta_o'\mathbb{E}_n[\x_i'\hat{\varepsilon}_{m^*,i}],\  \bm{b}_k'\mathbb{E}_n[\x_i'\x_i]\bm{b}_k \Big\} \geq \eta_\varepsilon p/4 \Big)\nonumber\\
        & \leq \mathbb{P}\Big(|\E_n[\hat{\varepsilon}_{m_o,i}^2 - \hat{\varepsilon}_{m^*,i}^2]| \geq \eta_\varepsilon p/4\Big)  + \mathbb{P}\Big(|\bbeta_o'\mathbb{E}_n[\x_i'\hat{\varepsilon}_{m^*,i}]| \geq \eta_\varepsilon p/8\Big) \nonumber\\ 
        & + \mathbb{P}\Big( |\bm{b}_k'\mathbb{E}_n[\x_i'\hat{\varepsilon}_{m_o,i}]| \geq \eta_\varepsilon p/8\Big) + \mathbb{P}\Big(\bm{b}_k'\mathbb{E}_n[\x_i'\x_i]\bm{b}_k \geq \eta_\varepsilon p/4\Big)\nonumber\\
        & \lesssim k(np)^{-1} + (np)^{-1} + ||\bm{b}_k||^2/(p^2n) + ||\bm{b}_k||^2/p \nonumber\\
        & \lesssim ||\bm{b}_k||^2/p + n^{-1}.\nonumber
\end{align}

\noindent The first equality follows by the definition of the objective function $Q_n(\m\hat{\bdelta}_n(\m))$, and the second equality is an algebraic expansion and re-arrangement of terms noting that $\mu_n(\m)$ is idempotent for any $\m\in\mathcal{M}_k$. The penultimate line follows from \Cref{Lem:Aux_mnconsistent}(a), \Cref{Lem:Aux_mnconsistent}(b), Assumptions \ref{Ass:Bounded_Support_B}, \ref{Ass:Bnds_x_eps}, and the Markov inequality. For the second and third terms of the penultimate line, observe that for $\mathcal{S}_p \ni \tilde{\uptau}_p:= \frac{\bbeta}{||\bbeta||} $,

\begin{align*}
    \frac{1}{p}\bbeta'\mathbb{E}_n[\x_i'\hat{\varepsilon}_{m^*,i}] = \tilde{\uptau}_p'\mathbb{E}_n[\x_i'\hat{\varepsilon}_{m^*,i}]/(p||\bbeta||^{-1}).
\end{align*}
The second term of the penultimate line further uses the boundedness condition on $\bbeta_o$ (\Cref{Ass:Bounded_Support_B}). For the last line, notice that $k(np)^{-1}\leq n^{-1}$ as $k\leq p$.

As $ \eta_m >0 $ is arbitrary, it can be chosen sufficiently small such that $ \sqrt{2} \geq \eta_m >0 $, then
	\begin{align*}
		||\hat{\m}_n-\m_o|| &= ||\hat{\m}_n-\m_o||\cdot\mathrm{I}(||\hat{\m}_n-\m_o||\geq\eta_m) + ||\hat{\m}_n-\m_o||\cdot\mathrm{I}(||\hat{\m}_n-\m_o||<\eta_m)\\
		&= ||\hat{\m}_n-\m_o||\cdot\mathrm{I}(||\hat{\m}_n-\m_o||\geq\eta_m)\\
		&\lesssim \mathrm{I}(||\hat{\m}_n-\m_o||\geq\eta_m) \ a.s.
	\end{align*}
\noindent The second equality follows because $ \m\in\mathcal{M}_k $ is a binary matrix with $ \{0,1\} $ entries whence $ ||\hat{\m}_n-\m_o||\cdot\mathrm{I}(||\hat{\m}_n-\m_o||<\sqrt{2}) = 0 $ by construction, i.e., $\mathrm{I}(||\hat{\m}_n-\m_o||<\sqrt{2})=1$ if and only if $||\hat{\m}_n-\m_o||=0$. The last line follows from the triangle inequality and the boundedness of group sizes from above (\Cref{Cond:m}). It follows from the foregoing that 
	\[\E[||\hat{\m}_n-\m_o||] \lesssim \mathbb{P}(||\hat{\m}_n-\m_o||\geq \eta_m) = O(||\bm{b}_k||^2/p) + O(n^{-1}).\]
 \noindent The conclusion follows from the Markov inequality.
\qed

\subsection{Proof of \Cref{Theorem:Consis_AsympNorm}}
Thanks to the decomposition
\begin{align*}
    \widehat{\bbeta}_n - \bbeta_o = \hat{\m}_n\hat{\bdelta}_n(\hat{\m}_n) - \hat{\m}_n\hat{\bdelta}_n(\m_o) + \hat{\m}_n\hat{\bdelta}_n(\m_o) - \hat{\m}_n\bdelta_o + \hat{\m}_n\bdelta_o - \m_o\bdelta_o + \m_o\bdelta_o - \bbeta_o,
\end{align*}
one obtains the following useful decomposition for the study of GPE's convergence rate:
\begin{align}\label{eqn:beta_decomp}
    \uptau_p'(\widehat{\bbeta}_n - \bbeta_o\big) = \uptau_p'\hat{\m}_n(\hat{\bdelta}_n(\m_o) - \bdelta_o) + \uptau_p'\hat{\m}_n(\hat{\bdelta}_n(\hat{\m}_n) - \hat{\bdelta}_n(\m_o)) + \uptau_p'(\hat{\m}_n -\m_o)\bdelta_o + \uptau_p'\bm{b}_k.
\end{align}

\noindent From the proof of \Cref{Lem:Conv_CLT_delta}, $ \hat{\bdelta}_n(\m_o) - \bdelta_o = \hat{A}_{m_o}^{-1}\m_o'\E_n[\x_i'\x_i]\bm{b}_k + \hat{A}_{m_o}^{-1}\m_o'\E_n[\x_i'\varepsilon_i]$. Applying this expression to further decompose \eqref{eqn:beta_decomp} gives:
\begin{align*}
    \uptau_p'(\widehat{\bbeta}_n - \bbeta_o) &= \uptau_p'\m_o(\hat{\bdelta}_n(\m_o) - \bdelta_o) + \uptau_p'(\hat{\m}_n-\m_o)(\hat{\bdelta}_n(\m_o) - \bdelta_o) \\
    & + \uptau_p'\hat{\m}_n(\hat{\bdelta}_n(\hat{\m}_n) - \hat{\bdelta}_n(\m_o)) + \uptau_p'(\hat{\m}_n -\m_o)\bdelta_o + \uptau_p'\bm{b}_k\\
    &= \uptau_p'\m_o\hat{A}_{m_o}^{-1}\m_o'\E_n[\x_i'\varepsilon_i] + \uptau_p'\widehat{\mathcal{B}}_k.
\end{align*}

\noindent Observe that
\begin{equation}\label{eqn:B_k_decomp}
    \begin{split}
        \widehat{\mathcal{B}}_k &= \m_o\hat{A}_{m_o}^{-1}\m_o'\E_n[\x_i'\x_i]\bm{b}_k + (\hat{\m}_n-\m_o)(\hat{\bdelta}_n(\m_o) - \bdelta_o) + \hat{\m}_n(\hat{\bdelta}_n(\hat{\m}_n) - \hat{\bdelta}_n(\m_o)) \\
        & + (\hat{\m}_n -\m_o)\bdelta_o + \bm{b}_k
    \end{split}
\end{equation}
\noindent simplifies to $\widehat{\mathcal{B}}_k = \widehat{\bbeta}_n - \m_o\hat{\bdelta}_n(\m_o) + (\mathrm{I}_p + \m_o\hat{A}_{m_o}^{-1}\m_o'\E_n[\x_i'\x_i])\bm{b}_k $ as given in the main text. 

\textbf{Part (a)}:	The proof of this part proceeds by studying the summands of \eqref{eqn:B_k_decomp} in turn. 

\noindent Step I: Under \Cref{Ass:Convergence_Am},
\[
    \uptau_p'\m_o\hat{A}_{m_o}^{-1}\m_o'\E_n[\x_i'\x_i]\bm{b}_k = O_p(||\bm{b}_k||)
\]
\noindent follows from the first part of the proof of \Cref{Lem:Conv_CLT_delta}.

\noindent Step II: $ \uptau_p'(\hat{\m}_n-\m_o)(\hat{\bdelta}_n(\m_o) - \bdelta_o) =0$ if $||\hat{\m}_n-\m_o||=0$ otherwise define $ \tilde{\uptau}_k:= (\hat{\m}_n-\m_o)\uptau_p/||(\hat{\m}_n-\m_o)\uptau_p|| $. Under the conditions of \Cref{Lem:Conv_CLT_delta}, \Cref{Theorem:Consis_Rate_M}, and \Cref{Prop:Rate_ApprxBias},
\begin{align*}
    \uptau_p'(\hat{\m}_n-\m_o)(\hat{\bdelta}_n(\m_o) - \bdelta_o) &= \tilde{\uptau}_k'(\hat{\bdelta}_n(\m_o) - \bdelta_o)\cdot ||\hat{\m}_n-\m_o|| \\
    &=O_p(||\bm{b}_k||^2/(p\sqrt{n})) + O_p(||\bm{b}_k||^3/p) + O_p(||\bm{b}_k||/n) + O_p(n^{-3/2})\\
    &= O_p(||\bm{b}_k||^2/(p\sqrt{n})) + O_p(||\bm{b}_k||/k^2) + O_p(||\bm{b}_k||/n) + O_p(n^{-3/2})
\end{align*}

\noindent since $||\bm{b}_k||^3/p \asymp p^{3/2}/(pk^3) = (\sqrt{p}/k)/k^2 \asymp ||\bm{b}_k||/k^2 $ by \Cref{Theorem:Consis_Rate_M} if $k<|\mathrm{supp}(\bbeta_o)|$ and zero otherwise.

\noindent Step III: Define $\tilde{\uptau}_k:= \hat{\m}_n\uptau_p/||\hat{\m}_n\uptau_p|| $, then
\begin{align*}
    |\uptau_p'\hat{\m}_n(\hat{\bdelta}_n(\hat{\m}_n) - \hat{\bdelta}_n(\m_o))| = |\tilde{\uptau}_k'(\hat{\bdelta}_n(\hat{\m}_n) - \hat{\bdelta}_n(\m_o))|\cdot ||\uptau_p'\hat{\m}_n|| \lesssim |\tilde{\uptau}_k'(\hat{\bdelta}_n(\hat{\m}_n) - \hat{\bdelta}_n(\m_o))|
\end{align*}
\noindent since group sizes are uniformly bounded from above (\Cref{Cond:m}). It follows from the conditions of \Cref{Lem:Rate_delmn_delmo} that 
$\uptau_p'\hat{\m}_n(\hat{\bdelta}_n(\hat{\m}_n) - \hat{\bdelta}_n(\m_o)) = O_p(||\bm{b}_k||^2/\sqrt{p}) + O_p(\sqrt{p}/n) $. 

\noindent Step IV: Define $\tilde{\uptau}_k:= (\hat{\m}_n -\m_o)\uptau_p/||(\hat{\m}_n -\m_o)\uptau_p|| $ if $\hat{\m}_n \neq \m_o$ else $|\uptau_p'(\hat{\m}_n -\m_o)\bdelta_o|=0$.
\begin{align*}
    |\uptau_p'(\hat{\m}_n -\m_o)\bdelta_o| = |\tilde{\uptau}_k'\bdelta_o|\cdot||\hat{\m}_n -\m_o||\lesssim \sqrt{k}||\hat{\m}_n -\m_o||\leq \sqrt{p}||\hat{\m}_n -\m_o||
\end{align*} under \Cref{Ass:Bounded_Support_B} since elements of $\bdelta_o$ are averages of sub-vectors of $\bbeta_o$ and $k\leq p$. Thus, $\uptau_p'(\hat{\m}_n -\m_o)\bdelta_o = O_p(||\bm{b}_k||^2/\sqrt{p}) + O_p(\sqrt{p}/n) $ by \Cref{Theorem:Consis_Rate_M}. 

Putting terms together, $\uptau_p'\widehat{\mathcal{B}}_k = O_p(||\bm{b}_k||) + O_p(||\bm{b}_k||^2/\sqrt{p})  + O_p(\sqrt{p}/n) = O_p(||\bm{b}_k||) + O_p(\sqrt{p}/n) $ by the argument in \eqref{eqn:bk_dom_bk2rp}. Further, $O_p(\sqrt{p}/n) = o(n^{-1/2})$ by \Cref{Ass:Rate_p_n}, whence $\uptau_p'\widehat{\mathcal{B}}_k = O_p(||\bm{b}_k||) + o_p(n^{-1/2}) $ as claimed.
 
\textbf{Part (b)}:	This part follows from the second part of the proof of \Cref{Lem:Conv_CLT_delta}, i.e., 
 \[\uptau_p'(\widehat{\bbeta}_n - \bbeta_o - \widehat{\mathcal{B}}_k) = \uptau_p'\m_o\hat{A}_{m_o}^{-1}\m_o'\E_n[\x_i'\varepsilon_i]  = O_p(n^{-1/2}).\]

\textbf{Part (c)}:	From part (b) above, \Cref{Ass:Convergence_Am}, the continuity of the inverse at a non-singular matrix, and the continuous mapping theorem,
\begin{equation*}
\sqrt{n}\uptau_p'(\widehat{\bbeta}_n - \bbeta_o - \widehat{\mathcal{B}}_k) = \uptau_p'\m_oA_{m_o}^{-1}\m_o'(\sqrt{n}\E_n[\x_i'\varepsilon_i]) + o_p(1).
\end{equation*}

\noindent Under \Cref{Ass:Sampling,Ass:ZeroCorr_x.eps}, 
$$
\mathrm{var}[\uptau_p'\m_oA_{m_o}^{-1}\m_o'(\sqrt{n}\E_n[\x_i'\varepsilon_i])] = \uptau_p'\m_oA_{m_o}^{-1}\m_o'\E[\x'\x\varepsilon^2]\m_oA_{m_o}^{-1}\m_o'\uptau_p=\sigma_\theta^2.
$$
\noindent The conclusion follows from \Cref{Lem:CLT}.
\qed

\subsection{Proof of \Cref{Theorem:Select_k}}
Recall the decomposition: $\widehat{\bbeta}_n - \bbeta_o = \widehat{\mathcal{A}}_k + \widehat{\mathcal{B}}_k $ used in \Cref{Theorem:Consis_AsympNorm}, and observe that under the DGP \eqref{eqn:basemodel}, the average squared GPE residuals can be expressed as
\begin{align}\label{eqn:RSS_decomp}
    \mathbb{E}_n[(y_i - \x_i\widehat{\bbeta}_n)^2] &= \mathbb{E}_n[\varepsilon_i^2] -2(\widehat{\bbeta}_n-\bbeta_o)'\mathbb{E}_n[\x_i'\varepsilon_i] + (\widehat{\bbeta}_n-\bbeta_o)'\mathbb{E}_n[\x_i'\x_i](\widehat{\bbeta}_n-\bbeta_o)\\ \nonumber
    &= \mathbb{E}_n[\varepsilon_i^2] - 2\widehat{\mathcal{A}}_k'\mathbb{E}_n[\x_i'\varepsilon_i] + \widehat{\mathcal{A}}_k'\mathbb{E}_n[\x_i'\x_i]\widehat{\mathcal{A}}_k + 2\widehat{\mathcal{B}}_k'\mathbb{E}_n[\x_i'\x_i]\widehat{\mathcal{A}}_k - 2\widehat{\mathcal{B}}_k'\mathbb{E}_n[\x_i'\varepsilon_i]\\ \nonumber
    &+ \widehat{\mathcal{B}}_k'\mathbb{E}_n[\x_i'\x_i]\widehat{\mathcal{B}}_k\\ \nonumber
    &= \mathbb{E}_n[\varepsilon_i^2] - \mathbb{E}_n[\varepsilon_i\x_i]\m_o\hat{A}_{m_o}^{-1}\m_o'\mathbb{E}_n[\x_i'\varepsilon_i]\\ \nonumber
    &- 2\widehat{\mathcal{B}}_k'(\mathrm{I}_p - \mathbb{E}_n[\x_i'\x_i]\m_o\hat{A}_{m_o}^{-1}\m_o')\mathbb{E}_n[\x_i'\varepsilon_i] + \widehat{\mathcal{B}}_k'\mathbb{E}_n[\x_i'\x_i]\widehat{\mathcal{B}}_k.
\end{align}

It follows from \eqref{eqn:RSS_decomp} and the conditions of \Cref{Lem:phi1_Op1} that 
\begin{align}\label{eqn:diff_RSS}
    &\mathbb{E}_n[(y_i - \x_i\widehat{\bbeta}_n^k)^2] - \mathbb{E}_n[(y_i - \x_i\widehat{\bbeta}_n^{k+1})^2] \nonumber \\ 
    &= \mathbb{E}_n[\varepsilon_i\x_i]\m_o^{(k+1)}\hat{A}_{m_o^{k+1}}^{-1}\m_o^{(k+1)\prime}\mathbb{E}_n[\x_i'\varepsilon_i] - \mathbb{E}_n[\varepsilon_i\x_i]\m_o^k\hat{A}_{m_o^k}^{-1}\m_o^{k\prime}\mathbb{E}_n[\x_i'\varepsilon_i] \nonumber \\ 
    &-2\Big(\widehat{\mathcal{B}}_k'\big(\mathrm{I}_p - \mathbb{E}_n[\x_i'\x_i]\m_o^k\hat{A}_{m_o^k}^{-1}\m_o^{k\prime}\big) -  \widehat{\mathcal{B}}_{k+1}'\big(\mathrm{I}_p - \mathbb{E}_n[\x_i'\x_i]\m_o^{(k+1)}\hat{A}_{m_o^{k+1}}^{-1}\m_o^{(k+1)\prime}\big)\Big)\mathbb{E}_n[\x_i'\varepsilon_i] \nonumber \\ 
    &+ \widehat{\mathcal{B}}_k'\mathbb{E}_n[\x_i'\x_i]\widehat{\mathcal{B}}_k - \widehat{\mathcal{B}}_{k+1}'\mathbb{E}_n[\x_i'\x_i]\widehat{\mathcal{B}}_{k+1} \nonumber \\
    &= O_p(n^{-1}) + O_p(n^{-1/2}\mathbb{E}[||\widehat{\mathcal{B}}_k||]) + O_p(\mathbb{E}[||\widehat{\mathcal{B}}_k||^2]) 
\end{align}

\noindent under \Cref{Ass:Bounded_Support_B,Ass:Bnds_x_eps,Ass:ZeroCorr_x.eps,Ass:Sampling} and that $||\bm{b}_k||$ is non-increasing in $k$ (\Cref{Prop:Rate_ApprxBias}) -- recall $\uptau_p'\widehat{\mathcal{B}}_k = O_p(||\bm{b}_k||) + o_p(n^{-1/2})$ by \Cref{Theorem:Consis_AsympNorm}(a). The first term of \eqref{eqn:diff_RSS} is always $O_p(n^{-1})$ under the conditions of \Cref{Lem:phi1_Op1}. 

\textbf{Part (a):}  Thus from \eqref{eqn:RSS_decomp}, \eqref{eqn:diff_RSS}, and the preceding argument,
\begin{align*}
    n\widehat{\varphi}_n(k) = \frac{O_p(1) + O_p(n^{1/2}\mathbb{E}[||\widehat{\mathcal{B}}_k||]) + O_p(n\mathbb{E}[||\widehat{\mathcal{B}}_k||^2])}{O_p(1) + O_p(n^{-1/2}\mathbb{E}[||\widehat{\mathcal{B}}_{k+1}||]) + O_p(\mathbb{E}[||\widehat{\mathcal{B}}_{k+1}||^2])}
\end{align*}where (for emphasis), the denominator, which is $\mathbb{E}_n[(y_i - \x_i\widehat{\bbeta}_n)^2]$ is positive $a.s.$ Therefore $n\widehat{\varphi}_n(k) = O_p(1)$ if $a_k\leq 2 $, i.e., if $\E[||\widehat{\mathcal{B}}_k||] \lesssim n^{-1/a_k} = n^{-1/2} $. This proves the ``if" part.

\textbf{Part (b):}  For the ``only if" part, rearrange terms in the expression of $n\widehat{\varphi}_n(k)$ using \eqref{eqn:RSS_decomp} and \eqref{eqn:diff_RSS} to get
\begin{align*}    & \underbrace{n\Big(\widehat{\mathcal{B}}_k'\mathbb{E}_n[\x_i'\x_i]\widehat{\mathcal{B}}_k - (1+\widehat{\varphi}_n(k))\widehat{\mathcal{B}}_{k+1}'\mathbb{E}_n[\x_i'\x_i]\widehat{\mathcal{B}}_{k+1}\Big)}_{\widehat{T}_{n,1}}\\
&-2\underbrace{n\Big(\widehat{\mathcal{B}}_k'\big(\mathrm{I}_p - \mathbb{E}_n[\x_i'\x_i]\m_o^k\hat{A}_{m_o^k}^{-1}\m_o^{k\prime}\big) - (1-\widehat{\varphi}_n(k)) \widehat{\mathcal{B}}_{k+1}'\big(\mathrm{I}_p - \mathbb{E}_n[\x_i'\x_i]\m_o^{(k+1)}\hat{A}_{m_o^{k+1}}^{-1}\m_o^{(k+1)\prime}\big)\Big)\mathbb{E}_n[\x_i'\varepsilon_i]}_{\widehat{T}_{n,2}}\\
&= n\widehat{\varphi}_n(k)\mathbb{E}_n[\varepsilon_i^2] - \underbrace{n\Big((1+\widehat{\varphi}_n(k))\mathbb{E}_n[\varepsilon_i\x_i]\m_o^{(k+1)}\hat{A}_{m_o^{k+1}}^{-1}\m_o^{(k+1)\prime}\mathbb{E}_n[\x_i'\varepsilon_i] - \mathbb{E}_n[\varepsilon_i\x_i]\m_o^k\hat{A}_{m_o^k}^{-1}\m_o^{k\prime}\mathbb{E}_n[\x_i'\varepsilon_i]\Big)}_{\widehat{T}_{n,3}}.
\end{align*}
Since $||\bm{b}_k||$ is non-increasing in $k$ (\Cref{Prop:Rate_ApprxBias}), it follows from \Cref{Theorem:Consis_AsympNorm}(a) and \Cref{Ass:Bnds_x_eps} that $\widehat{T}_{n,1}=O_p(n\mathbb{E}[||\widehat{\mathcal{B}}_k||^2)$ and $\widehat{T}_{n,2}=O_p(n^{1/2}\mathbb{E}[||\widehat{\mathcal{B}}_k||)$ if $n\widehat{\varphi}_n(k) = O_p(1) $, i.e., if $\widehat{\varphi}_n(k) = O_p(n^{-1})$. $\widehat{T}_{n,3}=O_p(1)$ if $\widehat{\varphi}_n(k) = O_p(n^{-1})$ under the conditions of \Cref{Lem:phi1_Op1}. Thus, the above equation (in $O_p(\cdot)$ notation) simplifies to
$$
O_p(n\mathbb{E}[||\widehat{\mathcal{B}}_k||^2]) + O_p(n^{1/2}\mathbb{E}[||\widehat{\mathcal{B}}_k||]) = O_p(1)
$$which further implies that $ n^{1/2}\mathbb{E}[||\widehat{\mathcal{B}}_k||] \lesssim 1 $, i.e., $a_k<2$ if $n\widehat{\varphi}_n(k) = O_p(1) $. The conclusion of this part follows by contraposition noting that $n\widehat{\varphi}_n(k)\geq 0$ by construction.

\qed

\section{Supporting Lemmata}\label{Sect:Useful_Lemmata}
Lemmata contained in this section are used in the proofs of the theoretical results in the main text.
\subsection{Convergence rate: non-stochastic $k$-means clustering}
Consider the $k$-means clustering problem for a non-stochastic vector $\mathscr{B}_p= (b_1,\ldots,b_p)'$ in $ \R^p $. The problem involves partitioning $ \mathscr{B}_p $ into $ k $ groups $ \bm{S} = \{S_1,\ldots,S_k\} $, $ k \geq 1 $, in order to minimise the sum of within-group sums of squared deviations: 

\[
\{g_1^*, \ldots,  g_p^*\} =  \displaystyle \argmin_{\{ g_1, \ldots, g_p\}} \sum_{j=1}^{p}(b_j - \tilde{b}(g_j))^2
\]

\noindent where $ \tilde{b}(l) $ is the mean of elements $ b_j $  in group $ S_{l}:= \{b_j:g_j=l\} $, and $ g_j \in \{1,\ldots,k\} $. Let $ \{ S_{l}^*, 1\leq l \leq k \} $ denote the sets of optimal group assignments and $ |S_l| $ denote the cardinality of $ S_l $. The following lemma establishes a worst-case bound on the maximum absolute within-group deviation and the average of within-group squared deviations in terms of $ k $.

\begin{lemma}\label{Apndix:Useful:Lem_Max_K_b}
Suppose $\mathscr{B}_p \in [-C_b,C_b]^p $, i.e., it has a bounded support in $ \R $ uniformly in $p$, $\displaystyle \{g_1^*, \ldots,  g_p^*\}:= \argmin_{\{ g_1, \ldots, g_p\}} \sum_{j=1}^{p}(b_j - \tilde{b}(g_j))^2  $ are the optimal group assignments where $ g_j \in \{1,\ldots,k\} $ for each $j\in \{1,\ldots,p\} $, group sizes are bounded above by $M<\infty$ and away from zero uniformly in $ p $, and $k<|\mathrm{supp}(\mathscr{B}_p)|$, i.e., $k$ is less than the number of support points of $\mathscr{B}_p$, then (a) $\displaystyle \max_{1 \leq j \leq p} |b_j - \tilde{b}(g_j^*)| \leq 2C_bM/k $ and (b) $\displaystyle \min_{\{ g_1, \ldots, g_p\}} \frac{1}{p} \sum_{j=1}^{p}(b_j - \tilde{b}(g_j))^2 \leq 4C_b^2M^2/k^2 $.
\end{lemma}

\paragraph{Proof:}
Uniformly in $p$, $\displaystyle \max_{1 \leq j \leq p}|b_j| \leq C_b$ by the boundedness of the support of $ \mathscr{B}_p $.
	\vspace{-.5cm}
	\begin{align*}\label{}
		\begin{split}
			\min_{\{ g_1, \ldots, g_p\}} \frac{1}{p} \sum_{j=1}^{p}(b_j - \tilde{b}(g_j))^2 &= \min_{\bm{S}} \frac{1}{p}\sum_{l=1}^{k}\sum_{b \in S_l }(b - \tilde{b}(l))^2\\ 
			&= \frac{1}{p}\sum_{l=1}^{k}\sum_{b \in S_l^* }(b - \tilde{b}(l))^2 \leq \sum_{l=1}^{k}\frac{|S_l^*|}{p}\max_{b \in S_l^* }|b - \tilde{b}(l)|^2\\
			& =: \sum_{l=1}^{k}\tilde{p}_l\tilde{r}_l^2
		\end{split}
	\end{align*}
	
\noindent where $\displaystyle \{S_1^*,\ldots,S_k^*\} = \argmin_{\bm{S}} \frac{1}{p}\sum_{l=1}^{k}\sum_{b \in S_l }(b - \tilde{b}(l))^2 $, $ \ \tilde{p}_l := \frac{|S_l^*|}{p}  $, and $\displaystyle \tilde{r}_l :=  \max_{b \in S_l^* }|b - \tilde{b}(l)| $ is the $l$'th group's maximum within-group deviation from the group mean. $ \sum_{l=1}^{k}\tilde{p}_l = 1 $ such that $ \tilde{p}_l > 0 $ for each $ l=1,\ldots,k $ as there are no empty groups.

\textbf{Part (a):}  Denote the diameter of group $ S_l^* $ by $ \displaystyle \mathrm{diam}(S_l^*) =  \max_{b_1, b_2 \in S_l^* }|b_1-b_2| $. At the minimised sum of within-group sums of squared deviations, the sum of group diameters cannot exceed the length of the support, i.e.,  $\displaystyle \sum_{l=1}^{k} \mathrm{diam}(S_l^*) \leq 2C_b $ because groups do not overlap -- see, e.g., \textcite{fisher1958grouping} on the contiguity of least squares partitions. Since maximum within-group deviation from the group mean cannot exceed the diameter of the group, i.e.,  $\displaystyle \max_{1 \leq l \leq k} \big(\tilde{r}_l - \mathrm{diam}(S_l^*)\big) \leq 0  $, it follows that $ \sum_{l=1}^{k}\tilde{r}_l \leq 2C_b $. 
	
Define a parameter vector $ \bm{r} := (r_1,\ldots,r_k)' \in \R_+^k$, then
	\begin{equation*}\label{key}
		\sum_{l=1}^{k}\tilde{p}_l\tilde{r}_l^2 \leq \max_{ \bm{r} } \sum_{l=1}^{k}\tilde{p}_lr_l^2 \text{ subject to } \sum_{l=1}^{k}r_l \leq 2C_b. 
	\end{equation*}
 
 \noindent The Lagrangian corresponding to the constrained maximisation problem above is given by 
	\[\mathcal{L}(\bm{r},\lambda) = \sum_{l=1}^{k}\tilde{p}_lr_l^2 -\lambda (\sum_{l=1}^{k}r_l - 2C_b)\] where $ \lambda \geq 0 $ is the Lagrange multiplier. Taking partial derivatives gives
	\begin{align}
		\label{eqn:FOC_r}\frac{\partial \mathcal{L}(\bm{r})}{\partial r_l} &:= 2\tilde{p}_lr_l - \lambda, \ l = 1,\ldots,k \text{ and} \\
		\label{eqn:FOC_lambda}\frac{\partial \mathcal{L}(\bm{r})}{\partial \lambda} &:= -(\sum_{l=1}^{k}r_l - 2C_b).
	\end{align}

\noindent The case of $ \lambda = 0 $ implies the maximiser of $ \mathcal{L}(\bm{r}) $  is $ r_l^*=0 $ for each $ l=1,\ldots,k $ from \eqref{eqn:FOC_r} but this rather minimises the objective function. This case is hence ruled out. For $ \lambda > 0 $, \eqref{eqn:FOC_lambda} gives $ \sum_{l=1}^{k}r_l - 2C_b = 0 $ which is equivalently expressed as $ r_k = 2C_b-\sum_{l=1}^{k-1}r_l  $. Setting the set of equations \eqref{eqn:FOC_r} to zero results in $ 2\tilde{p}_lr_l = \lambda $ for each $ l=1,\ldots,k $ and $ \tilde{p}_lr_l = \tilde{p}_kr_k$ for each $ l = 1,\ldots, k-1 $ with $ r_k = 2C_b-\sum_{l'=1}^{k-1}r_{l'} $. Substituting $ r_k = 2C_b-\sum_{l'=1}^{k-1}r_{l'} $ into the $ k-1 $ equations gives $ \tilde{p}_lr_l = \tilde{p}_k(2C_b-\sum_{l'=1}^{k-1}r_{l'}), \ l = 1,\ldots, k-1 $. Dividing through by $ \tilde{p}_k $ and rearranging terms gives $ \bm{A}\bm{r}_{-k} = 2\mathbbm{1}_{k-1}C_b $ where 
	
	\[ \bm{A} = \left( \begin{array}{cccc}
		1 + \frac{\tilde{p}_1}{\tilde{p}_k} & 1 & \ldots & 1 \\
		\vdots & \ddots & \ldots & \vdots\\
		1 & 1 & \ldots & 1 + \frac{\tilde{p}_{k-1}}{\tilde{p}_k} \\
	\end{array} \right)
	\], $ \bm{r}_{-k} = (r_1,\ldots,r_{k-1})' $, and $ \mathbbm{1}_{k-1} $ is a $ (k-1) \times 1 $ vector of ones. $ \bm{A}\bm{r}_{-k} = 2\mathbbm{1}_{k-1}C_b $ is solved by $ \bm{r}_{-k}^* := 2\bm{A}^{-1}\mathbbm{1}_{k-1}C_b $. 
 
 The next step is to obtain the explicit expression for $ \bm{A}^{-1} $. Note that $ \bm{A} = \bm{D} + \mathbbm{1}_{k-1}\mathbbm{1}_{k-1}' $ where $ \bm{D} $ is a diagonal matrix with diagonal entries $ \{D_{ll} := \frac{\tilde{p}_l}{\tilde{p}_k} $, $ 1 \leq l \leq k-1\} $. Applying the formula of \textcite{sherman1949adjustment}, see e.g., \textcite[eqn. 2]{bartlett1951inverse},
	
	\[ (\bm{A}^{-1})_{ll'} = \begin{cases} 
		\frac{\tilde{p}_k}{\tilde{p}_l} - \frac{\tilde{p}_k^2}{q\tilde{p}_l^2} & \text{ if } l=l' \\
		-\frac{\tilde{p}_k^2}{q\tilde{p}_l\tilde{p}_{l'}} & \text{ if } l\not=l'
	\end{cases}
	\] where $ q = 1 + \sum_{\ell=1}^{k-1}\frac{\tilde{p}_k}{\tilde{p}_\ell} $. It hence follows that for each $ l \in \{1,\ldots,k-1\} $,
	\begin{align*}
		r_l^* &= 2C_b\sum_{l'=1}^{k-1} (\bm{A}^{-1})_{ll'} = 2C_b\Big(\frac{\tilde{p}_k}{\tilde{p}_l} - \frac{\tilde{p}_k^2}{q\tilde{p}_l^2} -  \sum_{l'\not=l}^{k-1}\frac{\tilde{p}_k^2}{q\tilde{p}_l\tilde{p}_{l'}}\Big) = 2C_b\frac{\tilde{p}_k}{q\tilde{p}_l}\Big(q -   \sum_{l'=1}^{k-1}\frac{\tilde{p}_k}{\tilde{p}_{l'}}\Big)\\ 
		&= 2C_b\frac{\tilde{p}_k}{(1 + \sum_{l'=1}^{k-1}\frac{\tilde{p}_k}{\tilde{p}_{l'}})\tilde{p}_l} = 2C_b\frac{1}{ \sum_{l'=1}^{k}\frac{\tilde{p}_l}{\tilde{p}_{l'}}} = 2C_b\frac{1}{ \sum_{l'=1}^{k}\frac{|S_l^*|}{|S_{l'}^*|}} \leq 2C_bM/k
	\end{align*}
 
 \noindent under the condition that group sizes $ \{|S_{l}^*|, 1 \leq l \leq k \} $ are bounded above by $M$ and away from zero. Also, observe that $ r_k^* = 2C_b - \sum_{l=1}^{k-1}r_l^* = 2C_b\frac{1}{ \sum_{l'=1}^{k}\frac{\tilde{p}_k}{\tilde{p}_{l'}}} = 2C_b\frac{1}{ \sum_{l'=1}^{k}\frac{|S_k^*|}{|S_{l'}^*|}} \leq 2C_bM/k $. Thus, $ r_l^* \leq 2C_bM/k $ for all $ l \in \{1,\ldots,k\} $, and that concludes the proof of the first part. 
	
\textbf{Part (b):}   From part (a), note that $\displaystyle \min_{\{ g_1, \ldots, g_p\}} \frac{1}{p} \sum_{j=1}^{p}(b_j - \tilde{b}(g_j))^2 \leq \sum_{l=1}^{k}\tilde{p}_l\tilde{r}_l^2 \leq \max_{1 \leq l \leq k}\tilde{r}_l^2\sum_{l=1}^{k}\tilde{p}_l = \max_{1 \leq l \leq k}\tilde{r}_l^2 \leq 4C_b^2M^2/k^2 $ recalling that $ \sum_{l=1}^{k}\tilde{p}_l = 1 $.

\qed

\subsection{Triangular array CLT}
Define the triangular array $ \{S_{in}, 1\leq i\leq n, n \in\mathbb{N}\} $ where $ S_{in} = \tilde{S}_{in}/s_n $, $ \tilde{S}_{in} = \uptau_p'\x_i'\varepsilon_i $, $ s_n^2 = \sum_{i=1}^{n}\uptau_p'\E[\x_i'\x_i\varepsilon_i^2]\uptau_p $, and $\uptau_p \in \mathcal{S}_p$.
\begin{lemma}\label{Lem:CLT}
	Under \Cref{Ass:Sampling,Ass:ZeroCorr_x.eps,Ass:Bnds_x_eps}, $ \sum_{i=1}^{n} S_{in} \xrightarrow{d} \mathcal{N}(0,1)  $.
\end{lemma}
\paragraph{Proof:}
First, $ \E[\tilde{S}_{in}] = \uptau_p'\E[\x_i'\varepsilon_i] = \E[\x_i'\E[\varepsilon_i|\x_i]] = 0 $ by \Cref{Ass:ZeroCorr_x.eps} and the law of iterated expectations (LIE) for each $ i \in \{1,\ldots,n\} $. Second, under independent sampling (\Cref{Ass:Sampling}), $ \E[(\sum_{i=1}^{n} S_{in})^2] = \sum_{i=1}^{n} \E[S_{in}^2] = s_n^2/s_n^2 = 1 $. Third, it suffices to verify the Lyapunov condition $\displaystyle \lim_{n\rightarrow\infty} \sum_{i=1}^{n} \E[|S_{in}|^d] = 0 $ in order to complete the proof. 
 
By the LIE and \Cref{Ass:Bnds_x_eps},
\begin{equation}\label{eqn:Bnd_CLT_xeps}
    \E[\tilde{S}_{in}^2] = \uptau_p'\E[\x_i'\x_i\varepsilon_i^2]\uptau_p = \uptau_p'\E[\x_i'\x_i\E[\varepsilon_i^2|\x_i]]\uptau_p \leq B\rho_{\mathrm{\max}}(\E[\x'\x]) \lesssim 1
\end{equation}

\noindent for each $i\in\{1,\ldots,n\}$. This implies $ s_n = O(n^{1/2}) $. Since $ \E[|\uptau_p'\x_i'|^d] \leq B $ by \Cref{Ass:Bnds_x_eps}, it follows from the LIE and \Cref{Ass:Bnds_x_eps} that $ \E[|\tilde{S}_{in}|^d] = \E[|\uptau_p'\x_i'|^d\E[|\varepsilon_i|^d|\x_i]] \leq B^2<\infty $. Thus, $ \sum_{i=1}^{n} \E[|\tilde{S}_{in}|^d] = O(n) $. Combining the foregoing, $ \sum_{i=1}^{n} \E[|S_{in}|^d] = s_n^{-d}\sum_{i=1}^{n} \E[|\tilde{S}_{in}|^d] = O(n^{-(d/2-1)}) = o(1) $ since $ d>2 $ (\Cref{Ass:Bnds_x_eps}). The conclusion follows from \citet[Theorems 24.6 and 24.11]{davidson2021stochastic}.

\qed

\subsection{Auxiliary Result I}
\noindent The following lemma is used in the proof of \Cref{Theorem:Consis_Rate_M}. Let $ \hat{\varepsilon}_{m_o,i} $ and $ \hat{\varepsilon}_{m,i} $ denote the $i$'th residual from regressing $ \mathcal{E} $ on $ \X\m $ and $ \X\m_o $, respectively.
\begin{lemma}\label{Lem:Aux_mnconsistent}
Fix $ \m \in \mathcal{M}_k$, then under \Cref{Ass:ZeroCorr_x.eps,Ass:Bnds_x_eps,Ass:Sampling}, 
    (a) $ \mathbb{P}\Big(\Big|\sum_{i=1}^{n}(\hat{\varepsilon}_{m_o,i}^2 - \hat{\varepsilon}_{m,i}^2)\Big| \geq 1 \Big) \leq 2kB $; and 
    (b) $\mathbb{P}(|\uptau_p'\mathbb{E}_n[\x_i'\hat{\varepsilon}_{m,i}]| \geq 1) \leq 2B^2/n$ for any $\uptau_p \in \mathcal{S}_p $.
\end{lemma}

\paragraph{Proof:} Observe that $\X\m(\m\X'\X\m)^{-1}\m'\X'$ is a projection matrix with $k$ eigenvalues equal to one and the rest $n-k$ equal to zero \citep[Theorem 3.3]{hansen2022econometrics}. Let $h_{il}^{(m)}$ be $i$'th element of the eigenvector associated with the $l$'th eigenvalue of $\X\m(\m\X'\X\m)^{-1}\m'\X'$.

\textbf{Part (a):} Recall $ \hat{\varepsilon}_{m,i} = \varepsilon_i-\x_i\m\hat{A}_m^{-1}\m'\E_n[\x_i'\varepsilon_i] $. By the foregoing and the eigen-decomposition,    
	\begin{align*}
	\sum_{i=1}^{n}(\hat{\varepsilon}_{m_o,i}^2 - \hat{\varepsilon}_{m,i}^2) &= n\E_n[\varepsilon_i\x_i]\m\hat{A}_m^{-1}\m'\E_n[\x_i'\varepsilon_i] -n\E_n[\varepsilon_i\x_i]\m_o\hat{A}_{m_o}^{-1}\m_o'\E_n[\x_i'\varepsilon_i]\\
	 &= \mathcal{E}'\X\m(\m\X'\X\m)^{-1}\m'\X'\mathcal{E} - \mathcal{E}'\X\m_o(\m_o\X'\X\m_o)^{-1}\m_o'\X'\mathcal{E}\\
    &= \sum_{l=1}^k\big(\sum_{i=1}^n h_{il}^{(m)}\varepsilon_i\big)^2 - \sum_{l=1}^k\big(\sum_{i=1}^n h_{il}^{(m_o)}\varepsilon_i\big)^2.
\end{align*}
\noindent By the triangle inequality and the arguments similar to those of \Cref{Lem:phi1_Op1},
\begin{align*}
    \E\Big[\Big|\sum_{i=1}^{n}(\hat{\varepsilon}_{m_o,i}^2 - \hat{\varepsilon}_{m,i}^2)\Big|\Big] &\leq \sum_{l=1}^k\E\Big[\big(\sum_{i=1}^n h_{il}^{(m)}\varepsilon_i\big)^2\Big] + \sum_{l=1}^k\E\Big[\big(\sum_{i=1}^n h_{il}^{(m_o)}\varepsilon_i\big)^2\Big]\\
    &\leq 2kB
\end{align*}
\noindent under \Cref{Ass:Sampling,Ass:ZeroCorr_x.eps,Ass:Bnds_x_eps}. The conclusion of this part follows from the Markov inequality.

\textbf{Part (b):} $\uptau_p'\mathbb{E}_n[\x_i'\hat{\varepsilon}_{m,i}] = \uptau_p'\mathbb{E}_n[\x_i'\varepsilon_i] - \uptau_p'\mathbb{E}_n[\x_i'\x_i]\m\hat{A}_m^{-1}\m'\E_n[\x_i'\varepsilon_i]$, hence $(\uptau_p'\mathbb{E}_n[\x_i'\hat{\varepsilon}_{m,i}])^2 \leq 2(\uptau_p'\mathbb{E}_n[\x_i'\varepsilon_i])^2 + 2(\uptau_p'\mathbb{E}_n[\x_i'\x_i]\m\hat{A}_m^{-1}\m'\E_n[\x_i'\varepsilon_i])^2$ by the $c_r$-inequality. From \eqref{eqn:Bnd_CLT_xeps}, $\mathrm{var}[\uptau_p'\mathbb{E}_n[\x_i'\varepsilon_i]]\leq n^{-1}B^2$ under \Cref{Ass:Sampling,Ass:ZeroCorr_x.eps,Ass:Bnds_x_eps}. Next, define $\mathcal{P}_n(\m):= \X\m(\m'\X'\X\m)^{-1}\m\X'$ and $\tilde{\uptau}_p:=\X\uptau_p/||\X\uptau_p||$. 
    \begin{align*} \uptau_p'\mathbb{E}_n[\x_i'\x_i]\m\hat{A}_m^{-1}\m'\E_n[\x_i'\varepsilon_i] &= n^{-1} \uptau_p'\X'\X\m(\m'\X'\X\m)^{-1}\m'\X'\mathcal{E}\\
    &= n^{-1}\tilde{\uptau}_p'\X\m(\m'\X'\X\m)^{-1}\m\X'\mathcal{E}\cdot||\X\uptau_p||\\
    &=: n^{-1}\tilde{\uptau}_p'\mathcal{P}_n(\m)\mathcal{E}\cdot||\X\uptau_p||.
    \end{align*}

    Thus by the LIE, \Cref{Ass:ZeroCorr_x.eps,Ass:Sampling,Ass:Bnds_x_eps}, 
    \begin{align*}
        \mathrm{var}[\uptau_p'\mathbb{E}_n[\x_i'\x_i]\m\hat{A}_m^{-1}\m'\E_n[\x_i'\varepsilon_i]] &= n^{-2}\E\big[||\X\uptau_p||^2\uptau_p'\mathcal{P}_n(\m)\E[\mathcal{E}\mathcal{E}'|\X]\mathcal{P}_n(\m)\tilde{\uptau}_p\big]\\
        &\leq n^{-2}B\E\big[||\X\uptau_p||^2\tilde{\uptau}_p'\mathcal{P}_n(\m)\tilde{\uptau}_p\big]\\
        &\leq n^{-1}B\E\big[\uptau_p'\mathbb{E}_n[\x_i'\x_i]\uptau_p\big]\\
        &= n^{-1}B\uptau_p'\E[\x'\x]\uptau_p\\
        &\leq n^{-1}B^2.
    \end{align*}
    \noindent The first inequality uses the \emph{almost sure} bound on the conditional variance (\Cref{Ass:Bnds_x_eps}) and the idempotence property of $\mathcal{P}_n(\m)$. The second inequality follows because $1$ is the maximum eigenvalue of $\mathcal{P}_n(\m)$. The second equality holds by identical distribution (\Cref{Ass:Sampling}), and the last inequality follows from \Cref{Ass:Bnds_x_eps}.

   Taken together,
    \begin{align*}
        \mathbb{P}(|\uptau_p'\mathbb{E}_n[\x_i'\hat{\varepsilon}_{m,i}]| \geq 1) &\leq \mathrm{var}[\uptau_p'\mathbb{E}_n[\x_i'\hat{\varepsilon}_{m,i}]]\\
        &\leq 2\mathrm{var}[\uptau_p'\mathbb{E}_n[\x_i'\varepsilon_i]] + 2\mathrm{var}\big[\uptau_p'\mathbb{E}[\x'\x]\m A_m^{-1}\m'\E_n[\x_i'\varepsilon_i]\big]\\
        &\leq 2B^2/n
    \end{align*}
by the $c_r$- and Chebyshev's inequalities.
\qed

\subsection{Auxiliary Result II}
The following is essential, in addition to \Cref{Theorem:Consis_Rate_M}, in establishing the rate of convergence of GPE $ \widehat{\bbeta}_n:= \hat{\m}_n\hat{\bdelta}_n(\hat{\m}_n) $. Recall $ y_i = \x_i\m_o\bdelta_o + \x_i\bm{b}_k + \varepsilon_i $. Thus for any $\m\in\mathcal{M}_k$ and $\uptau_p\in\mathbb{S}_p$,
\begin{align*}
\uptau_p'\m(\hat{\bdelta}_n(\m_o) - \bdelta_o) = \uptau_p'\m\hat{A}_{m_o}^{-1}\m_o'\E_n[\x_i'\x_i]\bm{b}_k + \uptau_p'\m\hat{A}_{m_o}^{-1}\m_o'\E_n[\x_i'\varepsilon_i].
\end{align*}

\begin{lemma}\label{Lem:Conv_CLT_delta}
Let \Cref{Ass:Sampling,Ass:ZeroCorr_x.eps,Ass:Bnds_x_eps,Ass:Convergence_Am} hold, then, \begin{align*}
\uptau_p'\m(\hat{\bdelta}_n(\m_o) - \bdelta_o) = O_p(n^{-1/2}) + O_p(||\bm{b}_k||)
\end{align*} for any $\m\in\mathcal{M}_k$ and $\uptau_p\in\mathbb{S}_p$.
\end{lemma}

\paragraph{Proof:}

\noindent First, by the triangle inequality,

\begin{equation}\label{eqn:delhat_summands}
    |\uptau_p'\m\hat{A}_{m_o}^{-1}\m_o'\E_n[\x_i'\x_i]\bm{b}_k| \leq \E_n[|\tilde{\uptau}_p'\x_i'\x_i\bm{b}_k|]\cdot||\m A_{m_o}^{-1}\m_o'\uptau_p|| + \E_n[|\tilde{\uptau}_p'\x_i'\x_i\bm{b}_k|]\cdot||\m(\hat{A}_{m_o}^{-1}-A_{m_o}^{-1})\m_o'\uptau_p||
\end{equation}where $\tilde{\uptau}_p:=\m A_{m_o}^{-1}\m_o'\uptau_p/||\m A_{m_o}^{-1}\m_o'\uptau_p||$. Now, by \Cref{Ass:Bnds_x_eps}, \Cref{Ass:Sampling}, and the Cauchy-Schwartz inequality, 
\begin{align*}
    \E\big[\E_n[|\tilde{\uptau}_p'\x_i'\x_i\bm{b}_k|]\big] = \E[|\tilde{\uptau}_p'\x'\x\bm{b}_k|] \leq \sqrt{\E[(\x\tilde{\uptau}_p)^2]\E[(\x\bm{b}_k)^2]}\lesssim ||\bm{b}_k||.
\end{align*}
Thus, $\E_n[|\tilde{\uptau}_p'\x_i'\x_i\bm{b}_k|]\cdot||\m A_{m_o}^{-1}\m_o'|| = O_p(||\bm{b}_k||)$ by the Markov inequality, \Cref{Ass:Bnds_x_eps}, and that group sizes of $\m\in\mathcal{M}_k$ are uniformly bounded from above, viz. $||\m A_{m_o}^{-1}\m_o'||\lesssim 1 $. Observe that the second summand of \eqref{eqn:delhat_summands} is dominated by the first thanks to \Cref{Ass:Convergence_Am}, hence $\uptau_p\m\hat{A}_{m_o}^{-1}\m_o'\E_n[\x_i'\x_i]\bm{b}_k = O_p(||\bm{b}_k||)$ for any $\uptau_p \in \mathbb{S}_p $.

\noindent Second,
\begin{align*}
    \uptau_p'\m\hat{A}_{m_o}^{-1}\m_o'\E_n[\x_i'\varepsilon_i] \leq \tilde{\uptau}_p'\E_n[\x_i'\varepsilon_i]\cdot||\m_o A_{m_o}^{-1}\m'\uptau_p|| + \tilde{\uptau}_p'\E_n[\x_i'\varepsilon_i]\cdot||\m_o(\hat{A}_{m_o}^{-1} - A_{m_o}^{-1})\m'\uptau_p||
\end{align*}where $\tilde{\uptau}_p:=\m_o A_{m_o}^{-1}\m'\uptau_p/||\m_o A_{m_o}^{-1}\m'\uptau_p||$. Following arguments in the part above, $\uptau_p'\m\hat{A}_{m_o}^{-1}\m_o'\E_n[\x_i'\varepsilon_i] = O_p(n^{-1/2}) $ under \Cref{Ass:Sampling,Ass:Bnds_x_eps,Ass:ZeroCorr_x.eps,Ass:Convergence_Am}.
\qed

\subsection{Auxiliary Result III}
\begin{lemma}\label{Lem:Rate_delmn_delmo} 
Under \Cref{Ass:Bounded_Support_B,Ass:ZeroCorr_x.eps,Ass:Sampling,Ass:Bnds_x_eps},
$\uptau_k'(\hat{\bdelta}_n(\hat{\m}_n) - \hat{\bdelta}_n(\m_o)) = O_p(||\bm{b}_k||^2/\sqrt{p}) + O_p(\sqrt{p}/n) $ for all $\uptau_k \in \mathcal{S}_k$.
\end{lemma}

\paragraph{Proof:}
Fixing $k$ for both $\m_o$ and $\hat{\m}_n$, it is easy to see that $\hat{\m}_n$ is a row permutation of the rows of $\m_o$. Therefore, $\hat{\m}_n'\E_n[\x_i'\x_i]\hat{\m}_n = \m_o'\E_n[\x_i'\x_i]\m_o + \bm{\zeta}_m$, where $\bm{\zeta}_m$ is a $k\times k$ perturbation matrix of $\m_o'\E_n[\x_i'\x_i]\m_o$, namely,

\begin{equation}\label{eqn:Zeta_m}
\bm{\zeta}_m = (\hat{\m}_n-\m_o)'\E_n[\x_i'\x_i]\hat{\m}_n +\m_o'\E_n[\x_i'\x_i](\hat{\m}_n-\m_o).
\end{equation}

\noindent Notice that $(\m_o'\E_n[\x_i'\x_i]\m_o)^{-1} - (\hat{\m}_n'\E_n[\x_i'\x_i]\hat{\m}_n)^{-1} =: \hat{A}_{m_o}^{-1} - \hat{A}_{\hat{m}_n}^{-1} = \hat{A}_{\hat{m}_n}^{-1}\bm{\zeta}_m\hat{A}_{m_o}^{-1}$. One then obtains $\hat{\bdelta}_n(\hat{\m}_n) - \hat{\bdelta}_n(\m_o) = -\hat{A}_{\hat{m}_n}^{-1}\bm{\zeta}_m\hat{A}_{m_o}^{-1}\hat{\m}_n'\E_n[\x_i'y_i] + \hat{A}_{m_o}^{-1}(\hat{\m}_n - \m_o)'\E_n[\x_i'y_i].$ Note that $\uptau_p'\E_n[\x_i'y_i] = \uptau_p'\E_n[\x_i'\x_i]\bbeta_o + \uptau_p'\E_n[\x_i'\varepsilon_i] = O_p(\sqrt{p}) + O_p(n^{-1/2}) = O_p(\sqrt{p}) $ under \eqref{eqn:Bnd_CLT_xeps}, \Cref{Ass:Bnds_x_eps,Ass:ZeroCorr_x.eps,Ass:Bounded_Support_B,Ass:Sampling}.

First, since $||\hat{A}_{m_o}^{-1}|| = O_p(1) = ||\hat{A}_{\hat{m}_n}^{-1}||$ under \Cref{Ass:Bnds_x_eps} and that no group is empty, it follows from \eqref{eqn:Zeta_m}, \Cref{Ass:Bnds_x_eps}, and that group sizes are uniformly bounded above for all $\m\in\mathcal{M}_k$  that $||\hat{A}_{\hat{m}_n}^{-1}\bm{\zeta}_m\hat{A}_{m_o}^{-1}|| = O_p(\mathbb{E}[||\hat{\m}_n - \m_o||]) $. Define $\tilde{\uptau}_p:=\hat{\m}_n\hat{A}_{m_o}^{-1}\bm{\zeta}_m'\hat{A}_{\hat{m}_n}^{-1}\uptau_k/||\hat{\m}_n\hat{A}_{m_o}^{-1}\bm{\zeta}_m'\hat{A}_{\hat{m}_n}^{-1}\uptau_k||$, then since group sizes are bounded uniformly from above
\begin{align*}
|\uptau_k'\hat{A}_{\hat{m}_n}^{-1}\bm{\zeta}_m\hat{A}_{m_o}^{-1}\hat{\m}_n'\E_n[\x_i'y_i]| \lesssim |\tilde{\uptau}_p'\E_n[\x_i'y_i]|\cdot||\hat{A}_{m_o}^{-1}\bm{\zeta}_m'\hat{A}_{\hat{m}_n}^{-1}|| = O_p(\sqrt{p}\mathbb{E}[||\hat{\m}_n - \m_o||]).
\end{align*}

\noindent Second, by a similar token, $\uptau_k'\hat{A}_{m_o}^{-1}(\hat{\m}_n - \m_o)'\E_n[\x_i'y_i] = O_p(\sqrt{p}\mathbb{E}[||\hat{\m}_n - \m_o||])$. 

The conclusion follows from \Cref{Theorem:Consis_Rate_M} as $O_p(\sqrt{p}\mathbb{E}[||\hat{\m}_n - \m_o||])= O_p(||\bm{b}_k||^2/\sqrt{p}) + O_p(\sqrt{p}/n)$.

\qed

\subsection{Auxiliary Result IV}
Define the matrix 
\begin{equation}\label{eqn:Cnk}
    \mathcal{C}_n^k:= \big(\m_o^{(k+1)}(\m_o^{(k+1)\prime}\X'\X \m_o^{(k+1)})^{-1}\m_o^{(k+1)\prime} - \m_o^k(\m_o^{k\prime}\X'\X \m_o^k)^{-1}\m_o^{k\prime}\big)
\end{equation}
\noindent and observe that $\X\mathcal{C}_n^k\X'$ is the difference between two projection matrices. Recall the rank of a projection matrix equals its trace and the eigenvalues are either zero or one -- e.g., \citet[Theorem 3.3]{hansen2022econometrics}.

$n\widehat{\varphi}_n(k)$ is non-negative by design uniformly in $k$, thus it remains non-negative under $a_k<2$. This implies that the symmetric $p\times p$ matrix $\X\mathcal{C}_n^k\X'$ in $n\widehat{\varphi}_n(k) = \frac{1}{\mathbb{E}_n[\varepsilon_i^2]}\mathcal{E}'\X\mathcal{C}_n^k\X'\mathcal{E} + o_p(1)$ under $a_k<2$ (using matrix notation), is positive semi-definite $a.s.$ (up to some $o_p(1)$ term) since $\mathcal{E}$ is random. As $\X\m_o^k(\m_o^{k\prime}\X'\X \m_o^k)^{-1}\m_o^{k\prime}\X'$ is a projection matrix, observe that
\[
\mathrm{tr}[\X\mathcal{C}_n^k\X']
    = \mathrm{tr}\big[\X\m_o^{(k+1)}(\m_o^{(k+1)\prime}\X'\X \m_o^{(k+1)})^{-1}\m_o^{(k+1)\prime}\X'\big] - \mathrm{tr}\big[\X\m_o^k(\m_o^{k\prime}\X'\X \m_o^k)^{-1}\m_o^{k\prime}\X'\big] = 1.
\]

\noindent Combining both preceding arguments implies that all eigenvalues of $\X\mathcal{C}_n^k\X'$ are non-negative and they sum up to 1. The following result establishes the stochastic boundedness of $\mathcal{E}'\X\mathcal{C}_n^k\X'\mathcal{E}$.
\begin{lemma}\label{Lem:phi1_Op1}
    Suppose \Cref{Ass:ZeroCorr_x.eps,Ass:Sampling,Ass:Bnds_x_eps} hold, then $\mathcal{E}'\X\mathcal{C}_n^k\X'\mathcal{E} =O_p(1)$.
\end{lemma}

\paragraph{Proof:}
    By the eigen-decomposition, $\X\mathcal{C}_n^k\X' = \bm{H}_n\bm{\Lambda}_n\bm{H}_n'$ where $\mathrm{diag}(\bm{\Lambda}_n) = (\lambda_{1,n}, \lambda_{2,n}, \ldots,\lambda_{n,n})'$, $\mathrm{tr}(\bm{\Lambda}_n) = 1$, $\lambda_{1,n}\geq \lambda_{2,n} \geq \ldots,\lambda_{n,n}\geq 0$, and each column of $\bm{H}_n = (h_{i\ell})_{1\leq i,\ell\leq n} $ has unit Euclidean norm uniformly in $n$. In addition to the exogeneity, independent sampling, and dominance (on the conditional variance $\E[\varepsilon^2|\x]$) conditions  (\Cref{Ass:ZeroCorr_x.eps,Ass:Sampling,Ass:Bnds_x_eps}, respectively),
    \begin{align*}
        0\leq \E[\mathcal{E}'\X\mathcal{C}_n^k\X'\mathcal{E}] &= \sum_{\ell=1}^n\E\Big[\lambda_{\ell,n}(\sum_{i=1}^n h_{i\ell}\varepsilon_i)^2\Big] = \sum_{\ell=1}^n\E\Big[\E[\lambda_{\ell,n}\sum_{i=1}^n\sum_{i'=1}^n h_{i\ell}h_{i'\ell}\varepsilon_i\varepsilon_{i'}|\X]\Big]\\ &= \sum_{\ell=1}^n\E\Big[\lambda_{\ell,n}\sum_{i=1}^n\sum_{i'=1}^n h_{i\ell}h_{i'\ell}\E[\varepsilon_i\varepsilon_{i'}|\X]\Big] = \sum_{\ell=1}^n\E\Big[\lambda_{\ell,n}\sum_{i=1}^n h_{i\ell}^2\E[\varepsilon_i^2|\x_i]\Big]\\
        & \leq B.
    \end{align*}The conclusion follows from the Markov inequality.
\qed

\printbibliography
\end{refsection}

\newpage
\setcounter{page}{1}
	\appendix
	\renewcommand{\thetable}{S.\arabic{table}}
	\renewcommand{\thefigure}{S.\arabic{figure}}
	\renewcommand{\thesection}{S.\arabic{section}}
	\renewcommand{\theequation}{\thesection\arabic{equation}}

\begin{refsection}	
\begin{center}
		\Large{\bf Clustered Covariate Regression:  Online Appendix}
	\end{center}
	\begin{center}
	Abdul-Nasah Soale \& 	Emmanuel Selorm Tsyawo
\end{center}

The Online Appendix contains extra theoretical results, a discussion of models and extensions where GPE is applicable, a taxonomy of support configurations of $\bbeta_o$ as used in the main text, a discussion of the starting scheme employed in \Cref{Alg:Lloyd_GPE}, and extra simulation results.

\section{Sufficient conditions for \Cref{Ass:Convergence_Am}}
Let $\tilde{A}_{o,i}:= \m_o'(\x_i'\x_i - \mathbb{E}[\x_i'\x_i])\m_o $ and $ \{ r_n : n\geq 1 \} $ denote a non-stochastic sequence of positive numbers that satisfies $\displaystyle r_n \geq \max_{1\leq i \leq n} ||\tilde{A}_{o,i}||/\sqrt{p/\log(p)} \ a.s. $ We impose the following sufficient conditions for verifying \Cref{Ass:Convergence_Am}.

\begin{assumption}\label{Ass:Rate_max.x}
$r_n = o(\sqrt{n})$.
\end{assumption}

\begin{assumption}\label{ass:a_tile_bound}
$\E[|\x\uptau_p|^4] \leq B $ for all $\uptau_p \in \mathbb{S}_p$.
\end{assumption}

\noindent \Cref{Ass:Rate_max.x} controls the rate at which $\displaystyle \max_{1\leq i \leq n} ||\tilde{A}_{o,i}||$ is allowed to diverge. \Cref{Ass:Rate_max.x} thus allows $\x$ to have an unbounded support. \Cref{ass:a_tile_bound} strengthens \Cref{Ass:Bnds_x_eps} by requiring the existence of the fourth moment of $|\x\uptau_p|$ for all $\uptau_p \in \mathbb{S}_p$. Put together, \Cref{Ass:Rate_max.x,ass:a_tile_bound} are important sufficient conditions for applying the matrix weak law of large numbers of \citet{tropp2012user} -- cf. \citet[Corollary 4.1]{chen2015optimal}. This is shown in the following result.

\begin{lemma}\label{Lem:Consistency_Am}
\Cref{Ass:Convergence_Am} holds under  \Cref{Ass:Rate_p_n,Ass:Bnds_x_eps,Ass:Sampling,Ass:Rate_max.x,ass:a_tile_bound}.
\end{lemma}

\paragraph{Proof:}	
The proof commences by showing that $||\E[\tilde{A}_o\tilde{A}_o'] || \lesssim p$ under \Cref{ass:a_tile_bound}. Let $\tilde{\x}_o:=\x\m_o$ and $e_l \in \mathbb{S}_k $ whose $l$'th entry equals one and all else is zero. For any $\uptau_k\in\mathbb{S}_k$, define $\mathbb{S}_p \ni \uptau_{op}:= \m_o\uptau_k/||\m_o\uptau_k||$. Also, define $ \mathbb{S}_p \ni e_{ol}:= \m_oe_l/||\m_oe_l||$. Then,

\begin{align*}
    k^{-1}\uptau_k'\E[\tilde{A}_o\tilde{A}_o']\uptau_k &= k^{-1}\uptau_k'\E[\m_o'\x'\x\m_o\m_o'\x'\x\m_o]\uptau_k - k^{-1}\uptau_k'A_{m_o}'A_{m_o}\uptau_k\\
    &\leq k^{-1}\uptau_k'\E[\m_o'\x'\x\m_o\m_o'\x'\x\m_o]\uptau_k=: k^{-1}\uptau_k'\E[\tilde{\x}_o'\tilde{\x}_o\tilde{\x}_o'\tilde{\x}_o]\uptau_k\\
    &= k^{-1}\E[(\tilde{\x}_o\uptau_k)^2||\tilde{\x}_o||^2]\\ 
    &\leq k^{-1}\big( \E[(\tilde{\x}_o\uptau_k)^4] \big)^{1/2} \big( \E[||\tilde{\x}_o||^4] \big)^{1/2}\\
    &= \big( \E[(\tilde{\x}_o\uptau_k)^4] \big)^{1/2}\Big(k^{-2} \sum_{l=1}^k\sum_{l'=1}^k\E[(\tilde{\x}_oe_l)^2(\tilde{\x}_oe_{l'})^2] \Big)^{1/2}\\
    &= \big( \E[(\x\m_o\uptau_k)^4] \big)^{1/2}\Big(k^{-2} \sum_{l=1}^k\sum_{l'=1}^k\E[(\x\m_oe_l)^2(\x\m_oe_{l'})^2] \Big)^{1/2}\\
    &\lesssim \big( \E[(\x\uptau_{op})^4] \big)^{1/2}\Big(k^{-2} \sum_{l=1}^k\sum_{l'=1}^k\E[(\x e_{ol})^2(\x e_{ol'})^2] \Big)^{1/2}\\
    &\leq \big( \E[|\x\uptau_{op}|^4] \big)^{1/2}\Big(k^{-2} \sum_{l=1}^k\sum_{l'=1}^k\sqrt{\E[|\x e_{ol}|^4]\cdot \E[|\x e_{ol'}|^4]} \Big)^{1/2}\leq B.
\end{align*}

\noindent The first inequality follows because $A_{m_o}$ is positive definite under \Cref{Ass:Bnds_x_eps}. The second and last inequalities follow from the Cauchy-Schwartz inequality while the relation ``$\lesssim$" follows because $||\m_o\uptau_k||\lesssim 1$ since group sizes are uniformly bounded from above.

By identical sampling (\Cref{Ass:Sampling}) and \Cref{ass:a_tile_bound}, it follows from the foregoing that $ \varsigma_n^2:= ||\sum_{i=1}^{n} \mathbb{E}[\tilde{A}_{o,i}\tilde{A}_{o,i}']|| = n|| \mathbb{E}[\tilde{A}_o\tilde{A}_o']|| \lesssim np $ noting that $k$ is naturally bounded above by $p$. Let $R_n:=r_n\sqrt{p/\log(p)}$, and observe that $\displaystyle R_n \geq \max_{1\leq i \leq n} ||\tilde{A}_{o,i}|| \ a.s. $ by the definition of $r_n$. By \Cref{Ass:Rate_max.x,ass:a_tile_bound}, it follows from preceding arguments that
\begin{align*}
   \frac{R_n\sqrt{\log (2k)}}{\varsigma_n} \lesssim \frac{r_n\sqrt{p/\log(p)}\sqrt{\log (p)}}{\varsigma_n}  \lesssim \frac{r_n}{\sqrt{n}} = o(1).
\end{align*}
\noindent Combining the foregoing and Corollary 4.1 of \citet{chen2015optimal} implies 
\[||\hat{A}_{m_o} - A_{m_o}|| = ||\mathbb{E}_n[\tilde{A}_{o,i}]|| = O_p\Big(\sqrt{\frac{p\log p}{n}}\Big).\]
The conclusion then follows from \Cref{Ass:Rate_p_n}.

\qed

\section{Calibration of $C$}\label{SubSect:Calibration_gamma}

When $a_k<2$, $||\widehat{\mathcal{B}}_k||=o_p(n^{-1/2})$, and  
\[n\widehat{\varphi}_n(k) = n\Big(\mathbb{E}_n[\varepsilon_i\x_i]\m_o^{(k+1)}\hat{A}_{m_o^{k+1}}^{-1}\m_o^{(k+1)\prime}\mathbb{E}_n[\x_i'\varepsilon_i] - \mathbb{E}_n[\varepsilon_i\x_i]\m_o^k\hat{A}_{m_o^k}^{-1}\m_o^{k\prime}\mathbb{E}_n[\x_i'\varepsilon_i]\Big)/\E_n[\varepsilon_i^2] + o_p(1).
\] The crux of this exercise is to characterise the limiting distribution of $n\widehat{\varphi}_n(k)$ under plausible conditions for which it is asymptotically pivotal when $a_k<2$. $C$ can then be set to a suitable quantile of the resulting limiting distribution.  

The following conditions are imposed.

\begin{condition}[Non-random covariates]\label{Cond:Homoskedasticity}
    $\{\x_i\}_{i=1}^n$ is non-random.
\end{condition}

\noindent To ensure a pivotal limiting distribution, we impose the following condition.

\begin{condition}[Rank $\mathcal{C}_n^k$]\label{Cond:Rank_Ck}
    $\mathrm{rank}[\X\mathcal{C}_n^k\X'] = \mathrm{tr}[\X\mathcal{C}_n^k\X']$ uniformly in $n$.
\end{condition}

\noindent \Cref{Cond:Rank_Ck} is a property of projection matrices. In this particular case, a necessary and sufficient condition for \Cref{Cond:Rank_Ck} is that the principal eigenvectors of the summands $\X\m_o^{(k+1)}(\m_o^{(k+1)\prime}\X'\X \m_o^{(k+1)})^{-1}\m_o^{(k+1)\prime}\X' $ and $ \X\m_o^k(\m_o^{k\prime}\X'\X \m_o^k)^{-1}\m_o^{k\prime}\X'$ in $\X\mathcal{C}_n^k\X'$ are equal \citep[175]{knutson2001honeycombs}. \Cref{Cond:Rank_Ck} and the discussion preceding \Cref{Lem:phi1_Op1} imply that $\X\mathcal{C}_n^k\X'$ has only one non-zero eigenvalue which equals one.

\begin{lemma}\label{Lem:Calibration_gamma}
    Suppose \Cref{Ass:Sampling,Ass:ZeroCorr_x.eps,Ass:Bnds_x_eps} hold and $a_k<2$, i.e.,  $||\widehat{\mathcal{B}}_k||=o_p(n^{-1/2})$. (a) Under \Cref{Cond:Homoskedasticity}, $n\widehat{\varphi}_n(k) \xrightarrow{d} \sum_{i=1}^\infty \lambda_iZ_i^2 $ where $\{Z_i:i\geq 1\}$ is a sequence of standard normally distributed random variables and $\{\lambda_i:i\geq 1\}$ is a sequence of non-negative constants that satisfy $\sum_{i=1}^\infty \lambda_i=1$. (b) Further, suppose \Cref{Cond:Rank_Ck} holds, then $n\widehat{\varphi}_n(k) \xrightarrow{d} \chi_1^2 $.
\end{lemma}

The $\chi_1^2$ limiting distribution of $n\widehat{\varphi}_n(k)$ under $||\widehat{\mathcal{B}}_k||=o_p(n^{-1/2})$ does not depend on $k$. We set $C$ to the $90$'th percentile of the $\chi_1^2$ distribution, namely $C=2.7$ (rounded to the first decimal place).

\paragraph{Proof:}
(a) Following arguments in \Cref{Lem:phi1_Op1} and \Cref{Cond:Homoskedasticity}, the eigen-decomposition
$\X\mathcal{C}_n^k\X' = \bm{H}_n\bm{\Lambda}_n\bm{H}_n'$ holds where $\mathrm{tr}(\bm{\Lambda}_n) = \sum_{i=1}^n \lambda_{i,n} = 1$, each eigenvalue is non-negative, i.e., $\lambda_{1,n}\geq\lambda_{2,n}\geq\ldots,\lambda_{n,n}\geq 0$, and each column of $\bm{H}_n = (h_{ii'})_{1\leq i,i'\leq n} $ has unit Euclidean norm uniformly in $n$. Thus, by \Cref{Cond:Homoskedasticity}, $\mathcal{E}'\X\mathcal{C}_n^k\X'\mathcal{E} = \sum_{i'=1}^n\lambda_{i',n}(\sum_{i=1}^n h_{ii'}\varepsilon_i)^2 $ where $\sigma^{-1}\sum_{i=1}^n h_{ii'}\varepsilon_i \rightarrow_d \mathcal{N}(0,1) $ for each $i'\in\{1,\ldots,n\}$ under the conditions of \Cref{Lem:CLT} (by replacing $\uptau_p'\x_i/s_n$ with $\sigma^{-1}h_{ii'}$). The foregoing implies that under $a_k<2$,
\[
    n\widehat{\varphi}_n(k) \xrightarrow{d} \sum_{i=1}^\infty \lambda_iZ_i^2.
\]

(b) Combining arguments in \Cref{Lem:phi1_Op1} and \Cref{Cond:Rank_Ck}, $\lambda_{1,n} = 1$ and $\lambda_{i,n}=0$ for all $i\geq 2$ uniformly in $n$. The conclusion follows from part (a) above.
\qed

\section{Applicable models and extensions}
As \eqref{eqn:basemodel} is quite generic, it is instructive to outline a few models, extensions, and empirical contexts where GPE is applicable. These are scenarios where high dimensionality commonly arises.
\paragraph{Non-parametric functions:} Consider the conditional mean function $\E[y|\z]\approx P(\z)\bbeta_o$ with a dictionary of transformations $\x:=P(\z)$ of elementary variables $\z$, e.g., the demand for gasoline as a function of price, income, and household characteristics. \Cref{Sect:Empirical_Application} of this paper, \citet{semenova2021debiased,chernozhukov2022biased} provide flexible versions of the demand function estimated in  \citet{yatchew2001household}.

\paragraph{Treatment effect heterogeneity:} \citet[Sect. 7.6]{imbens2015causal}, \citet{sun2022estimation}, and \citet{semenovainference} consider specification \eqref{eqn:basemodel} with $\x: = (D,D\cdot \z, P(\z))$ which comprises a binary treatment variable $D$, a set of (transformations of) individual characteristics $\z$, and their interaction with the binary treatment variable in order to accommodate heterogeneity in the effect of treatment. 

\paragraph{Fixed effects in panel data models:} In large $N$ and large $T$ two-way fixed effects panel data models, the number of fixed effects easily becomes large. \citet{bonhomme2015grouped} considers this kind of problem using the clustering-based Grouped Fixed Effects estimator. GPE is applicable to this scenario if grouped elements in $\x$ comprise only unit and time dummies.

\section{A taxonomy of support configurations}\label{App:Sect:Taxonomy_Config}
This subsection elaborates a taxonomy of the various configurations of $ \mathrm{supp}(\bbeta_o)$, i.e., the support of $\bbeta_o$, for reference purposes. The configuration of $ \mathrm{supp}(\bbeta_o)$ considered in this paper is defined along two dimensions: (non)-sparsity and heterogeneity. The degree of (non)-sparsity can be (1) sparse, (2) approximately sparse, or (3) non-sparse while the degree of heterogeneity can be (1) discrete, (2) mixed, or (3) continuous. 

The degrees of parameter sparsity are defined as the following.
\begin{enumerate}
    \item \emph{Sparse}: $\mathrm{supp}(\bbeta_o)$ contains a zero atom, i.e., a non-vanishing fraction of elements in $\bbeta_o$ are zero uniformly in $p$.
    \item \emph{Approximately sparse}: zero is either a limit point or it falls within the lower and upper bounds of $\mathrm{supp}(\bbeta_o)$ without being an atom. 
    \item \emph{Non-sparse}: $\mathrm{supp}(\bbeta_o)$ is bounded away from zero.
\end{enumerate}

The degrees of parameter heterogeneity are defined as the following.
\begin{enumerate}
    \item \emph{Discrete}: the number of support points of $\bbeta_o$ is fixed even if $p\rightarrow\infty$.
    \item \emph{Mixed}: $\bbeta_o$ contains at least one atom although the number of support points of $\bbeta_o$ is increasing in $p$.
    \item \emph{Continuous}: $\bbeta_o$ contains no atom.
\end{enumerate}
The following table provides examples of configurations  of $\mathrm{supp}(\bbeta_o)$.

\begin{table}[!htbp]
	\setlength{\tabcolsep}{2pt}
	\caption{Configuration of $ \mathrm{supp}(\bbeta_o)$ - Examples}
\begin{tabular}{@{}lllll@{}}
    &  & \multicolumn{3}{c}{Sparsity} \\ \cmidrule(l){3-5}
    &  & Sparse & Approximately Sparse & Non-sparse \\ \cmidrule(l){3-5}
    \parbox[t]{3mm}{\multirow{5}{*}{\rotatebox[origin=c]{90}{Heterogeneity}}}
    & \multicolumn{1}{|l|}{Discrete}   & $\beta_{oj}=\mathrm{I}(j\leq\bar{s})$        & $\beta_{oj}=\mathrm{I}(j\leq\bar{s}) + \alpha^p$ & $\beta_{oj}=\mathrm{I}(j\leq\bar{s}) + \underline{b}$ \\
\multicolumn{1}{l}{} & \multicolumn{1}{|l|}{} &  & & \\
\multicolumn{1}{l}{} & \multicolumn{1}{|l|}{Mixed} & $\beta_{oj}=\mathrm{I}(j\leq s)\cdot\Phi^{-1}(\tau_j)$ & $\beta_{oj}=\mathrm{I}(j\leq s)\cdot|\Phi^{-1}(\tau_j)| + \alpha^p $ & $\beta_{oj}=\mathrm{I}(j\leq s)\cdot|\Phi^{-1}(\tau_j)| + \underline{b}$ \\
\multicolumn{1}{l}{} & \multicolumn{1}{|l|}{} & & & \\ 
\multicolumn{1}{l}{} & \multicolumn{1}{|l|}{Continuous} &  & $\beta_{oj}=\alpha^{j-1}$ or $\beta_{oj}=\Phi^{-1}(\tau_j)$ & $\beta_{oj}=\bar{b}(j-1)/(p-1)+\underline{b}$ \\
\bottomrule
\end{tabular}
\label{Tab:Config_beta}

{\footnotesize
    \textit{Notes}: $\bar{s}<p$ is a fixed natural number. The natural number $s$ satisfies $0<s/p<1$. $\Phi^{-1}(\tau_j)$ is the $\tau_j$'th quantile of the standard normal distribution, $ \tau_j = 0.9(j-1)/(p-1) + 0.05 $, $0<\alpha<1$, and $\bar{b} > \underline{b} > 0$.}
\end{table}

\Cref{Tab:Config_beta} illustrates different relevant configurations in a $p\rightarrow\infty$ setting. Observe that a continuous-sparse configuration does not exist as sparsity requires that zero be an atom.

\section{Starting schemes of \Cref{Alg:Lloyd_GPE}}\label{SubSect:Starting_Schemes}
Instead of employing random starting schemes, our experiments show a faster and better convergence when informative preliminary estimates are used as starting values. Specifically, we use a model-based subgroup analysis in which a penalty function is applied to the pairwise differences of the model parameters. A modified alternating direction method of multipliers (ADMM) algorithm of \citet{boyd2011distributed} is used to solve the constrained optimisation to generate the starting values for \Cref{Alg:Lloyd_GPE}. To find the initial subgroups, we minimise the objective function
\begin{align*} \label{admm_obj}
    &\cfrac{1}{2} \displaystyle\sum_{i=1}^n (y_i - \x_i\bbeta )^2 + \displaystyle\sum_{1\leq j < j' \leq n} p_\gamma (\eta_{jj'}, \lambda) \text{ subject to } \beta_j - \beta_{j'} = \eta_{jj'},
    \numberthis
\end{align*}
where $p_\gamma (\eta_{jj'}, \lambda)$ is a minimax concave penalty (MCP) and $\eta_{jj'}$ is a splitting variable. For $t > 0$ and $\gamma > 1$, we define 
\begin{align*}
    p_\gamma (t, \lambda) = \lambda \int_{0}^t \left(1 - \cfrac{u}{\gamma\lambda} \right )_{+} du,
\end{align*}
where $(a)_+ = \max\{0,a\}$ and $\gamma$ controls the level of concavity. As $\gamma \to \infty$, MCP converges to the $\ell_1$ penalty. Therefore, like LASSO, MCP initially applies the same rate of penalisation and then smoothly relaxes the rate down to zero as the absolute value of the input increases. Hence, MCP avoids unnecessary shrinkage of large coefficients.

We solve \ref{admm_obj} through the augmented Lagrangian
\begin{align*}\label{admm_lag}
    \bm{L}_\omega(\bbeta, \bm\eta, \bm\nu) &= \cfrac{1}{2} \sum_{i=1}^n (y_i - \x_i\bbeta )^2 + \sum_{1\leq j < j' \leq p} p_\gamma (\eta_{jj'}, \lambda) + \sum_{1\leq j < j' \leq p} \nu_{jj'}(\beta_j - \beta_{j'} - \eta_{jj'})\\
    &+ \cfrac{\omega}{2} \sum_{1\leq j < j' \leq p} \lVert \beta_j - \beta_{j'} - \eta_{jj'} \rVert^2,
    \numberthis
\end{align*}
where $\bm\nu \in \R^d$ is the dual variable and $\omega \geq 0$ is a tuning parameter. Let $\bm\eta = \{\eta_{jj'}, 1\leq j < j' \leq p\} \in \R^{d}$ with $d = p(p-1)/2$, $e_j \in \R^p$ be a vector whose $j$th element is 1 and 0 elsewhere, and denote $\bm A = \{e_j - e_{j'}, 1 \leq j < j' \leq p \}$.

ADMM minimises the augmented Lagrangian using the following algorithm:
\begin{itemize}
    \item Step 1: Set the starting values $\widehat{\bbeta}^{(0)} = (\hat\beta^{(0)}_1,\ldots, \hat\beta^{(0)}_p)$, $\hat\eta_{jj'}^{(0)} = \hat\beta^{(0)}_j - \hat\beta^{(0)}_{j'}$ and $\widehat{\bm\nu}^{(0)} = 0$.

    \item Step 2: Given $\widehat\bbeta^{(s)}, \widehat{\bm\eta}^{(s)}$, and $\widehat{\bm\nu}^{(s)}$, in the $(s+1)$th iteration, update 
    \begin{align*}
        \widehat\bbeta^{(s+1)} &= \argmin_{\bbeta} \bm{L}_\omega(\bbeta, \widehat{\bm\eta}^{(s)}, \widehat{\bm\nu}^{(s)}) = (\omega \bm A'\bm A + \X'\X)^{-1}[\omega\bm A'(\widehat{\bm\eta}^{(s)} - \omega^{-1}\widehat{\bm\nu}^{(s)})  + \X'Y],  \\
        \widehat{\bm\eta}^{(s+1)} &= \argmin_{\bm\eta} \bm{L}_\omega(\widehat{\bm\beta}^{(s+1)}, \bm\eta, \widehat{\bm\nu}^{(s)}) = \begin{cases}
             \cfrac{\gamma}{\gamma - 1/\omega}\tilde{f}\big(\widehat{\bm\xi}^{(s+1)},\lambda/\omega\big), \text{ if } \lVert \widehat{\bm\xi}^{(s+1)} \rVert \leq \gamma\lambda \\
             \widehat{\bm\xi}^{(s+1)}, \text{ otherwise},
         \end{cases},\\
         &\text{where } \tilde{f}(z,a) = z(1-a/\lVert z \rVert)_+ \text{ and } \widehat{\xi}^{(s+1)}_{jj'} = \hat\beta^{(s+1)}_j - \hat\beta^{(s+1)}_{j'} -\omega^{-1}\hat{\nu}_{jj'}^{(s)}, \\
        \widehat{\bm{\nu}}^{(s+1)} &= \widehat{\bm{\nu}}^{(s)} + \omega(\hat\beta^{(s+1)}_j - \hat\beta^{(s+1)}_{j'}).
    \end{align*}

    \item Step 3: Stop when $\lVert \bm A\widehat\bbeta^{(s+1)} - \widehat{\bm\eta}^{(s+1)}\rVert < tol$ for some small $tol>0$ else return to Step 2. 
    
\end{itemize}

\noindent We set $(\gamma,\lambda,\omega) = (2,1,1)$ throughout the implementation of the above algorithm in both simulations and the empirical application. The comparable performance of GPE and Orac.GPE in \Cref{Tab:DGP_CaS1,Tab:DGP_CaS2,Tab:DGP_CnS,Tab:DGP_D_S1,Tab:DGP_D_S2,Tab:DGP_DaS2,Tab:DGP_DnS,Tab:DGP_M_S,Tab:DGP_MnS} suggests the starting scheme is reliable and robust to different configurations of $\bbeta_o$.

\section{Additional Simulation Results}\label{Sect:Additional_Details}

This section provides mixed and discrete heterogeneity DGPs to complement those in \Cref{Sect_MC_Experiments} of the main text. 

\begin{enumerate}
	\item[] DGP MnS: $ \beta_{oj} = \mathrm{I}(j\leq 5)|\Phi^{-1}(\tau_j)| + 0.1$;
        \item[] DGP M-S: $ \beta_{oj} = \mathrm{I}(j\leq 5)\Phi^{-1}(\tau_j)$;
        \item[] DGP DnS: $\beta_{oj} = \mathrm{I}(j\leq 5) + 0.1 $; 
        \item[] DGP D-S$_2$: $\beta_{oj} = \mathrm{I}(j\leq \ceil{p/2}) $; and 
	\item[] DGP $\mathrm{DaS_2}$: $ \beta_{oj} = \mathrm{I}(j\leq 5) + F_{\chi_1^2}^{-1}(\tau_j)/\sqrt{2n} $ where $F_{\chi_1^2}^{-1}(\tau)$ denotes the $\tau$'th quantile of the $\chi_1^2$ distribution.
\end{enumerate}

\noindent A few details on the above DGPs are worth highlighting. First, although sparse, DGP D-S$_2$ may contain too many relevant covariates relative to sample size $n$, especially at small $n$. The goal is to examine the finite sample performance of pLASSO, pGDS, and GPE under such a configuration. Second, DGP $\mathrm{DaS_2}$ introduces a $O(n^{-1/2})$ perturbation term in $\bbeta_o$ in order to evaluate the performance of estimators, including GPE, and the sensitivity of the GPE selection rule to ``local" violations of discrete heterogeneity. This exercise also serves to check the sensitivity of pLASSO and pGDS to such ``local" violations.

\begin{table}[!htbp]
	\setlength{\tabcolsep}{4pt}
	\caption{DGP MnS}
	\footnotesize
	\centering
	\begin{minipage}{.5\linewidth}
		\label{Tab:DGP_MnS}
		\begin{tabular}{lccccc}
			\toprule
			& \multicolumn{5}{c}{$ p = 75 $}  \\ \cmidrule(rl){2-6}
			$ n=100 $ & MnB & MAD & RMSE & Rej. & $ \mathrm{med}(\hat{k}_n) $ \\
			\midrule
GPE   &0.034 &0.037 &0.061 &0.040 & 4    \\ 
pLASSO &0.120  &0.781  &0.773  &1.000  & 5     \\ 
pGDS  &0.141 &0.304 &0.319 &0.910 &28    \\ 
OLS   &0.259 &0.070 &0.118 &0.000 &75    \\ 
Orac.OLS &0.085    &0.028    &0.039    &0.019    &75       \\ 
Orac.GPE &0.030    &0.036    &0.061    &0.039    & 3       \\
                
			\midrule
			$ n=400 $ &  &  &  &  &  \\
			\midrule
GPE   &0.022 &0.020 &0.030 &0.042 & 3    \\ 
pLASSO &0.117  &0.605  &0.610  &1.000  &10     \\ 
pGDS  &0.085 &0.083 &0.095 &0.686 &52    \\ 
OLS   &0.071 &0.023 &0.033 &0.030 &75    \\ 
Orac.OLS &0.038    &0.012    &0.018    &0.039    &75       \\ 
Orac.GPE &0.017    &0.020    &0.030    &0.039    & 4       \\

			\bottomrule
		\end{tabular}
	\end{minipage}%
	\begin{minipage}{.5\linewidth}
		\begin{tabular}{rccccc}
			\toprule
			& \multicolumn{5}{c}{$ p = 150 $}  \\ \cmidrule(rl){2-6}
			& MnB & MAD & RMSE & Rej. & $ \mathrm{med}(\hat{k}_n) $ \\
			\midrule
			
 &0.260 &0.099 &0.146 &0.058 &  2   \\ 
 &0.129  &1.113  &1.106  &1.000  &  6    \\ 
 &0.149 &0.663 &0.665 &0.999 & 35   \\ 
 &--  &--  &--  &--  &--  \\ 
 &0.104    &0.031    &0.048    &0.004    &150      \\ 
 &0.017    &0.038    &0.058    &0.037    &  2      \\

			\midrule
			&  &  &  &  &  \\
			\midrule
			
 &0.015 &0.020 &0.029 &0.053 &  3   \\ 
 &0.121  &0.861  &0.863  &1.000  & 17    \\ 
 &0.099 &0.275 &0.283 &1.000 & 76   \\ 
 &0.081 &0.025 &0.037 &0.006 &150   \\ 
 &0.040    &0.012    &0.018    &0.038    &150      \\ 
 &0.011    &0.020    &0.029    &0.055    &  3      \\

			\bottomrule
		\end{tabular}
	\end{minipage}
\end{table}

\begin{table}[!htbp]
	\setlength{\tabcolsep}{4pt}
	\caption{DGP M-S}
	\footnotesize
	\centering
	\begin{minipage}{.5\linewidth}
		\label{Tab:DGP_M_S}
		\begin{tabular}{lccccc}
			\toprule
			& \multicolumn{5}{c}{$ p = 75 $}  \\ \cmidrule(rl){2-6}
			$ n=100 $ & MnB & MAD & RMSE & Rej. & $ \mathrm{med}(\hat{k}_n) $ \\
			\midrule
GPE   &0.037 &0.037 &0.062 &0.042 & 4    \\ 
pLASSO &0.032  &0.013  &0.021  &0.046  & 5     \\ 
pGDS  &0.058 &0.029 &0.047 &0.023 &16    \\ 
OLS   &0.259 &0.070 &0.118 &0.000 &75    \\ 
Orac.OLS &0.085    &0.028    &0.039    &0.019    &75       \\ 
Orac.GPE &0.030    &0.037    &0.062    &0.046    & 3       \\
                
			\midrule
			$ n=400 $ &  &  &  &  &  \\
			\midrule
GPE   &0.017 &0.020 &0.030 &0.041 & 4    \\ 
pLASSO &0.016  &0.007  &0.010  &0.044  & 5     \\ 
pGDS  &0.031 &0.015 &0.023 &0.037 &21    \\ 
OLS   &0.071 &0.023 &0.033 &0.030 &75    \\ 
Orac.OLS &0.038    &0.012    &0.018    &0.039    &75       \\ 
Orac.GPE &0.017    &0.020    &0.030    &0.043    & 4       \\

			\bottomrule
		\end{tabular}
	\end{minipage}%
	\begin{minipage}{.5\linewidth}
		\begin{tabular}{rccccc}
			\toprule
			& \multicolumn{5}{c}{$ p = 150 $}  \\ \cmidrule(rl){2-6}
			& MnB & MAD & RMSE & Rej. & $ \mathrm{med}(\hat{k}_n) $ \\
			\midrule
			
 &0.208 &0.076 &0.112 &0.044 &  3   \\ 
 &0.023  &0.009  &0.016  &0.055  &  5    \\ 
 &0.049 &0.025 &0.042 &0.039 & 23   \\ 
 &--  &--  &--  &--  &--  \\ 
 &0.104    &0.031    &0.048    &0.004    &150      \\ 
 &0.015    &0.037    &0.058    &0.039    &  3      \\

			\midrule
			&  &  &  &  &  \\
			\midrule
			
 &0.014 &0.020 &0.029 &0.052 &  3   \\ 
 &0.011  &0.005  &0.007  &0.039  &  5    \\ 
 &0.023 &0.011 &0.018 &0.034 & 27   \\ 
 &0.081 &0.025 &0.037 &0.006 &150   \\ 
 &0.040    &0.012    &0.018    &0.038    &150      \\ 
 &0.009    &0.020    &0.029    &0.052    &  3      \\

			\bottomrule
		\end{tabular}
	\end{minipage}
\end{table}

\begin{table}[!htbp]
	\setlength{\tabcolsep}{4pt}
	\caption{DGP DnS}
	\footnotesize
	\centering
	\begin{minipage}{.5\linewidth}
		\label{Tab:DGP_DnS}
		\begin{tabular}{lccccc}
			\toprule
			& \multicolumn{5}{c}{$ p = 75 $}  \\ \cmidrule(rl){2-6}
			$ n=100 $ & MnB & MAD & RMSE & Rej. & $ \mathrm{med}(\hat{k}_n) $ \\
			\midrule
GPE   &0.014 &0.040 &0.059 &0.047 & 2    \\ 
pLASSO &0.122  &0.783  &0.778  &1.000  & 5     \\ 
pGDS  &0.140 &0.350 &0.374 &0.984 &24    \\ 
OLS   &0.263 &0.080 &0.121 &0.001 &75    \\ 
Orac.OLS &0.086    &0.027    &0.040    &0.029    &75       \\ 
Orac.GPE &0.014    &0.040    &0.059    &0.043    & 2       \\ 
                
			\midrule
			$ n=400 $ &  &  &  &  &  \\
			\midrule
GPE   &0.007 &0.018 &0.029 &0.053 & 2    \\ 
pLASSO &0.117  &0.603  &0.610  &1.000  &10     \\ 
pGDS  &0.085 &0.080 &0.091 &0.668 &53    \\ 
OLS   &0.071 &0.021 &0.033 &0.024 &75    \\ 
Orac.OLS &0.038    &0.012    &0.018    &0.050    &75       \\ 
Orac.GPE &0.007    &0.019    &0.029    &0.053    & 2       \\
			\bottomrule
		\end{tabular}
	\end{minipage}%
	\begin{minipage}{.5\linewidth}
		\begin{tabular}{rccccc}
			\toprule
			& \multicolumn{5}{c}{$ p = 150 $}  \\ \cmidrule(rl){2-6}
			& MnB & MAD & RMSE & Rej. & $ \mathrm{med}(\hat{k}_n) $ \\
			\midrule
			
 &0.197 &0.108 &0.161 &0.065 &  2   \\ 
 &0.132  &1.109  &1.106  &1.000  &  6    \\ 
 &0.148 &0.672 &0.677 &1.000 & 33   \\ 
 &--  &--  &--  &--  &--  \\ 
 &0.105    &0.034    &0.049    &0.007    &150      \\ 
 &0.009    &0.042    &0.060    &0.048    &  2      \\

			\midrule
			&  &  &  &  &  \\
			\midrule
			
 &0.004 &0.021 &0.030 &0.057 &  2   \\ 
 &0.121  &0.860  &0.865  &1.000  & 17    \\ 
 &0.099 &0.275 &0.282 &1.000 & 76   \\ 
 &0.081 &0.027 &0.039 &0.019 &150   \\ 
 &0.040    &0.011    &0.017    &0.027    &150      \\ 
 &0.005    &0.021    &0.030    &0.057    &  2      \\

			\bottomrule
		\end{tabular}
	\end{minipage}
\end{table}

\begin{table}[!htbp]
	\setlength{\tabcolsep}{4pt}
	\caption{DGP D-S$_2$}
	\footnotesize
	\centering
	\begin{minipage}{.5\linewidth}
		\label{Tab:DGP_D_S2}
		\begin{tabular}{lccccc}
			\toprule
			& \multicolumn{5}{c}{$ p = 75 $}  \\ \cmidrule(rl){2-6}
			$ n=100 $ & MnB & MAD & RMSE & Rej. & $ \mathrm{med}(\hat{k}_n) $ \\
			\midrule
GPE   &0.016 &0.040 &0.059 &0.046 & 3    \\ 
pLASSO &0.393  &0.440  &0.497  &0.933  &32     \\ 
pGDS  &0.376 &0.354 &0.399 &0.706 &36    \\ 
OLS   &0.263 &0.080 &0.121 &0.001 &75    \\ 
Orac.OLS &0.086    &0.027    &0.040    &0.029    &75       \\ 
Orac.GPE &0.010    &0.040    &0.059    &0.048    & 2       \\
                
			\midrule
			$ n=400 $ &  &  &  &  &  \\
			\midrule
GPE   &0.006 &0.019 &0.029 &0.051 & 2    \\ 
pLASSO &0.048  &0.014  &0.021  &0.037  &38     \\ 
pGDS  &0.058 &0.019 &0.029 &0.038 &57    \\ 
OLS   &0.071 &0.021 &0.033 &0.024 &75    \\ 
Orac.OLS &0.038    &0.012    &0.018    &0.050    &75       \\ 
Orac.GPE &0.005    &0.019    &0.029    &0.049    & 2       \\

			\bottomrule
		\end{tabular}
	\end{minipage}%
	\begin{minipage}{.5\linewidth}
		\begin{tabular}{rccccc}
			\toprule
			& \multicolumn{5}{c}{$ p = 150 $}  \\ \cmidrule(rl){2-6}
			& MnB & MAD & RMSE & Rej. & $ \mathrm{med}(\hat{k}_n) $ \\
			\midrule
			
 &0.556 &0.361 &0.564 &0.053 &  3   \\ 
 &0.621  &1.505  &1.567  &1.000  & 48    \\ 
 &0.617 &1.328 &1.382 &0.954 & 61   \\ 
 &--  &--  &--  &--  &--  \\ 
 &0.105    &0.034    &0.049    &0.007    &150      \\ 
 &0.008    &0.042    &0.060    &0.055    &  2      \\

			\midrule
			&  &  &  &  &  \\
			\midrule
			
 &0.004 &0.021 &0.030 &0.058 &  2   \\ 
 &0.050  &0.015  &0.024  &0.037  & 75    \\ 
 &0.069 &0.025 &0.044 &0.104 & 85   \\ 
 &0.081 &0.027 &0.039 &0.019 &150   \\ 
 &0.040    &0.011    &0.017    &0.027    &150      \\ 
 &0.004    &0.021    &0.030    &0.057    &  2      \\

			\bottomrule
		\end{tabular}
	\end{minipage}
\end{table}

\begin{table}[!htbp]
	\setlength{\tabcolsep}{4pt}
	\caption{DGP DaS$_2$}
	\footnotesize
	\centering
	\begin{minipage}{.5\linewidth}
		\label{Tab:DGP_DaS2}
		\begin{tabular}{lccccc}
			\toprule
			& \multicolumn{5}{c}{$ p = 75 $}  \\ \cmidrule(rl){2-6}
			$ n=100 $ & MnB & MAD & RMSE & Rej. & $ \mathrm{med}(\hat{k}_n) $ \\
			\midrule
GPE   &0.239 &0.083 &0.122 &0.080 & 2    \\ 
pLASSO &0.112  &0.477  &0.468  &1.000  & 5     \\ 
pGDS  &0.109 &0.206 &0.222 &0.887 &17    \\ 
OLS   &0.263 &0.080 &0.121 &0.001 &75    \\ 
Orac.OLS &0.086    &0.027    &0.040    &0.029    &75       \\ 
Orac.GPE &0.050    &0.048    &0.071    &0.054    & 4       \\ 
                
			\midrule
			$ n=400 $ &  &  &  &  &  \\
			\midrule
GPE   &0.033 &0.023 &0.033 &0.056 & 2    \\ 
pLASSO &0.049  &0.247  &0.243  &1.000  & 5     \\ 
pGDS  &0.049 &0.071 &0.080 &0.750 &22    \\ 
OLS   &0.071 &0.021 &0.033 &0.024 &75    \\ 
Orac.OLS &0.038    &0.012    &0.018    &0.050    &75       \\ 
Orac.GPE &0.026    &0.021    &0.031    &0.053    & 3       \\ 

			\bottomrule
		\end{tabular}
	\end{minipage}%
	\begin{minipage}{.5\linewidth}
		\begin{tabular}{rccccc}
			\toprule
			& \multicolumn{5}{c}{$ p = 150 $}  \\ \cmidrule(rl){2-6}
			& MnB & MAD & RMSE & Rej. & $ \mathrm{med}(\hat{k}_n) $ \\
			\midrule
			
 &0.202 &0.099 &0.162 &0.070 &  2   \\ 
 &0.113  &0.637  &0.637  &1.000  &  6    \\ 
 &0.113 &0.337 &0.348 &0.991 & 28   \\ 
 &--  &--  &--  &--  &--  \\ 
 &0.105    &0.034    &0.049    &0.007    &150      \\ 
 &0.043    &0.051    &0.079    &0.049    &  5      \\ 

			\midrule
			&  &  &  &  &  \\
			\midrule
			
 &0.032 &0.024 &0.035 &0.052 &  2   \\ 
 &0.048  &0.343  &0.336  &1.000  &  5    \\ 
 &0.048 &0.120 &0.125 &0.975 & 34   \\ 
 &0.081 &0.027 &0.039 &0.019 &150   \\ 
 &0.040    &0.011    &0.017    &0.027    &150      \\ 
 &0.026    &0.022    &0.033    &0.050    &  3      \\

			\bottomrule
		\end{tabular}
	\end{minipage}
\end{table}

\Cref{Tab:DGP_MnS,Tab:DGP_M_S,Tab:DGP_DnS,Tab:DGP_D_S2,Tab:DGP_DaS2} present simulation results corresponding to the DGPs introduced above. The results generally agree with those in \Cref{Sect_MC_Experiments} of the main text. GPE is the only feasible estimator that controls size meaningfully while having low bias across all DGPs. Sparsity-dependent pLASSO and pGDS perform well under sparsity (\Cref{Tab:DGP_M_S}) while their performance deteriorates even under ``local" $O(n^{-1/2})$ violations of sparsity (\Cref{Tab:DGP_DaS2}) -- pLASSO, for example, has empirical rejection rates equal to 1 for all $(n,p)$ pairs in \Cref{Tab:DGP_DaS2} although the tuning parameter appears robust to the $O(n^{-1/2})$ perturbation as it selects the correct number of relevant variables with non-decaying parameters. Although GPE's selection rule also does not appear to be sensitive to the $O(n^{-1/2})$-perturbation, GPE's performance, unlike pLASSO or pGDS, remains robust. \Cref{Tab:DGP_D_S2} suggests that a very large number of relevant covariates at small sample sizes can hurt the performance of pLASSO and pGDS as the rate condition $s^2(\log(p\vee n))^2/n = o(1), s = \ceil{p/2} $ is a poor approximation at $n=100$ and $p\in\{75,150\}$ -- see \citet[p. 2398]{belloni2012sparse} for a similar discussion.

\printbibliography
\end{refsection}
\end{document}